\documentclass[12pt]{article}
\pdfoutput=1
\usepackage{amsmath,amssymb,amsfonts}
\usepackage{a4wide,epsfig,psfrag,scalefnt}
\usepackage[dvipsnames]{xcolor}
\usepackage{braket}
\usepackage{placeins}
\usepackage{breqn}
\usepackage{slashed}
\usepackage{enumitem}
\usepackage[numbers,sort&compress]{natbib}
\usepackage{caption}
\usepackage{subcaption}

\allowdisplaybreaks

\begin{document}

\title{\vskip-3cm{\baselineskip14pt
    \begin{flushleft}
     \normalsize LTH 1392, P3H-25-007, PSI-PR-25-02, TTP25-001, ZU-TH 06/25 
    \end{flushleft}} \vskip1.5cm
  Analytic next-to-leading order Yukawa and Higgs boson self-coupling
  corrections to $gg\to HH$ \\ at high energies}
 
\author{
  Joshua Davies$^{a}$,
  Kay Sch\"onwald$^{b}$,
  Matthias Steinhauser$^{c}$,
  Hantian Zhang$^{d}$
  \\
  {\small\it (a) Department of Mathematical Sciences, University of
    Liverpool,}
  {\small\it Liverpool, L69 3BX, UK}
  \\
  {\small\it (b) Physik-Institut, Universit\"at Z\"urich, Winterthurerstrasse 190,}\\
  {\small\it 8057 Z\"urich, Switzerland}
  \\
  {\small\it (c) Institut f{\"u}r Theoretische Teilchenphysik,
    Karlsruhe Institute of Technology (KIT),}\\
  {\small\it Wolfgang-Gaede Stra\ss{}e 1, 76131 Karlsruhe, Germany}
  \\
  {\small\it (d) PSI Center for Neutron and Muon Sciences, 5232 Villigen PSI, Switzerland}\\
}

\date{}

\maketitle

\thispagestyle{empty}

\begin{abstract}

  We consider electroweak corrections to Higgs boson pair production, taking
  into account the top quark Yukawa and Higgs boson self couplings. Using
  differential equations we compute a deep expansion of all master integrals
  in the high-energy limit and present analytic results for the two-loop
  box-type form factors.  We show that precise numerical results can be obtained
  even for relatively small values of the Higgs boson transverse momentum. We
  compare against recent numerical results and find good agreement.

\end{abstract}


\newpage


\section{Introduction}

Higgs boson pair production is one of the most promising processes to obtain
more insight into the scalar sector of the Standard Model.  For this reason it
will play a central role in the Large Hadron Collider's (LHC) physics program
in the coming years.  From the theory side it is important to provide precise
predictions for this process.

At the LHC, Higgs boson pair production is dominated by the gluon fusion
production process. The NLO QCD
corrections~\cite{Borowka:2016ehy,Borowka:2016ypz,Baglio:2018lrj,Davies:2019dfy}
turn out to be large, with a $K$ factor close to two.  In this work we compute
a subset of the electroweak corrections and construct analytic expansions in
the high-energy limit with explicit dependence on the kinematic variables and
mass parameters.  Our results have the advantage that their numerical
evaluation is fast and flexible; in particular, it is straightforward to use
different renormalization schemes and change the values of the particle
masses.

There are a number of publications in the literature in which electroweak
corrections to Higgs boson pair production have been considered. Leading
$m_t^4$ and $m_t^2$ terms have been considered in
Ref.~\cite{Muhlleitner:2022ijf}.  The contribution to $gg\to HH$ with four top
quark Yukawa couplings has been computed in Ref.~\cite{Davies:2022ram} in the
high-energy limit. It has been shown that the inclusion of about 100 expansion
terms provides stable results, even for rather small values of the Higgs boson
transverse momentum.  The first complete calculation of the top quark
contribution to $gg\to HH$, including all Standard Model contributions, has
been performed in Ref.~\cite{Davies:2023npk} in the large-$m_t$ limit, where
five expansion terms were presented. Numerical results for all Standard Model
contributions are available from Ref.~\cite{Bi:2023bnq}.  Analytic results for
all factorizing contributions have been published in Ref.~\cite{Zhang:2024rix}
and in Refs.~\cite{Heinrich:2024dnz,Li:2024iio} the contributions from all
diagrams with top quarks and Higgs bosons have been computed numerically.  In
this work we consider the same set of contributions as in
Ref.~\cite{Heinrich:2024dnz} and compute the form factors in the high-energy
limit, including terms up to $m_t^{120}$.
We restrict the discussion in this paper to the more complicated
box-type form factors.

In the past, deep expansions in the high-energy limit have proven to be
valuable ingredients for a number of processes. For example, the QCD
corrections to $gg\to HH$~\cite{Davies:2018ood,Davies:2018qvx} have been
combined with the numerical approach of
Refs.~\cite{Borowka:2016ehy,Borowka:2016ypz} in Ref.~\cite{Davies:2019dfy} and
with a forward-scattering approximation in
Refs.~\cite{Bellafronte:2022jmo,Davies:2023vmj}.  High-energy expansions of
$gg\to ZH$~\cite{Davies:2020drs} and $gg\to ZZ$~\cite{Davies:2020lpf} have
also been successfully combined with expansions around the forward-scattering
limit~\cite{Bellafronte:2022jmo,Degrassi:2022mro,Chen:2022rua,Degrassi:2024fye}.
In each case the high-energy expansion was a crucial input to be able to
describe the whole phase space.

Concerning electroweak corrections to $gg\to HH$, such expansions have been
computed in Ref.~\cite{Davies:2022ram} for the leading Yukawa
contributions. Two approaches concerning the hierarchies of the scales were
considered: (A) $s,t \gg m_t^2 \gg (m_H^{\rm int})^2, (m_H^{\rm ext})^2$ and
(B) $s,t \gg m_t^2 \approx (m_H^{\rm int})^2 \gg (m_H^{\rm ext})^2$, where
$m_H^{\rm ext}$ denotes the mass of the final-state Higgs bosons and
$m_H^{\rm int}$ denotes the mass of the Higgs bosons propagating in the loops.
For practical reasons, in this work only approach (B) can be applied to the
contributions involving Higgs boson self couplings.

\begin{figure}[t]
  \centering
  \begin{tabular}{ccc}
    \includegraphics[width=0.3\textwidth]{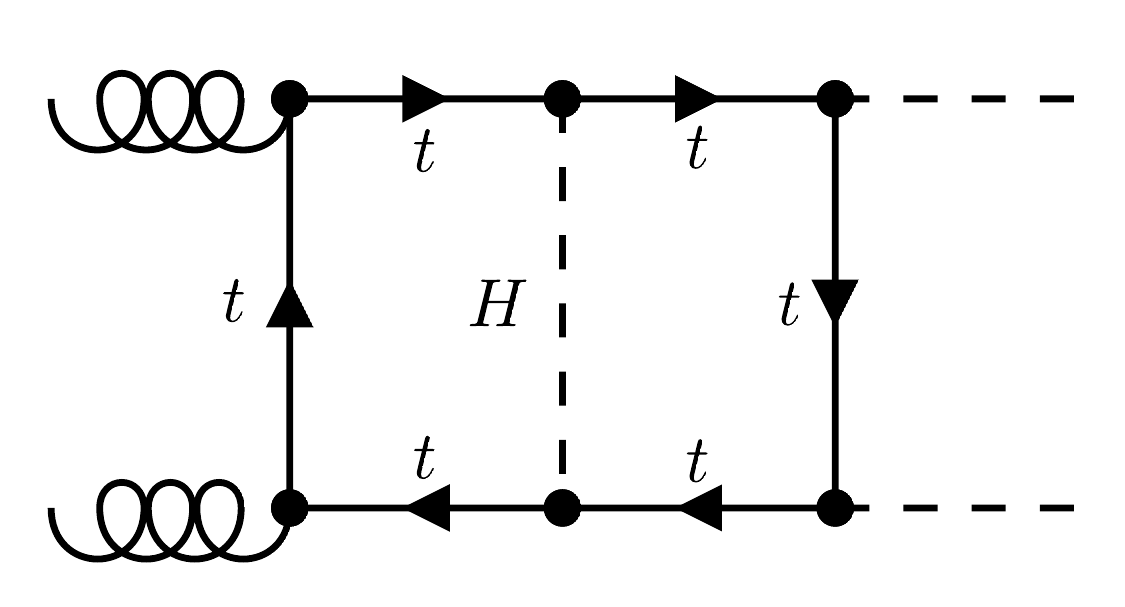} &
    \includegraphics[width=0.3\textwidth]{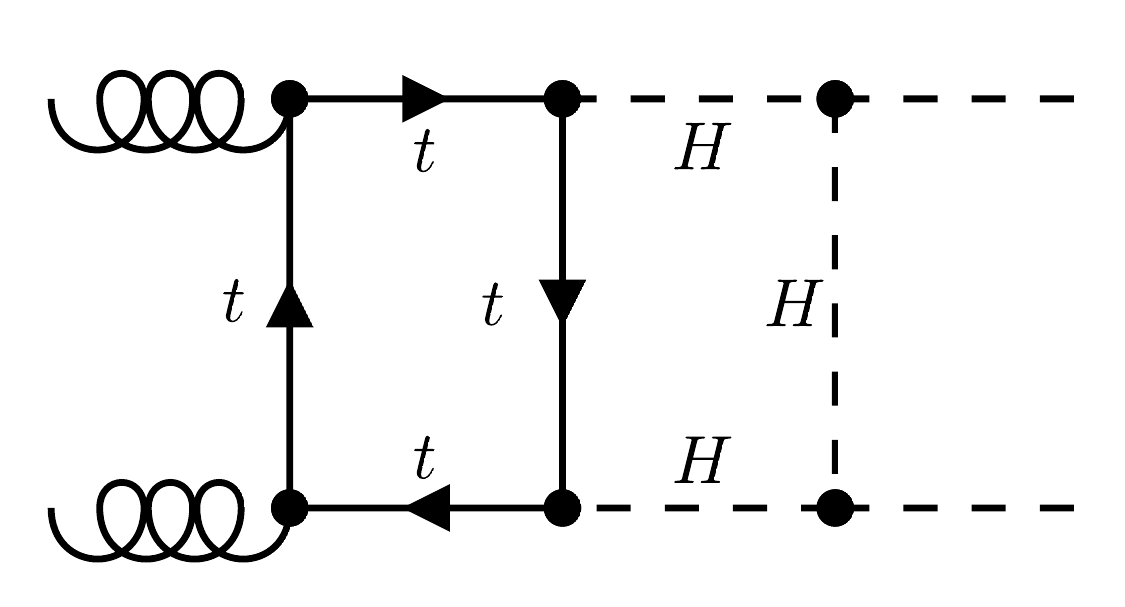} &
    \includegraphics[width=0.3\textwidth]{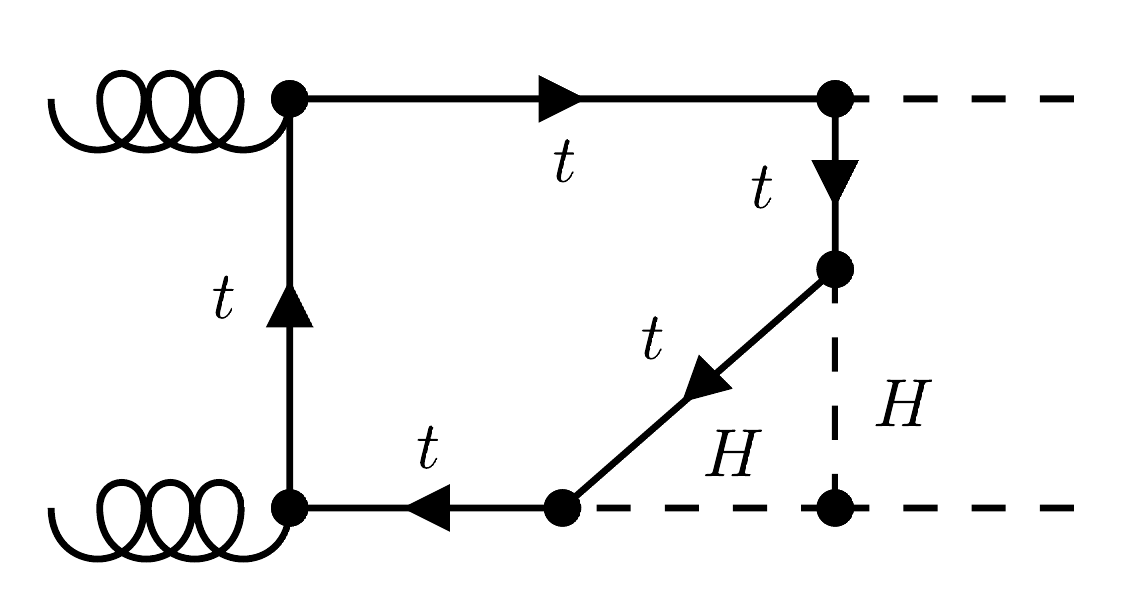} 
    \\
    \includegraphics[width=0.3\textwidth]{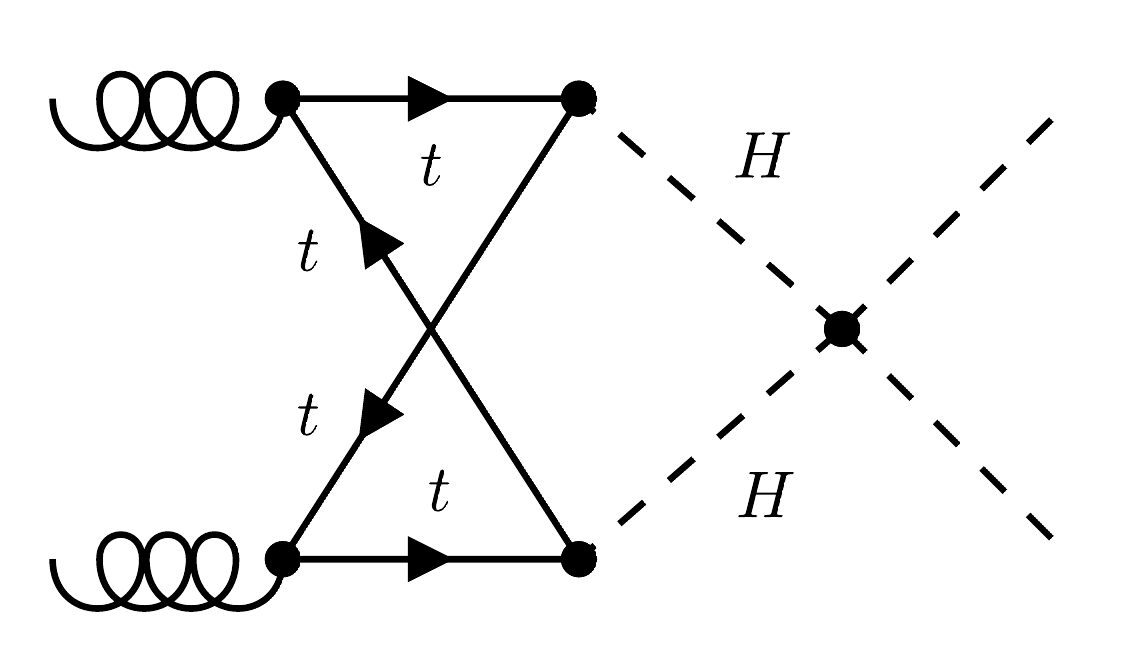} &
    \includegraphics[width=0.3\textwidth]{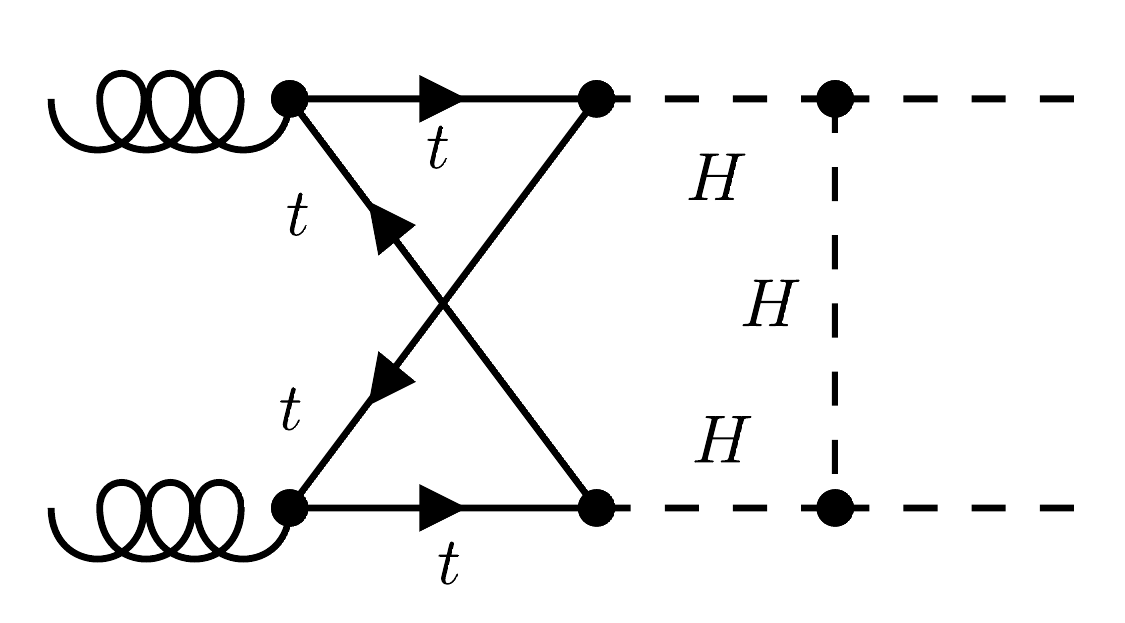} &
    \includegraphics[width=0.28\textwidth]{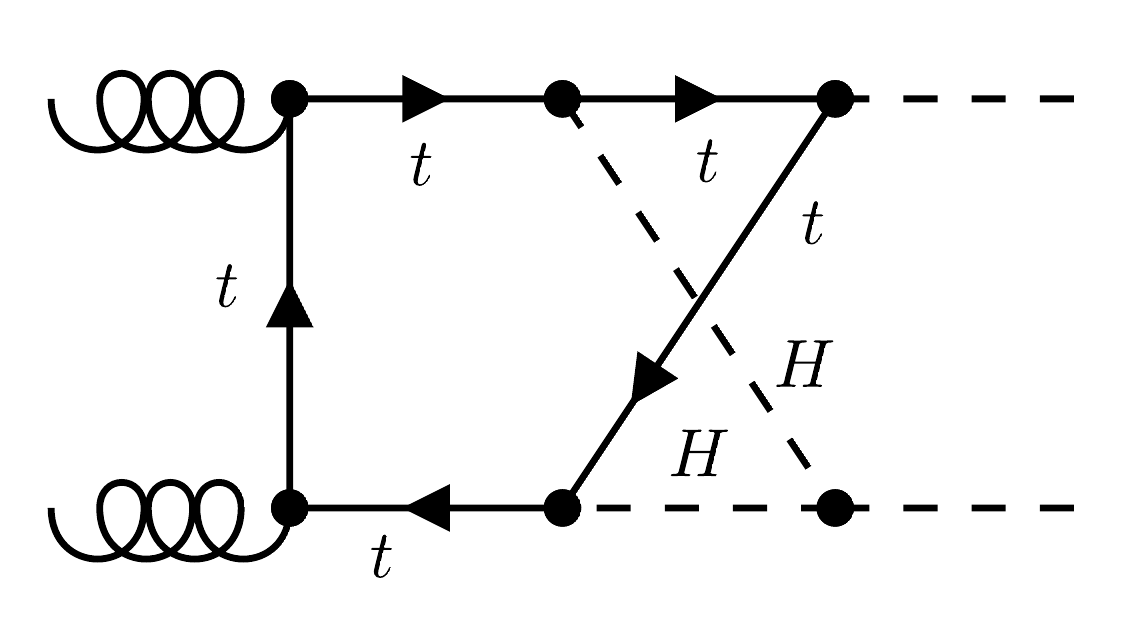} 
  \end{tabular}
  \caption{\label{fig::diags}Sample Feynman diagrams. Solid, dashed and curly
    lines represent top quarks, Higgs bosons and gluons, respectively.}
\end{figure}

In this paper we consider the contributions to the form factors from the
two-loop box diagrams involving top quarks and a Higgs boson.  Sample
Feynman diagrams are shown in Fig.~\ref{fig::diags}, and we can classify the diagrams according to their coupling
structure: $y_t^4$, $y_t^2 g_3^2$, $y_t^3 g_3$, $y_t^2 g_4$, where
$y_t, g_3$ and $g_4$ denote the 
Yukawa coupling to the top quark and the triple and quartic Higgs boson self couplings, respectively.  These
contributions have been computed in Ref.~\cite{Heinrich:2024dnz} using
numerical methods.  Below we will perform detailed comparisons of the
individual contributions.  The high-energy expansion of the $y_t^4$
contribution has already been computed in Ref.~\cite{Davies:2022ram}.

The outline of this paper is as follows. In Sec.~\ref{sec::conv} we discuss
the conventions used in the present paper.  Afterwards, in
Sec.~\ref{sec::tech}, we present technical details about the generation of the
amplitude and the approximations we apply, and the calculation of the master
integrals in the high-energy expansion.  In Sec.~\ref{sec::res} we show
numerical results for the master integrals and form factors, analyse the
convergence of the expansions, and compare our results to those of
Ref.~\cite{Heinrich:2024dnz}.  We conclude in Sec.~\ref{sec::conc}.

\section{Conventions}
\label{sec::conv}
The amplitude for the process 
$g(q_1)g(q_2)\to H(q_3)H(q_4)$
can be defined by two scalar matrix
elements ${\cal M}_1$ and ${\cal M}_2$ via
\begin{eqnarray}
  {\cal M}^{ab} &=& 
  \varepsilon_{1,\mu}\varepsilon_{2,\nu}
  {\cal M}^{\mu\nu,ab}
  \,\,=\,\,
  \varepsilon_{1,\mu}\varepsilon_{2,\nu}
  \delta^{ab}
  \left( {\cal M}_1 A_1^{\mu\nu} + {\cal M}_2 A_2^{\mu\nu} \right)
  \,,
\end{eqnarray}
where $a$ and $b$ are adjoint colour indices and the two Lorentz structures
are given by
\begin{eqnarray}
  A_1^{\mu\nu} &=& g^{\mu\nu} - {\frac{1}{q_{12}}q_1^\nu q_2^\mu
  }\,,\nonumber\\
  A_2^{\mu\nu} &=& g^{\mu\nu}
                   + \frac{1}{p_T^2 q_{12}}\left(
                   q_{33}    q_1^\nu q_2^\mu
                   - 2q_{23} q_1^\nu q_3^\mu
                   - 2q_{13} q_3^\nu q_2^\mu
                   + 2q_{12} q_3^\mu q_3^\nu \right)\,,
    \label{eq::LorStr}
\end{eqnarray}
with $q_{ij} = q_i\cdot q_j$.
The scalar matrix elements can furthermore be split into ``triangle'' and ``box'' form factors
\begin{eqnarray}
  {\cal M}_1 &=& X_0 \, s \, \left(\frac{3 m_H^2}{s-m_H^2} F_{\rm tri} + F_{\rm box1}\right)
                 \,,\nonumber\\
  {\cal M}_2 &=& X_0 \, s \, F_{\rm box2}
                 \,,
                 \label{eq::calM}
\end{eqnarray}
with
\begin{eqnarray}
  X_0 &=& \frac{G_F}{\sqrt{2}} \frac{\alpha_s(\mu)}{2\pi} T \,,
\end{eqnarray}
where $T=1/2$, $\mu$ is the renormalization scale and $G_F$ is the Fermi constant.

All momenta are defined incoming,
so the Mandelstam variables 
are given by
\begin{eqnarray}
  {s}=(q_1+q_2)^2\,,\qquad {t}=(q_1+q_3)^2\,,\qquad {u}=(q_2+q_3)^2\,,
  \label{eq::stu}
\end{eqnarray}
with
\begin{eqnarray}
  q_1^2=q_2^2=0\,,\qquad  q_3^2=q_4^2=m_H^{2}\,,\qquad
  {s}+{t}+{u}=2m_H^{2}\,.
  \label{eq::q_i^2}
\end{eqnarray}
Later we will also use the scattering angle $\theta$ and the transverse
momentum of the Higgs bosons in the center-of-mass frame, which are given by
\begin{eqnarray}
\label{eqn:thetadef}
  p_T^2 &=& \frac{{u}{t}-m_H^4}{{s}}\,,
          \nonumber\\
  t &=& m_H^2 - \frac{s}{2}\left(1-\cos\theta
        \,\sqrt{1-\frac{4m_H^2}{s}}\right)\,.
        \label{eq::pT_costhe}
\end{eqnarray}

We extend the perturbative expansion of the form factors
given in Eq.~(8) of Ref.~\cite{Davies:2022ram} to
\begin{eqnarray}
  F_{\rm box} &=& F_{\rm box}^{(0)} 
        + \frac{\alpha_s(\mu)}{\pi} F_{\rm box}^{(1,0)} 
        + \frac{\alpha_t}{\pi}               F_{\rm box}^{(0,y_t^4)} 
        + \sqrt{\frac{\alpha_t}{\pi}\frac{\hat\lambda}{\pi}} F_{\rm box}^{(0,y_t^3 g_3)}
        + \frac{\hat\lambda}{\pi}           F_{\rm box}^{(0,y_t^2 g_3^2)} 
        + \frac{\hat\lambda}{\pi}          F_{\rm box}^{(0,y_t^2 g_4)} 
        + \cdots
                  \nonumber\\
  &=&
      F_{\rm box}^{(0)} + \frac{\alpha_s(\mu)}{\pi} F_{\rm box}^{(1,0)} 
      + \frac{\alpha_t}{\pi}   F_{\rm box}^{(0,\alpha_t)} 
  \,,
  \label{eq::F}
\end{eqnarray}
with
\begin{eqnarray}
  F_{\rm box}^{(0,\alpha_t)} 
  &=&
      F_{\rm box}^{(0,y_t^4)} 
      + \frac{m_H}{m_t} F_{\rm box}^{(0,y_t^3 g_3)} 
      + \frac{m_H^2}{m_t^2} F_{\rm box}^{(0,y_t^2 g_3^2)} 
      + \frac{m_H^2}{m_t^2} F_{\rm box}^{(0,y_t^2 g_4)}
    \label{eq::Fsplit}
\end{eqnarray}
where 
\begin{eqnarray}
  \hat\lambda &=& \frac{\lambda}{\pi} \:=\: \frac{1}{\pi} \frac{m_H^2}{2v^2}\,,\nonumber\\
  \alpha_t &=& \frac{\alpha \, m_t^2}{2 \, \sin^2\theta_W\, m_W^2}\,.
               \label{eq::alpha}
\end{eqnarray}
$\alpha$ is the fine structure constant, $\theta_W$ is the
weak mixing angle and $v$ is the Higgs vacuum expectation
value.  In Eq.~(\ref{eq::F}), $F^{(0)}$ is the leading one-loop contribution
(which contains two Yukawa couplings) and $F^{(1,0)}$ represents the two-loop
QCD corrections. The remaining terms refer to the different contributions
considered in this paper.
We have introduced the quantities $y_t$, $g_3$ and
$g_4$ to distinguish the contributions from the Higgs-top quark Yukawa and the
three- and four-Higgs boson couplings. 
In this paper we only consider the one-particle irreducible box diagrams which
is indicated by the subscript ``box''.
We will use the $F_{\rm box1}$ and $F_{\rm box2}$ notation of Eq.~(\ref{eq::calM})
to distinguish between the Lorentz structures for these form factors.
From the technical point of view the
triangle and reducible contributions are much simpler and are not
considered in this paper.
Note that we count $F_{\rm box}^{(0,y_t^2 g_4)}$ as a ``box'' contribution since
it does not have an intermediate Higgs propagator. However, from the
technical point of view it is a triangle contribution; in particular it
has no dependence on $p_T$. For this reason it only contributes to $F_1$ but
not to $F_2$.

Of the four electroweak two-loop
contributions defined in Eq.~(\ref{eq::F}), only $F_{\rm box}^{(0,y_t^4)}$
develops ultra-violet $1/\epsilon$ poles.\footnote{In Table~3 of
  Ref.~\cite{Heinrich:2024dnz} there are poles for ``$g_3 g_t^3$'' since there also
  triangle contributions are included.} These cancel against the
contribution from the top quark mass counterterm (see Ref.~\cite{Davies:2022ram}
for details); the remaining three contributions are finite
after summing all bare two-loop diagrams. Note that there
are no infrared singularities present in any of these coupling structures.


\section{Technicalities}
\label{sec::tech}


\subsection{Generation of the amplitude}

For the generation of the analytic expressions for the form factors in terms
of Lorentz-scalar Feynman integrals we use a well-tested setup with is based
on {\tt qgraf}~\cite{Nogueira:1991ex} for the generation of the amplitude,
{\tt tapir}~\cite{Gerlach:2022qnc} for the translation to {\tt
  FORM}~\cite{Ruijl:2017dtg} notation and the generation of auxiliary files
for processing the integral topologies, and on {\tt
  exp}~\cite{Harlander:1998cmq,Seidensticker:1999bb} for the mapping of each
Feynman diagram onto a minimal subset of the integral topologies from {\tt
  tapir}.  In all integral families all internal lines have the same mass:
$m_t$. The internally propagating Higgs bosons are expanded around $m_t$ as a
series in $\delta^\prime= 1 - (m_H^{\rm int}/m_t)^2$ as described in
Ref.~\cite{Davies:2022ram}.  In the final results we switch to the expansion
parameter $\delta = 1 - m_H^{\rm int}/m_t$ by using the relation
\begin{align}
    \delta^\prime = \delta \left( 2 - \delta \right) ~, 
\end{align}
and expanding in $\delta$ again,
since it shows better convergence properties (as e.g.~described 
in Ref.~\cite{Fael:2022frj}).

The computation of the amplitude is performed with the in-house {\tt FORM} code ``{\tt
  calc}'', which applies projectors to obtain the form factors, takes traces and
expresses the final result in terms of scalar integrals of the
integral families selected by {\tt exp}.  In this step we also perform the expansion
of all Higgs boson propagators such that in the final results we have
terms up to order $\delta^3$.  At this point the scalar integrals
depend on the variables $\{s,t,m_H^{\rm ext},m_t\}$; $\delta$ is only
present in the integral coefficients.

The next step is to expand in the final state Higgs boson mass which is
encoded in the relation of the external momenta $\{q_1,q_2,q_3\}$ to
the Mandelstam variables and $m_H^{\rm ext}$. For the expansion at the
level of the integrands it is convenient to use {\tt
  LiteRed}~\cite{Lee:2012cn,Lee:2013mka} and to perform the expansion
for each of the scalar integrals.  We generate {\tt FORM} {\tt id}
statements which are applied to the summed expressions for the form
factors. Now the remaining scalar integrals depend only on $\{s,t,m_t\}$.

Next, we perform a reduction to master integrals within each topology using Kira~\cite{Maierhofer:2017gsa,Klappert:2020nbg}, followed by an additional reduction
of all 1224 master integrals among all topologies to yield a final minimal set of 168 master
integrals, 28 of which are non-planar. We made some effort to optimize the integral basis
to remove spurious $\epsilon$ poles (where $D=4-2\epsilon$) in the master integral
coefficients in the amplitude, particularly for master integrals in the top-level
sectors. This was only partially successful, thus in intermediate steps it was
necessary to compute some master integrals to order $\epsilon^1$, in terms of
some transcendental weight 5 objects which ultimately cancel in the amplitude. This
provides a strong consistency check of our reduction and basis minimization procedures.


\subsection{Computation of the master integrals}

The analytic calculation of the master integrals in the high-energy region is
more involved compared to the QCD-like master integrals considered in
Refs.~\cite{Davies:2018ood,Mishima:2018olh,Davies:2018qvx,Davies:2023vmj},
due to the larger number of massive propagators.  Nevertheless, for
their calculation similar methods can be used. In total we find $168$ master
integrals, 140 of these are planar integrals which have already been
considered in Ref.~\cite{Davies:2022ram} for the leading Yukawa corrections to
double Higgs production.  For the current calculation some of the planar
master integrals from the lower sectors had to be extended to
$\mathcal{O}(\epsilon)$ due to spurious poles in the amplitude which we were
not able to remove.  The remaining 28 master integrals are non-planar
integrals with six and seven massive internal lines and their calculation is
significantly more complex than the calculation of the planar master
integrals.

Let us briefly recap the strategy for the calculation of the planar master
integrals used in Ref.~\cite{Davies:2022ram} applied to all 168 master integrals, before discussing the modifications 
which were necessary for the non-planar case.  
The first step is to insert a power-log ansatz for each master integral, assuming
the hierarchy $s,|t| \gg m_t^2$,
\begin{align}
I_n = \sum_{i,j,k} C^{(n)}_{ijk} (s,t) \, \epsilon^i \, \left( m_t^2 \right)^j \, \log(m_t^2)^k
\label{eq:ansatzMT}
\end{align}
into the system of differential equations with respect to $m_t^2$.
\begin{align}
    \partial_{m_t^2} \mathbf{I} = \mathbf{A}(s,t,m_t^2,\epsilon) \, \mathbf{I}\,.
    \label{eq:mtdiffeq}
\end{align}
This yields a system of linear equations for the expansion coefficients
$C^{(n)}_{ijk} (s,t)$, which are still functions of $s$ and $t$.  By solving
the system of linear equations with \texttt{Kira}~\cite{Klappert:2020nbg} and
\texttt{FireFly}~\cite{Klappert:2019emp,Klappert:2020aqs}, we can reduce the
large number of independent coefficients to a small number of boundary
conditions which need to be fixed by a direct calculation.  In total, for the
planar and non-planar master integrals, we need to fix $543$ boundary
conditions (i.e.~$543$ functions of the type $C^{(n)}_{ijk} (s,t)$).  In the
planar cases we obtain them by means of Mellin-Barnes integrals and symbolic
summation~\cite{Davies:2022ram}.  This approach has recently been partially
automated in the package \texttt{AsyInt}~\cite{Zhang:2024fcu}, which employs
the packages \texttt{Asy}~\cite{Jantzen:2012mw},
\texttt{MB}~\cite{Czakon:2005rk},
\texttt{HarmonicSums}~\cite{Blumlein:1998if,Vermaseren:1998uu,Blumlein:2009ta,Ablinger:2009ovq,Ablinger:2011te,Ablinger:2012ufz,Ablinger:2013eba,Ablinger:2013cf,Ablinger:2014bra,Ablinger:2014rba,Ablinger:2015gdg,Ablinger:2018cja}
and \texttt{Sigma}~\cite{sigmaI,sigmaII}.

As for the case of the QCD-like master integrals, in the fully-massive case
the non-planar master integrals also develop both even and odd power terms in
the expansion for small $m_t$ (the high-energy region), the latter are not
covered by the ansatz in Eq.~\eqref{eq:ansatzMT}.  Since the equation systems
for the even and odd power expansions completely decouple, we treat them
separately.  In fact, for the odd-power contribution we can apply the same
strategy as for the planar master integrals and compute the boundary functions
$C^{(n)}_{ijk} (s,t)$ directly.  In contrast to the planar master integrals,
here we encounter square roots in the Mandelstam variables ($\sqrt{-t/s}$ and
$\sqrt{1+t/s}$).  Although the dependence on $t$ of the odd power expansion is
rather simple, we find involved numerical coefficients.  One of them is given
by
\begin{align}
    c_Z = & \int_{0}^{\infty} \int_{0}^{\infty} \, \frac{\mathrm{d} \alpha_1 \, \mathrm{d}  \alpha_2}{\sqrt{\alpha _1 \, \alpha _2 \, \big( \alpha _1+\alpha _2+1\big) \, \big( \alpha _2 \alpha _1+\alpha _1+\alpha _2\big) } }
    \nonumber \\
    &= 
    \sum_{k=0}^{\infty} \frac{\Gamma^4(k+1/2)}{\pi \Gamma^2(k+1)\Gamma(2k+1)}
    \biggl( 
        8 \log(2) + 6 S_{-1}(2k)  
    \biggr)
    \nonumber\\
    &= 4 \sqrt{3} \, K^2\Big(\frac{1}{2}-\frac{\sqrt{3}}{4}\Big)
      \nonumber\\  
    &= 17.695031908454309764234228747255048751062059438637\dots  \,, 
      \label{eq:mbsum1}
\end{align}
where $S_{-1}(n)=\sum_{i=1}^{n} (-1)^i/i$ and $K(z)$ is the complete elliptic
function of the first kind given by
\begin{eqnarray}
  K(z) &=& \int\limits_{0}^{1} {\rm d} t \frac{1}{\sqrt{1-t^2}\sqrt{1-z
           t^2}}\,.
\end{eqnarray}
We were able to
determine the closed-form solution with the help of \texttt{HarmonicSums} and
the \texttt{PSLQ} algorithm~\cite{PSLQ}, where the former was able to perform
the symbolic summation on parts of Eq.~\eqref{eq:mbsum1}.  In the amplitude
also a second elliptic constant appears
\begin{align}
        c_{Z_2} 
                &= 
                \sum_{k=0}^{\infty} 
                \frac{\Gamma^4(k+1/2)}{\pi \Gamma^2(k+1) \Gamma(2k+1)} 
                \biggl( 
                    \frac{k \ln(2)}{2(1+k)}
                    - \frac{1}{16(1+k)^2}
                    + \frac{3k}{8(1+k)}S_{-1}(2k)
                \biggr)
                \nonumber\\
                & = \Big(1+\frac{1}{\sqrt{3}}\Big) \pi + \Big(2+\frac{17}{4 \sqrt{3}}\Big) K^2\Big(\frac{1}{2}-\frac{\sqrt{3}}{4}\Big) -4 \sqrt{3} E^2\Big(\frac{1}{2}-\frac{\sqrt{3}}{4}\Big)  
                \nonumber\\
                &= -0.189114824528715301590514265622885511401742116540 \dots \, ,
\end{align}
where $E(z)$ is the complete elliptic integral of the second kind which reads
\begin{eqnarray}
  E(z) &=& \int\limits_{0}^{1} {\rm d} t \frac{\sqrt{1-z t^2}}{\sqrt{1-t^2}}\,.
\end{eqnarray}
The most difficult boundary
constant for the odd-$m_t$ terms is a seven-line non-planar master integral, evaluated up to $\mathcal{O}(\epsilon \, m_t)$.

For the even-power contributions of the non-planar master integrals this
strategy, however, becomes too complicated since a large number of non-trivial
one-, two- and three-dimensional Mellin-Barnes integrals have to be solved.  In
the following we describe the strategy which we have applied for these
contributions.

We use the expression of the even-power expansion of the master integrals in terms of the $543$ 
boundary conditions and insert them into the differential equation with respect to $t$
\begin{align}
    \partial_{t} \mathbf{I} = \mathbf{M}(s,t,m_t^2,\epsilon) \, \mathbf{I}\,,
    \label{eq:mtdiffeq_t}
\end{align}
where $\mathbf{M}(s,t,m_t^2,\epsilon)$ is still exact in all variables.
After expanding this equation in terms of $m_t^2$ and $\epsilon$, and comparing 
coefficients on both sides of the equation, we arrive at a
system of coupled linear differential equations with respect to $t$ for the $543$ boundary
conditions, which we group into the vector $\vec{C}(t)$ (here and in the
following we set $s=1$, since its dependence can be recovered using dimensional
analysis):
\begin{align}
\label{eq::diffeq}
    \partial_t \vec{C}(t) &= \mathbf{M}_t(t) \vec{C}(t) ~.
\end{align}
The $543 \times 543$ dimensional matrix $\mathbf{M}_t(t)$ now only depends on
$t$, since expansions for small $\epsilon$ and $m_t^2$ have already been
performed.  We solve this system of differential equations as outlined in
Ref.~\cite{Ablinger:2018zwz}, i.e.~we decouple the system of differential
equations into higher order differential equations for individual boundary
conditions and solve these with the help of \texttt{HarmonicSums}.  In order
to fully solve the differential equation we need to provide integration
constants for the limit $t \to 0$. In contrast to our previous approach these
are just numbers, and depend neither on $m_t$ nor on $t$.  We obtain
them with Mellin-Barnes techniques, which simplify since we do not need to
reconstruct the full $t$ dependence.  Note that there are non-trivial boundary
integrals that require computing the exact $u$-dependence even in the $t\to 0$
limit.  These are solved using the \texttt{Expand\&Fit} module of
\texttt{AsyInt}.

Before actually performing the calculation of the integration constants, we use the full
$t$ dependence of $\vec{C}(t)$ and use the crossing symmetries between the
master integrals to reduce the number of necessary integration constants.  It
turns out that this step leads to a reduction by a factor of 3 in the number of
integration constants which need to be calculated explicitly.  For example, it
turns out that deeper expansion terms in $m_t$ are not needed and it is
sufficient to provide the seven-line non-planar master integral up to
$\mathcal{O}(\epsilon \, m_t^0)$ instead of $\mathcal{O}(\epsilon \, m_t^2)$, which 
was necessary in the previous approach. 
Note that we also recalculate all planar master integrals in this way and 
find complete agreement with our previously obtained
results~\cite{Davies:2022ram}. We also reproduce the
${\cal O}(\epsilon^0 m_t^0)$ terms of the seven-line non-planar master
integral from Ref.~\cite{Zhang:2024fcu}.

We provide the analytic expressions for the master integrals expanded up to 
$m_t^{100}$ in an ancillary file to this publication~\cite{progdata}.
It turns out that also in
the fully massive case the function space of the high-energy expansion is
spanned by a subset of harmonic polylogarithms with the letters $0$ and $1$.
During the calculation we had to introduce several integration constants 
for which we were not able to determine closed expressions:
\begin{align} 
    c_6 = & \, -66.606101552505094835462532159285032068889118084828 \dots \,, \nonumber\\
    c_7 = &  \, \hphantom{-}\:\:\:\: 8.245227873742566544491578437011715020871425138609 \dots \,, \nonumber \\
    c_{Z}^{\prime} = & \, -46.834387660276172083477753650852669635341902837429 \dots \,, \nonumber\\
    c_{Z_2}^{\prime} = & \, -\:\:0.127899599664440859067122151592372323466385155017 \dots \,.
\end{align}
Note that the constants $c_6$, $c_7$, $c_Z^{\prime}$ and $c_{Z_2}^{\prime}$ enter the master integrals only 
at $\mathcal{O}(\epsilon)$ and cancel in the physical amplitudes. 
In the ancillary file we provide numerical evaluations with 600 significant digits for completeness. 
The physical amplitudes thus turn out to depend 
on the following set of transcendental numbers 
\begin{align}
    \bigg\{ \, \sqrt{3}, \, \log(3), \, \pi, \, c_Z, \, c_{Z_2}, \, \psi ^{(1)}\Big(\frac{1}{3}\Big), \, \zeta_3 , \, 
    \text{Im}\Big[\text{Li}_3\Big( \frac{i}{\sqrt{3}}\Big)\Big] \, \bigg\} \,,
\end{align}
while individual master integrals additionally depend on
\begin{align}
    \bigg\{ \, 
    \text{Im}\Big[\text{Li}_3\Big(\frac{i \sqrt{3}+1}{4}\Big)\Big] , \,  
    \text{Im}\left( G_{0, 0, 0, r_2}(1) \right), 
    \text{Im}\left( G_{0, 1, 1, r_4}(1) \right), \, \zeta_{5}, \,
    c_6, \, c_7, \, c_{Z}^{\prime}, \, c_{Z_2}^{\prime} \,
    \bigg\} \,,
\end{align}
where $G$ denotes multiple polylogarithms, see e.g.~Ref.~\cite{Goncharov:1998kja}, with the 
additional letters $r_2=(1-\sqrt{3} i)/2$, $r_4=-r_2^*$ of the sixth root of 
unity. 
A full determination of the basis up to weight six can be found in Refs.~\cite{Ablinger:2011te,Henn:2015sem}.



\section{Results}
\label{sec::res}


\subsection{\label{sub::MIs}Master integrals}

\begin{figure}[htb]
  \centering
  \begin{tabular}{ccc}
    \includegraphics[width=0.30\textwidth]{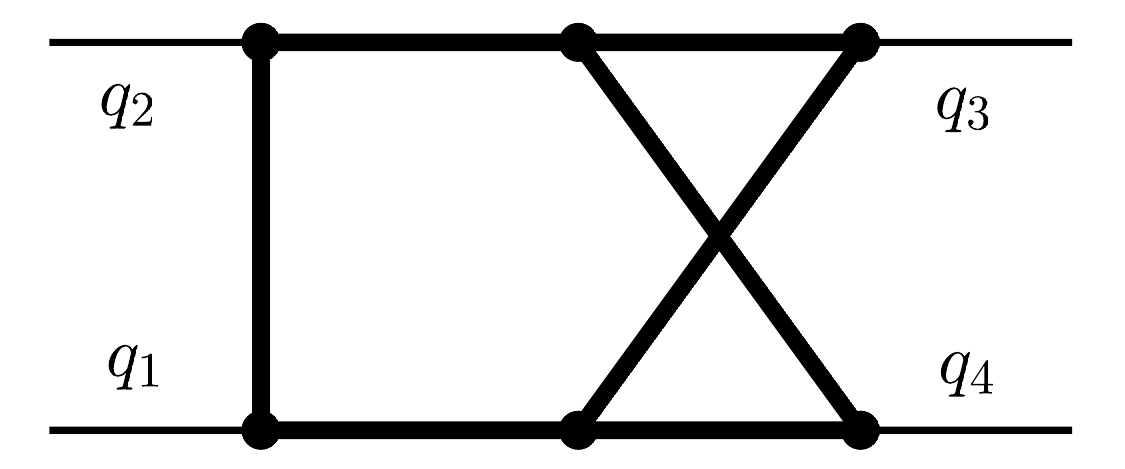}
    & 
    \includegraphics[width=0.30\textwidth]{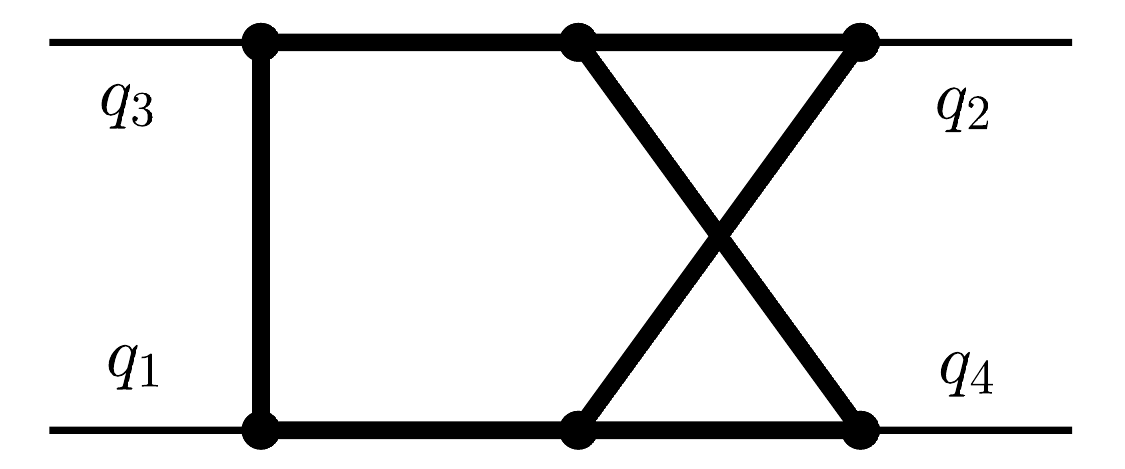}
    & 
    \includegraphics[width=0.30\textwidth]{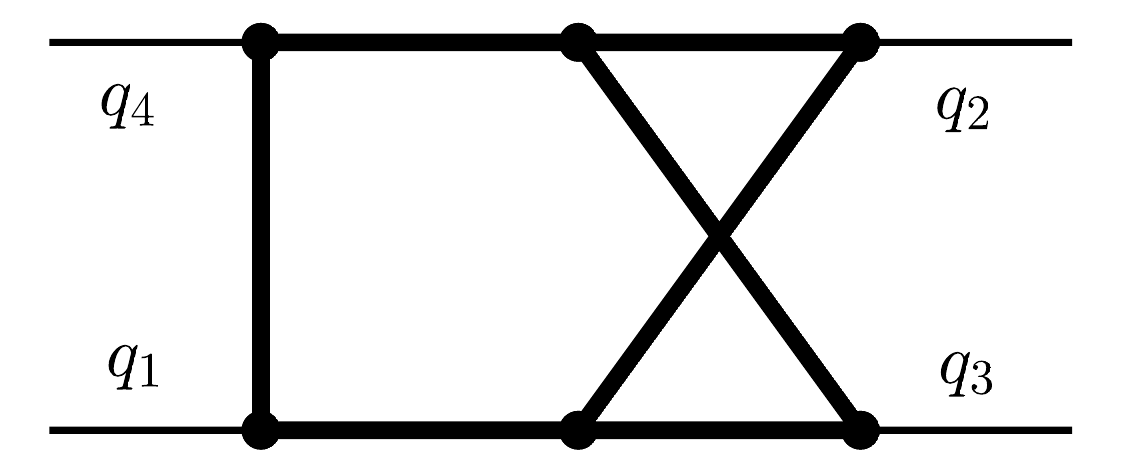} \\
    {\scriptsize $I_{150}=G_{52}(1,1,1,1,1,1,1,0,0)$}
    & {\scriptsize $I_{157}=G_{53}(1,1,1,1,1,1,1,0,0)$}
    & {\scriptsize $I_{164}=G_{54}(1,1,1,1,1,1,1,0,0)$}
  \end{tabular}
  \caption{\label{fig::masters}Sample top-level master integrals, where all
    internal lines have the same mass $m_t$ and all external lines are
    massless.  }
\end{figure}

We first discuss the quality of the high-energy expansion at the
level of individual master integrals. As  examples, we use the
seven-line (top sector) master integrals of the three non-planar integral families shown in
Fig.~\ref{fig::masters}. All internal lines have the mass $m_t$ and
all external particles are massless. For all integrals we have an
expansions up to $m_t^{120}$.

\begin{figure}[t]
  \centering
  \begin{tabular}{ccc}
    \includegraphics[width=0.30\textwidth]{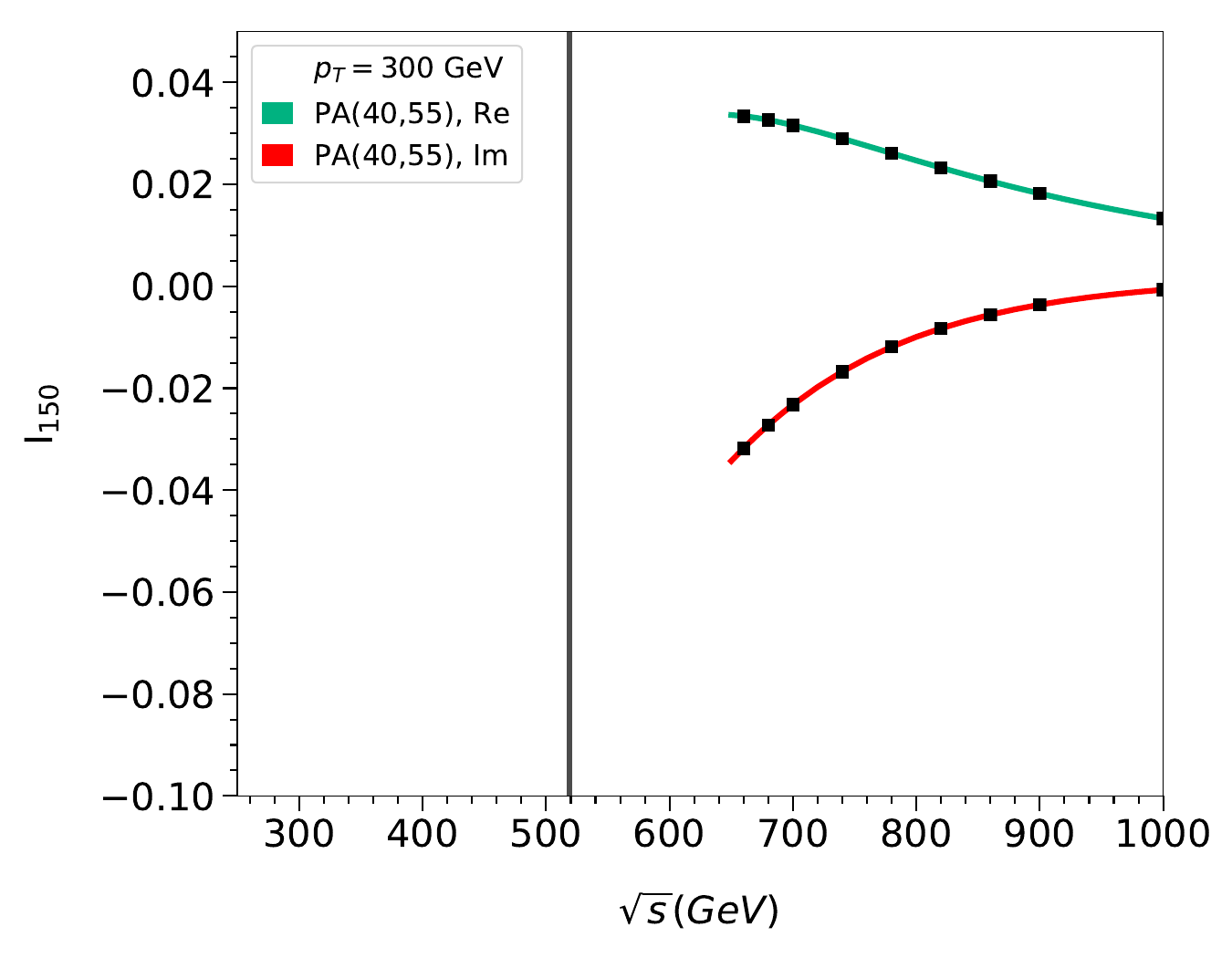}
    & 
    \includegraphics[width=0.30\textwidth]{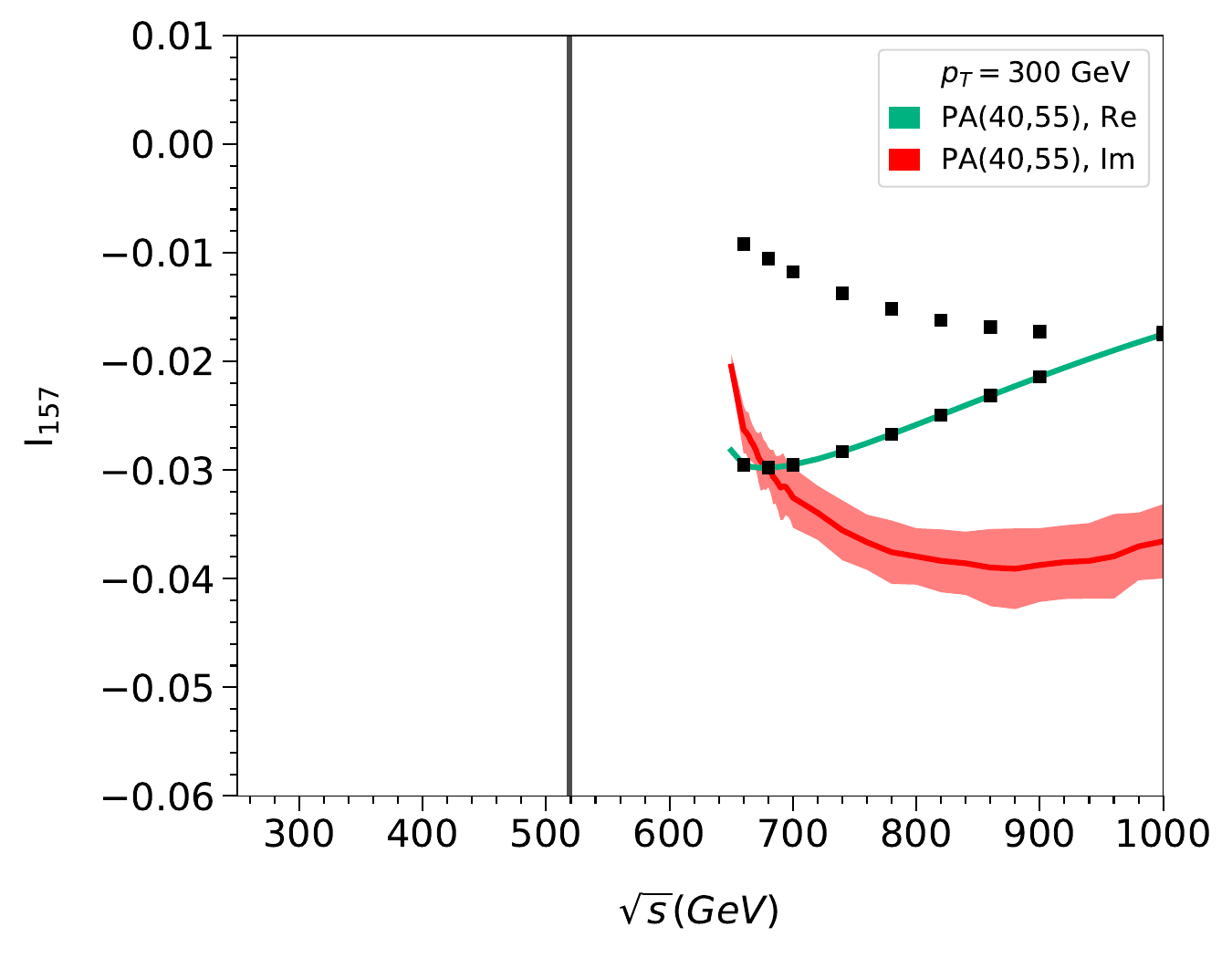}
    & 
    \includegraphics[width=0.30\textwidth]{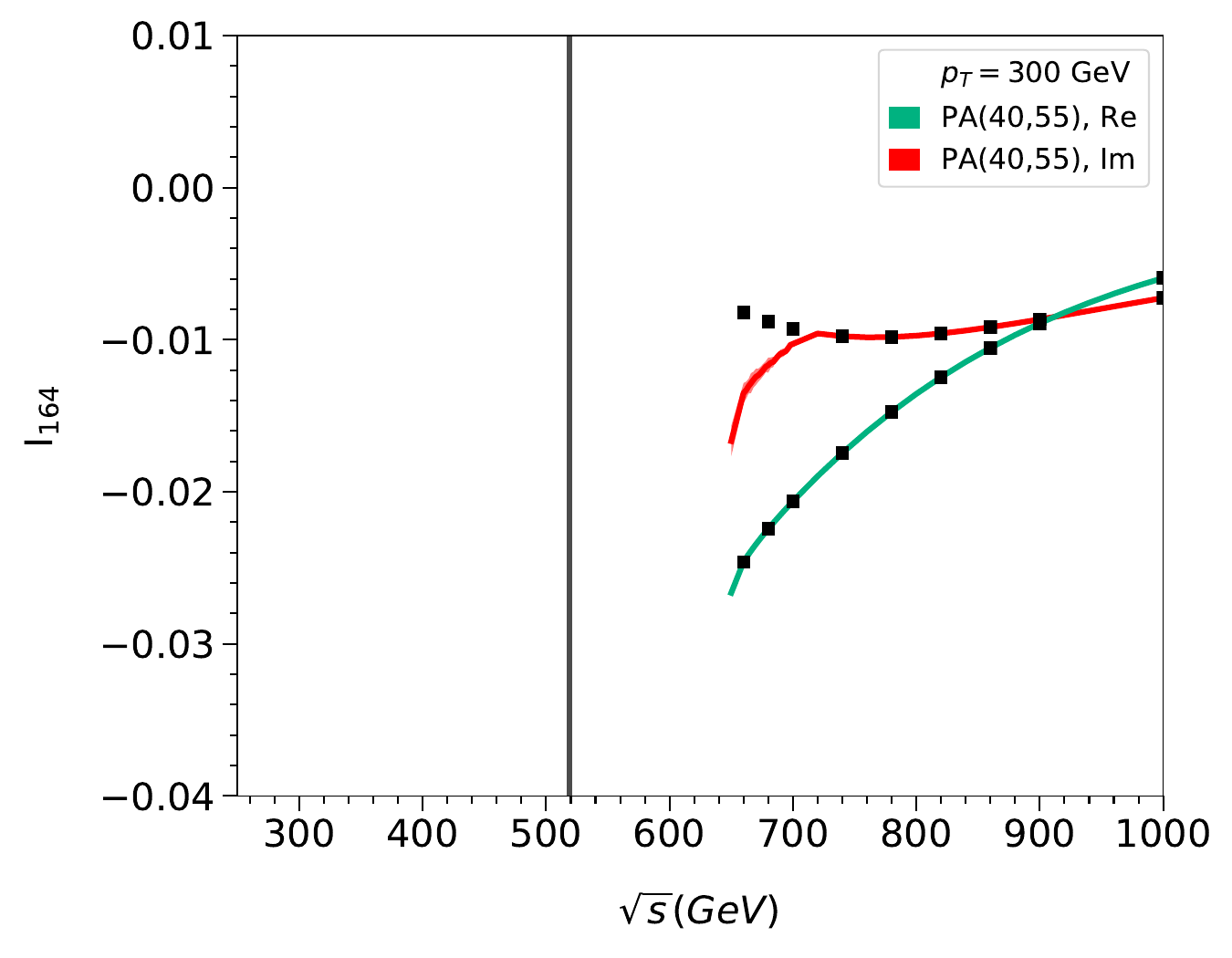}
    \\
    \includegraphics[width=0.30\textwidth]{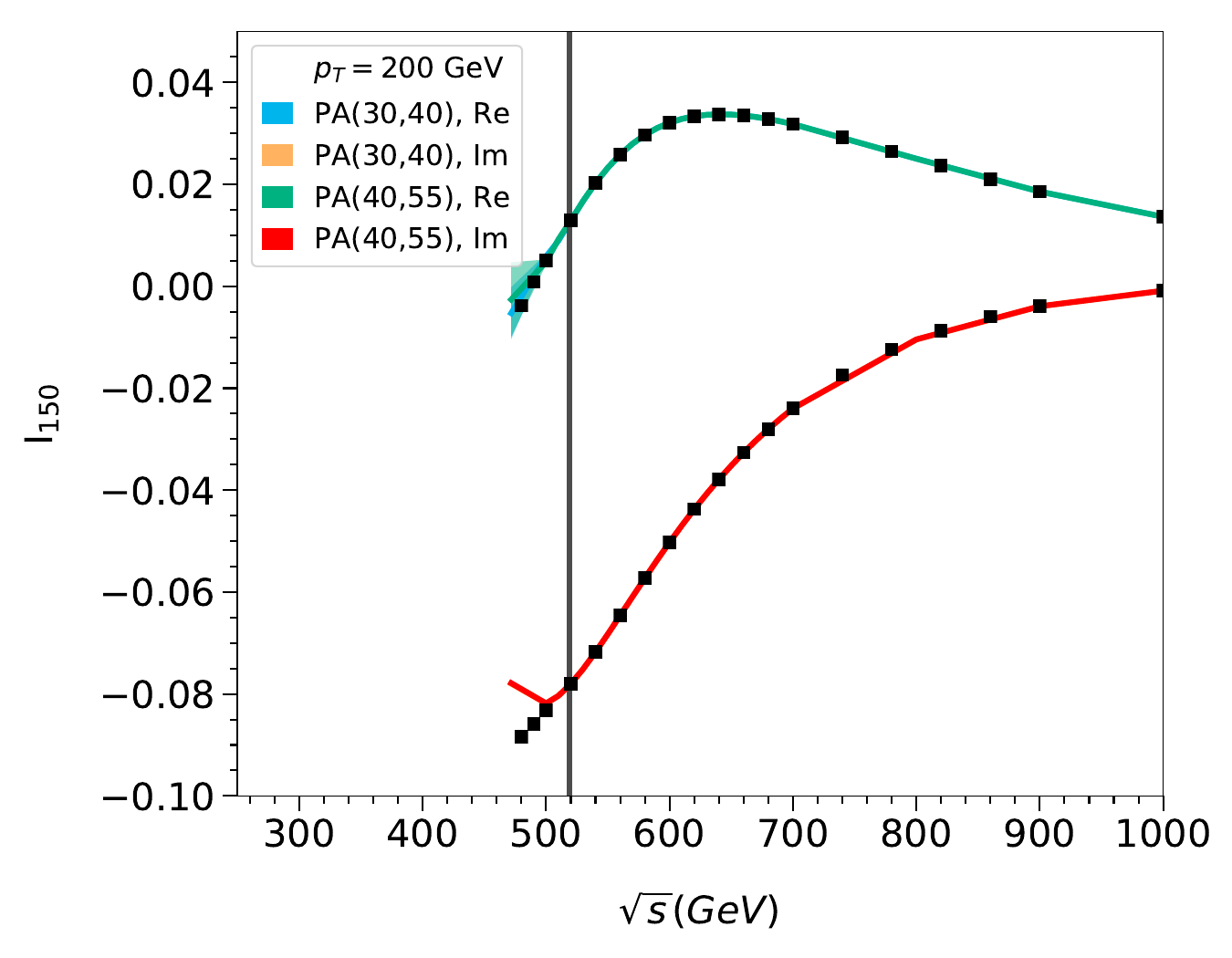}
    & 
    \includegraphics[width=0.30\textwidth]{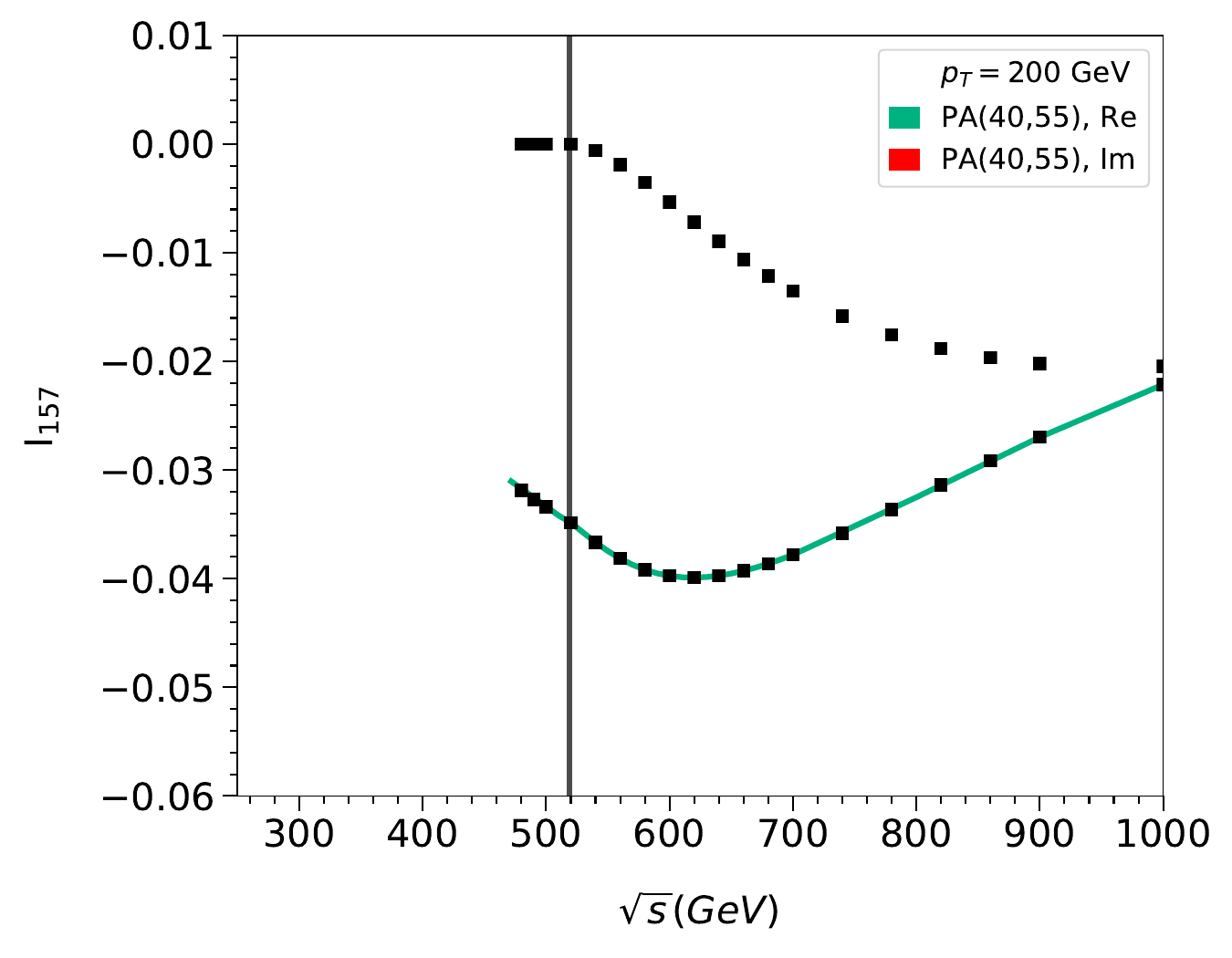}
    & 
    \includegraphics[width=0.30\textwidth]{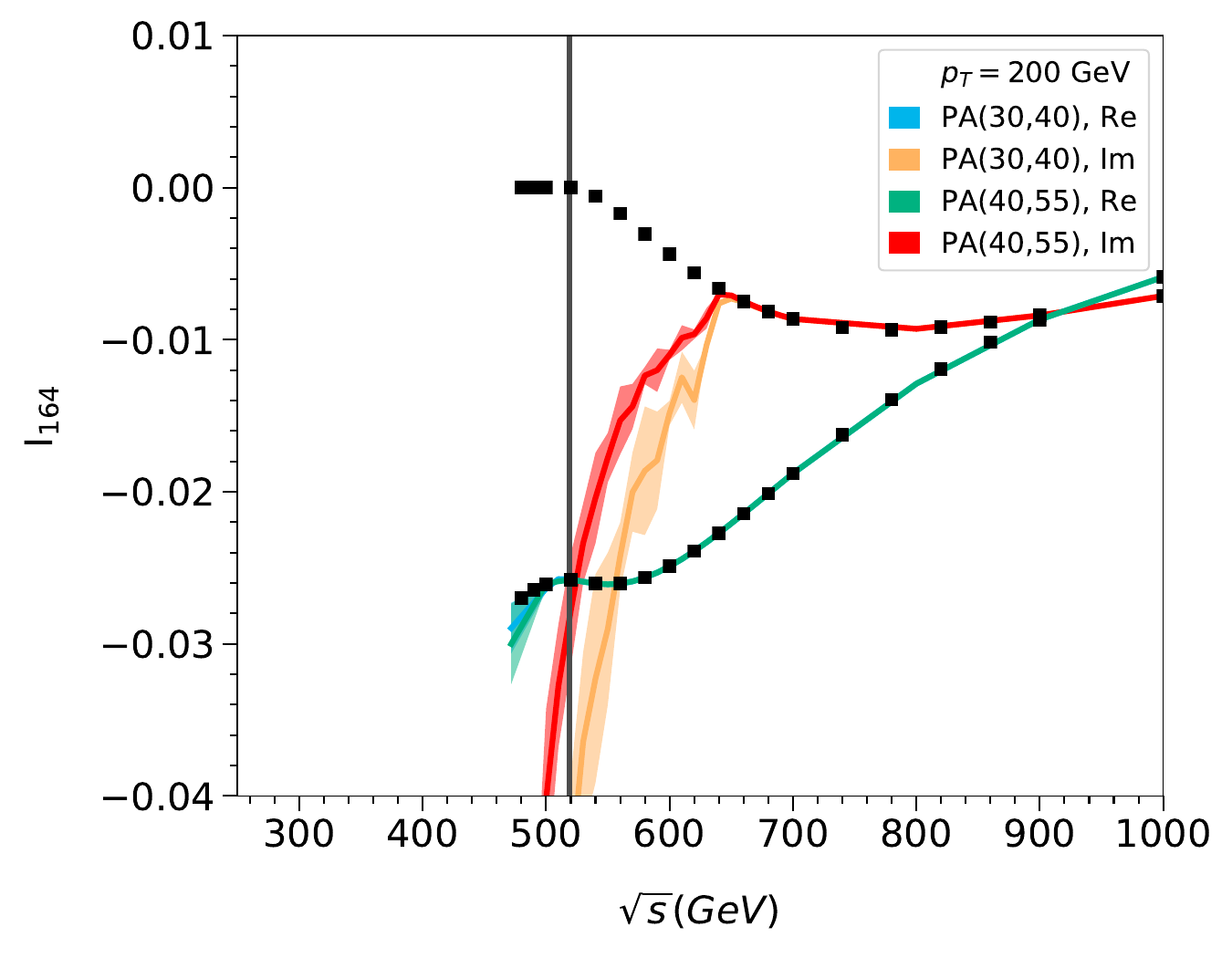}
    \\
    \includegraphics[width=0.30\textwidth]{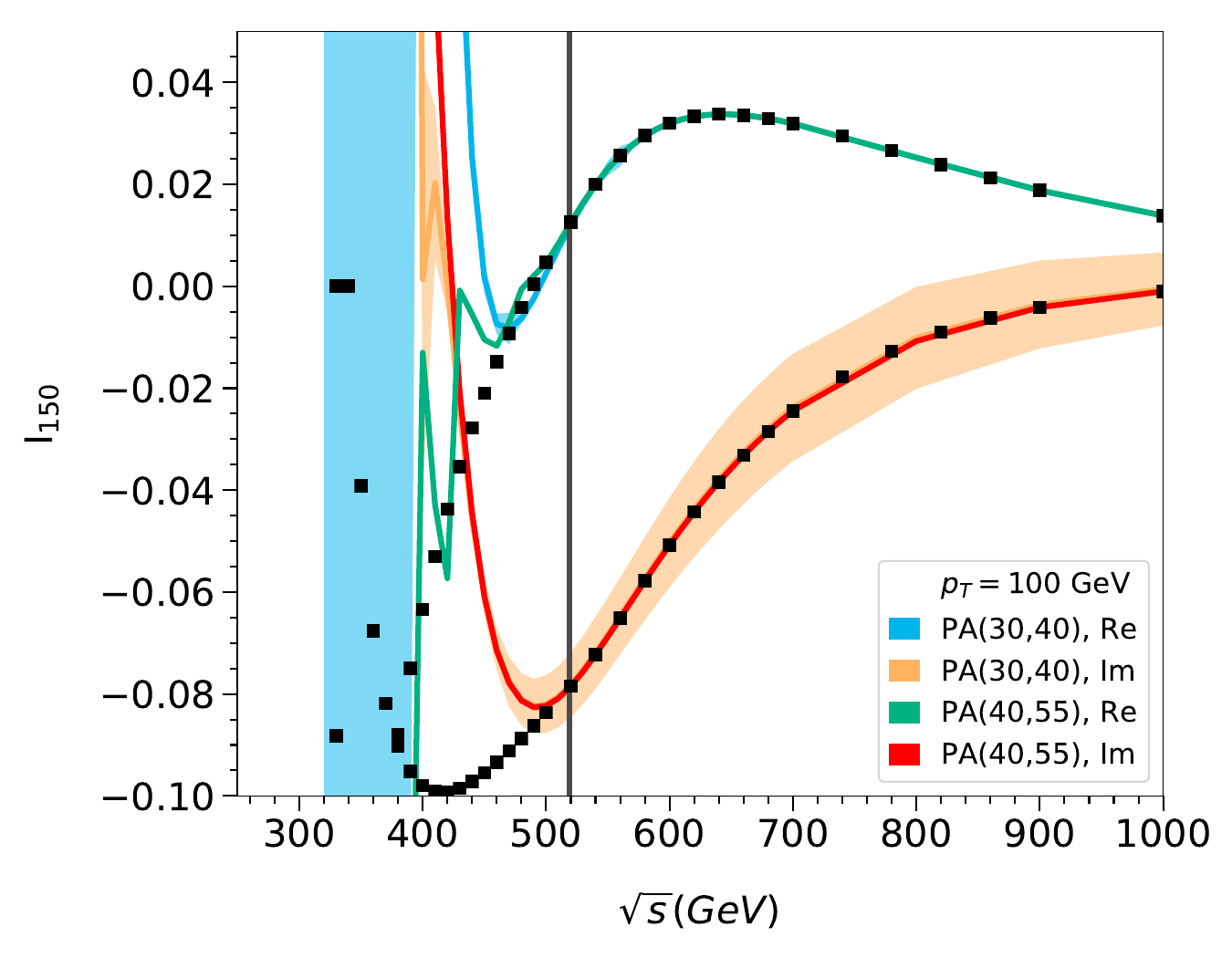}
    &
    \includegraphics[width=0.30\textwidth]{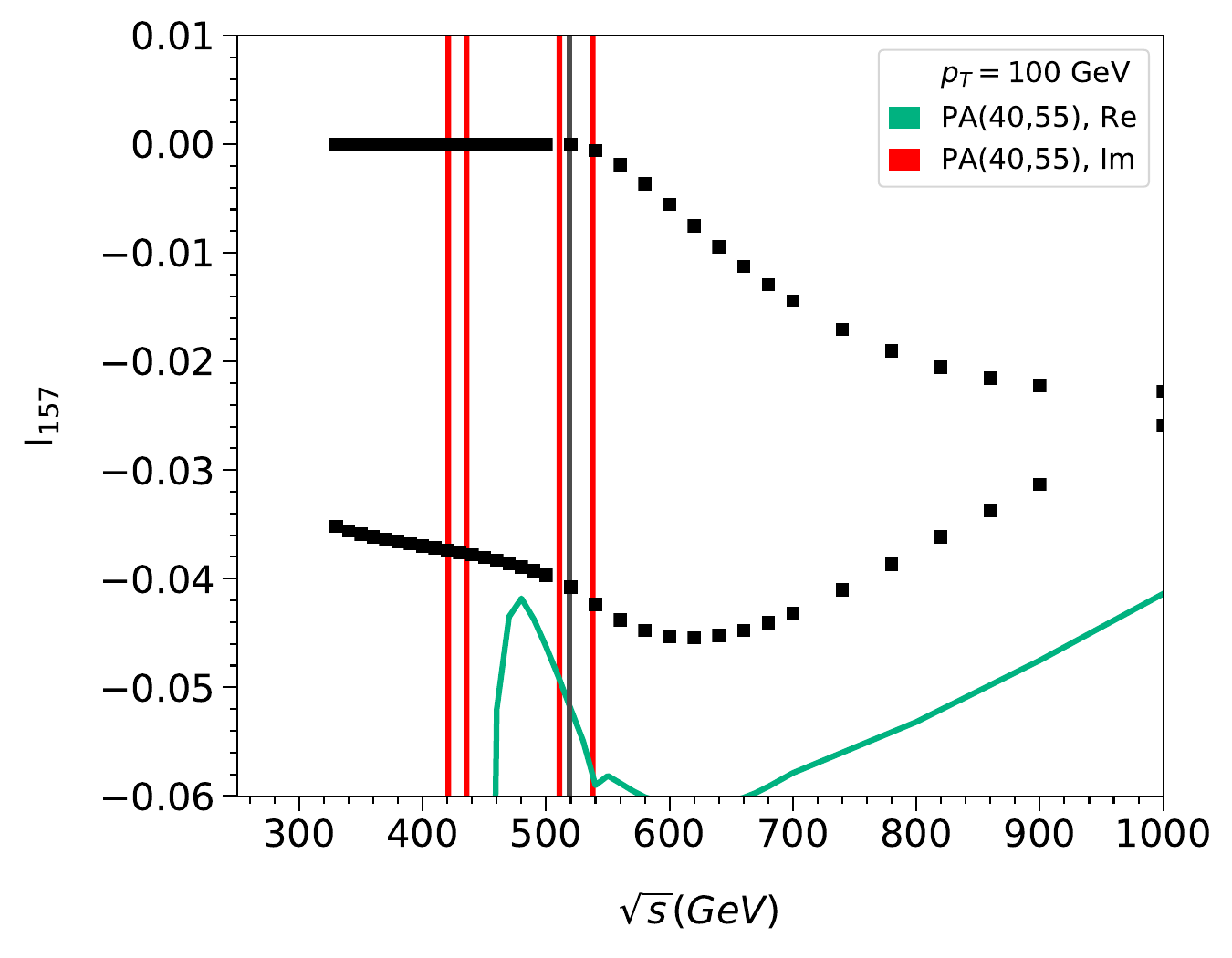} 
    & 
    \includegraphics[width=0.30\textwidth]{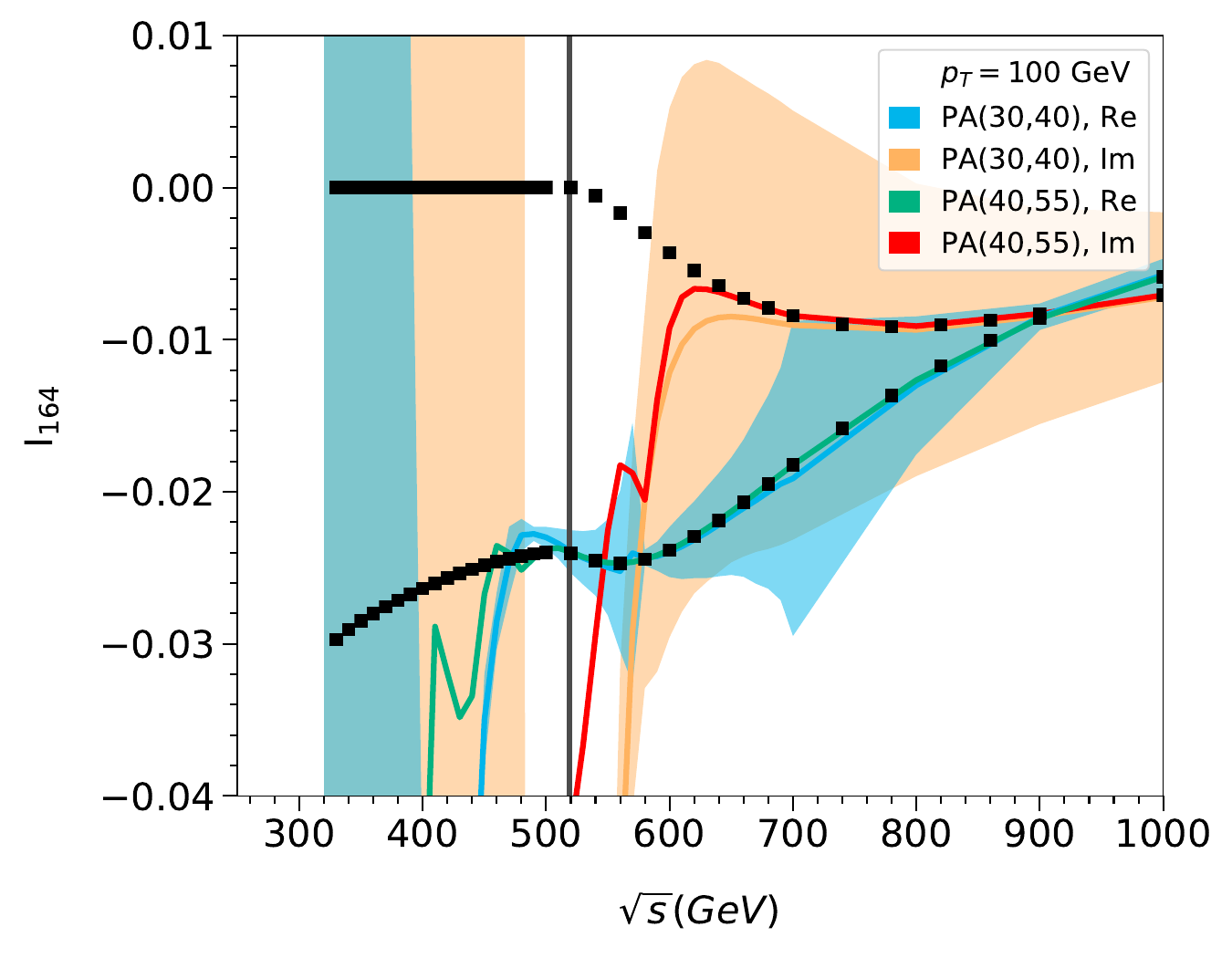}
    \end{tabular}
    \caption{\label{fig::masters_res} Dependence on $\sqrt{s}$ for real and
      imaginary parts for the three top-level master integrals shown in
      Fig.~\ref{fig::masters} for $p_T=300$~GeV (top row) $p_T=200$~GeV
      (middle row) and $p_T=100$~GeV (bottom row).  We multiply the integrals
      by $m_t^{6}$ such they are dimensionless.  The coloured lines and bands
      correspond to the results based on the Pad\'e-improved high-energy
      expansion and the black dots to the numerical results obtained with {\tt
        AMFlow}. No uncertainty bands are shown for $I_{157}$ for
      $p_T=100$~GeV and $p_T=200$~GeV. The gray vertical line indicates the
      threshold at $3m_t=519$~GeV.  }
\end{figure}

In Fig.~\ref{fig::masters_res} we show results for the $\epsilon^0$
coefficient of these integrals as a function of $\sqrt{s}$, for fixed values
of $p_T=300$~GeV, $p_T=200$~GeV and $p_T=100$~GeV.  For the internal mass
we use $m_t=173$~GeV and the renormalization scale is set to $\mu^2=s$.
Although the external masses are zero we use $m_H=125$~GeV in the relations
between $s, t, u$ and $p_T$. We follow Ref.~\cite{Davies:2020lpf} and
construct the so-called ``pole distance re-weighted'' Pad\'e approximants and
the corresponding uncertainties (see Section~4 of~\cite{Davies:2020lpf} for a
detailed discussion). We combine odd and even powers of
$m_t$~\cite{Davies:2023vmj} and construct all possible Pad\'e approximants
$[n/m]$ in the variable $m_t^2$ which include at least terms of order
$(m_t^2)^{N_{\rm low}}$ and at most of order $(m_t^2)^{N_{\rm high}}$, i.e.~we
have
\begin{eqnarray}
  	N_{\rm low}\le n+m \le N_{\rm high}\quad
	\textnormal{and}\quad N_{\rm low} \le n + m - | n - m | \,.
	\label{eq::N_low_high}
\end{eqnarray}

For demonstration purposes we show in Fig.~\ref{fig::masters_res} results for
the real and imaginary parts of the three master integrals for
$\{N_{\rm low},N_{\rm high}\}=\{30,40\}$ and
$\{N_{\rm low},N_{\rm high}\}=\{40,55\}$. We show both the central values
and the uncertainty bands as estimated by the Pad\'e procedure. The results are
compared to numerical results obtained with {\tt
  AMFlow}~\cite{Liu:2022chg}, which have no visible uncertainties.  We
plot the results from the kinematically allowed threshold up to
$\sqrt{s}=1000$~GeV.

For $I_{150}$ (left-hand column) we observe good agreement between the Pad\'e
approximation and the numerical results for each value of $p_T$. Deviations
are only observed for $p_T=100$~GeV when $\sqrt{s}$ is below the
three-particle cut at $3m_t= 519$~GeV, shown as a vertical grey line.  It is
impressive that for even lower values of $p_T$ (not shown in
Fig.~\ref{fig::masters_res}) good results are obtained for
$\sqrt{s}\gtrsim 3 m_t$.

In the case of $I_{164}$ (right-hand column) good agreement is observed for
the real part for all values of $p_T$ displayed, which has a similar quality
as in the case of $I_{150}$. For larger values of $\sqrt{s}$ the imaginary
part also shows a good behaviour, however for $\sqrt{s}\lesssim 650$~GeV we
observe a significant deviation of the Pad\'e result, though its uncertainty
band remains rather small.  This value of $\sqrt{s}$ is not directly connected
to any of the cuts of the diagram which are at $2m_t=346$~GeV for $\sqrt{-u}$
and $3 m_t=519$~GeV in all three Mandelstam variables. It appears that the
non-planar diagram has pseudo-thresholds which cause this behaviour.  The
advantage of the deep expansions in $m_t$ is clearly seen in the bottom-right
plot where $p_T=100$~GeV has been chosen. The sizeable uncertainty bands which
are still present for lower-order Pad\'e results largely disappear for the
high-order Pad\'e approximations.

For $I_{157}$ we observe good agreement for the real part for $p_T=300$~GeV
and $p_T=200$~GeV. However, already for $p_T=300$~GeV the imaginary part shows
an unstable behaviour and does not agree with the numerical results.  The
situation becomes worse for lower values of $p_T$ where the uncertainty bands
fill the whole plot area. For this reason we refrain from showing it.  For
$p_T=100$~GeV also the real part of the Pad\'e approximant does not agree with
the numerical result.

Although at first sight the three master integrals shown in
Fig.~\ref{fig::masters_res} look similar, their convergence properties are
very different.  The three diagrams each have three-particle cuts in all of
the Mandelstam variables $s,t$ and $u$.  However, the main difference is that
each of them has only one two-particle cut either in $\sqrt{s}$ ($I_{150}$),
$\sqrt{-t}$ ($I_{157}$) or $\sqrt{-u}$ ($I_{164}$), which seems to be
responsible for the observed convergence behaviour. For example, for the
kinematic values chosen in Fig.~\ref{fig::masters_res} $\sqrt{-t}$ is close to,
or even below, $2m_t$.  This might be the reason for the poor convergence
behaviour of $I_{157}$.  For lower $p_T$ and lower $\sqrt{s}$ additionally
$\sqrt{-u}$ becomes close to $3 m_t$, which might explain the observed behaviour
of $I_{164}$.

\begin{figure}[t]
  \centering
  \begin{tabular}{ccc}
    \includegraphics[width=0.30\textwidth]{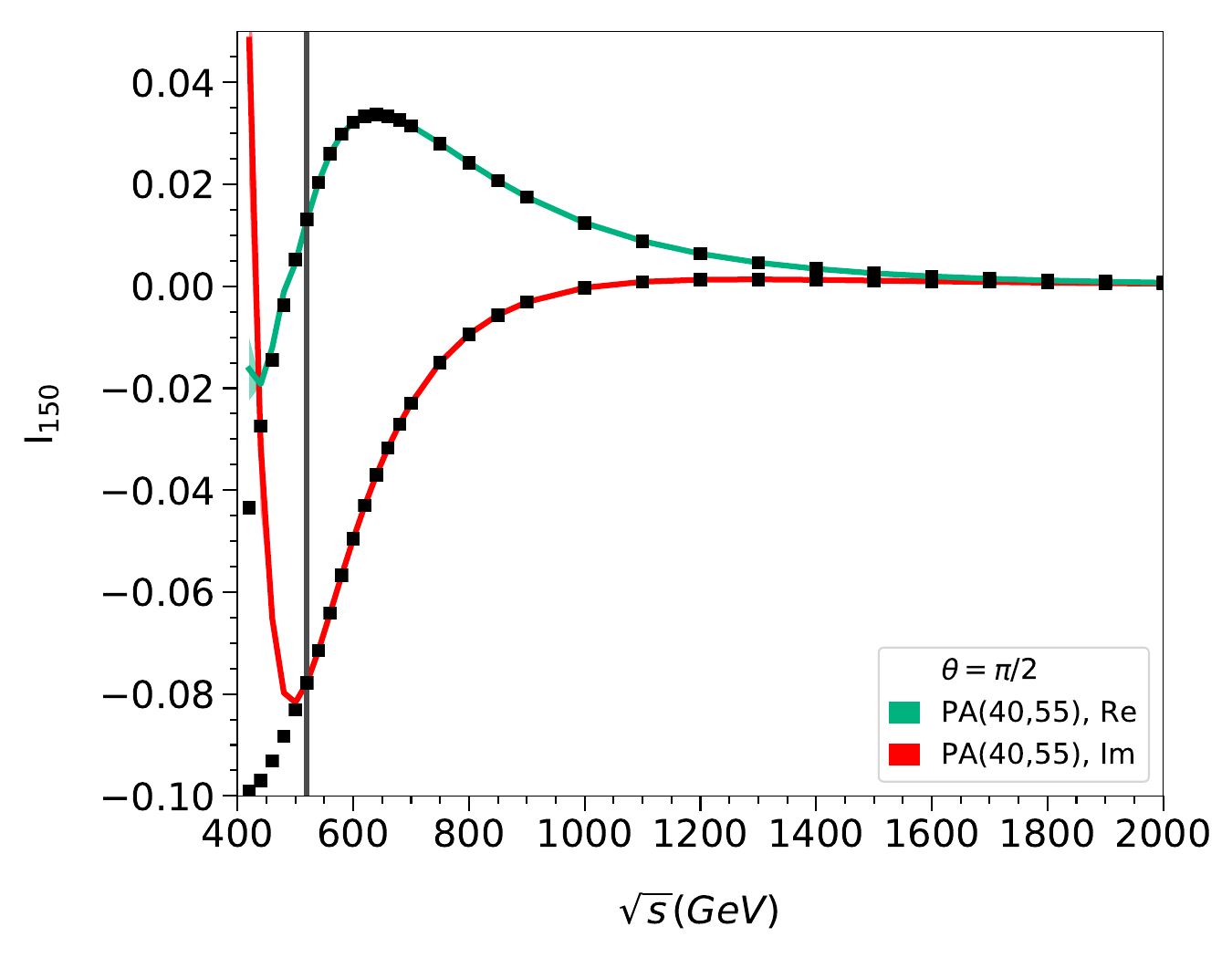}
    & 
    \includegraphics[width=0.30\textwidth]{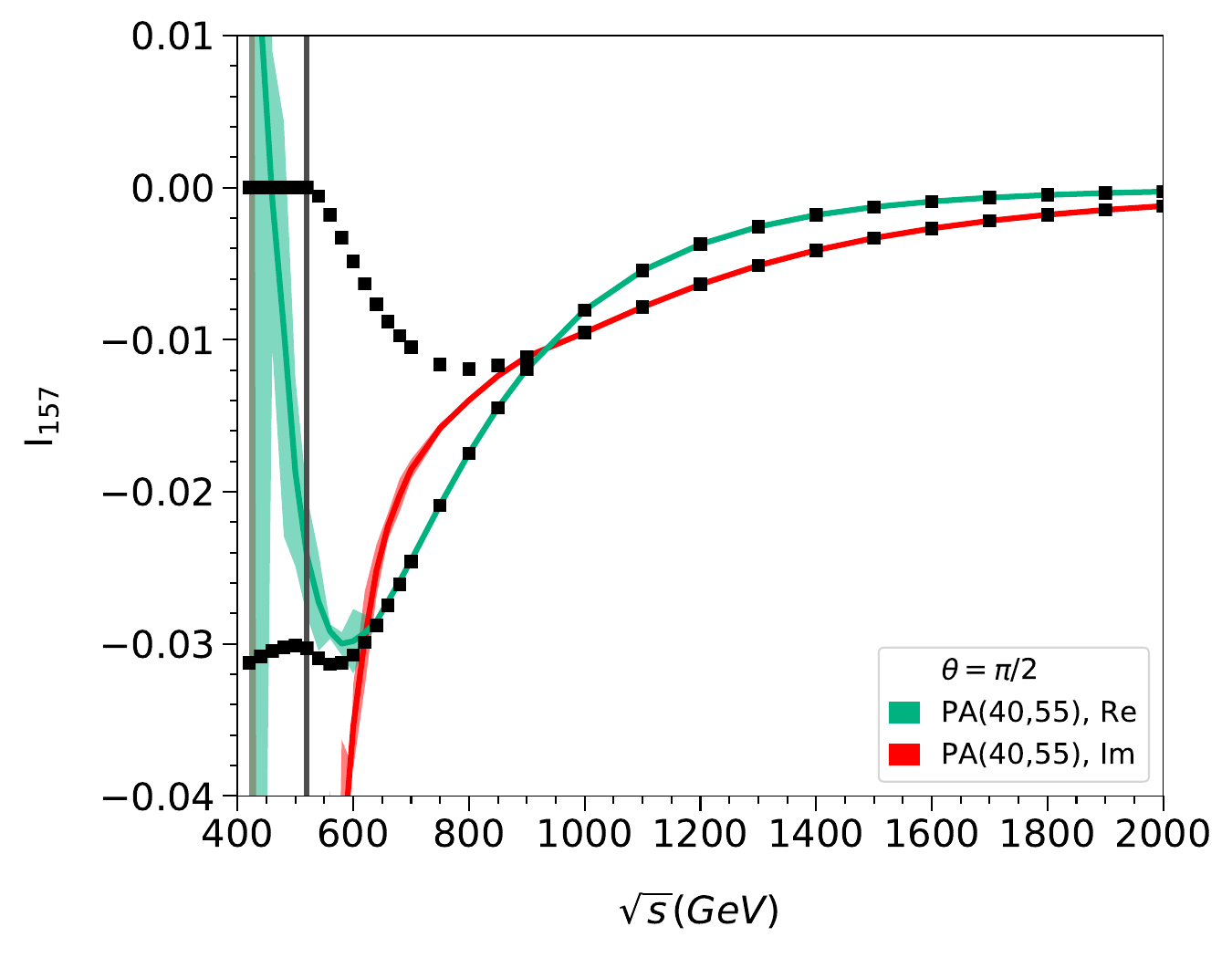}
    & 
    \includegraphics[width=0.30\textwidth]{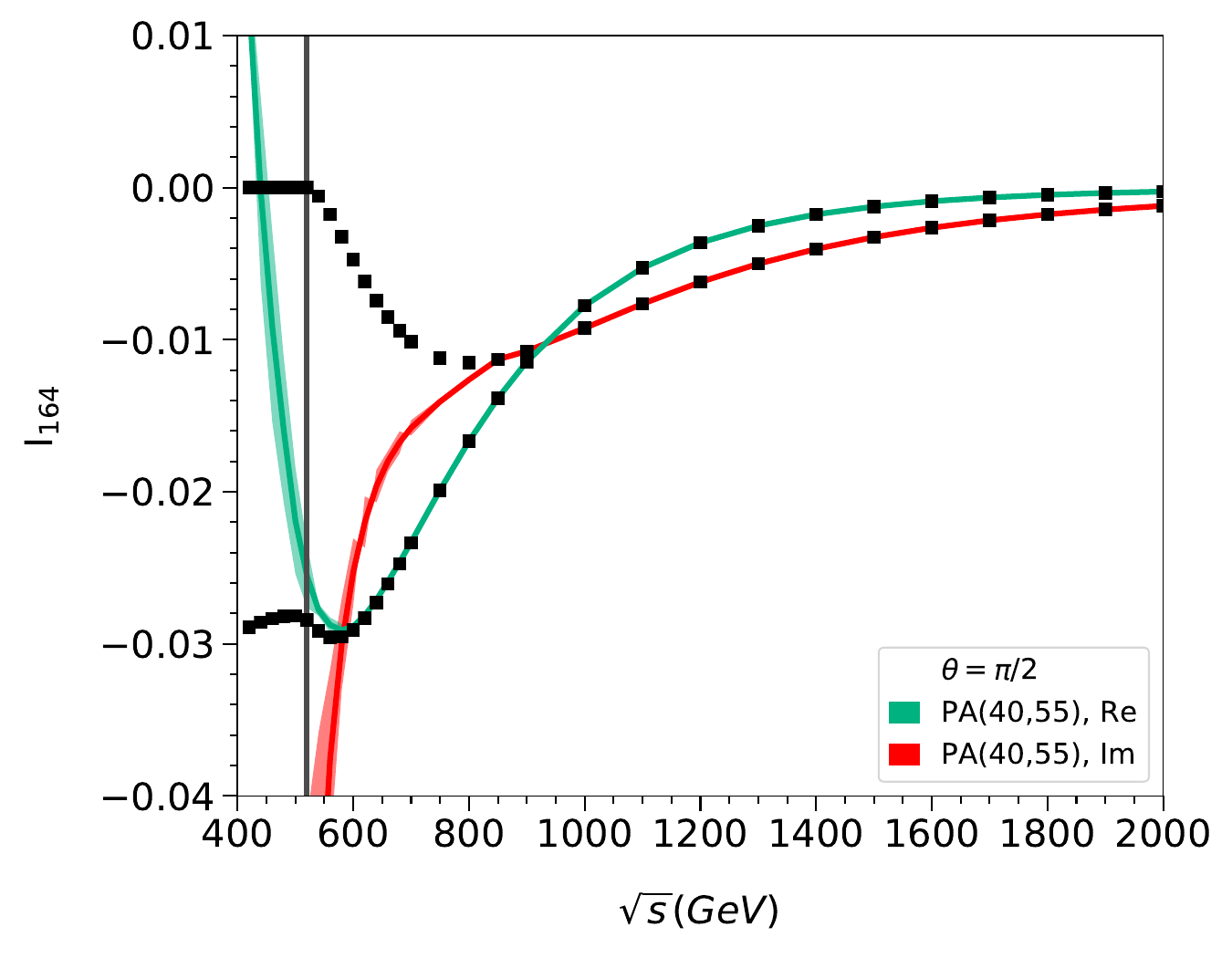}
    \\
    \includegraphics[width=0.30\textwidth]{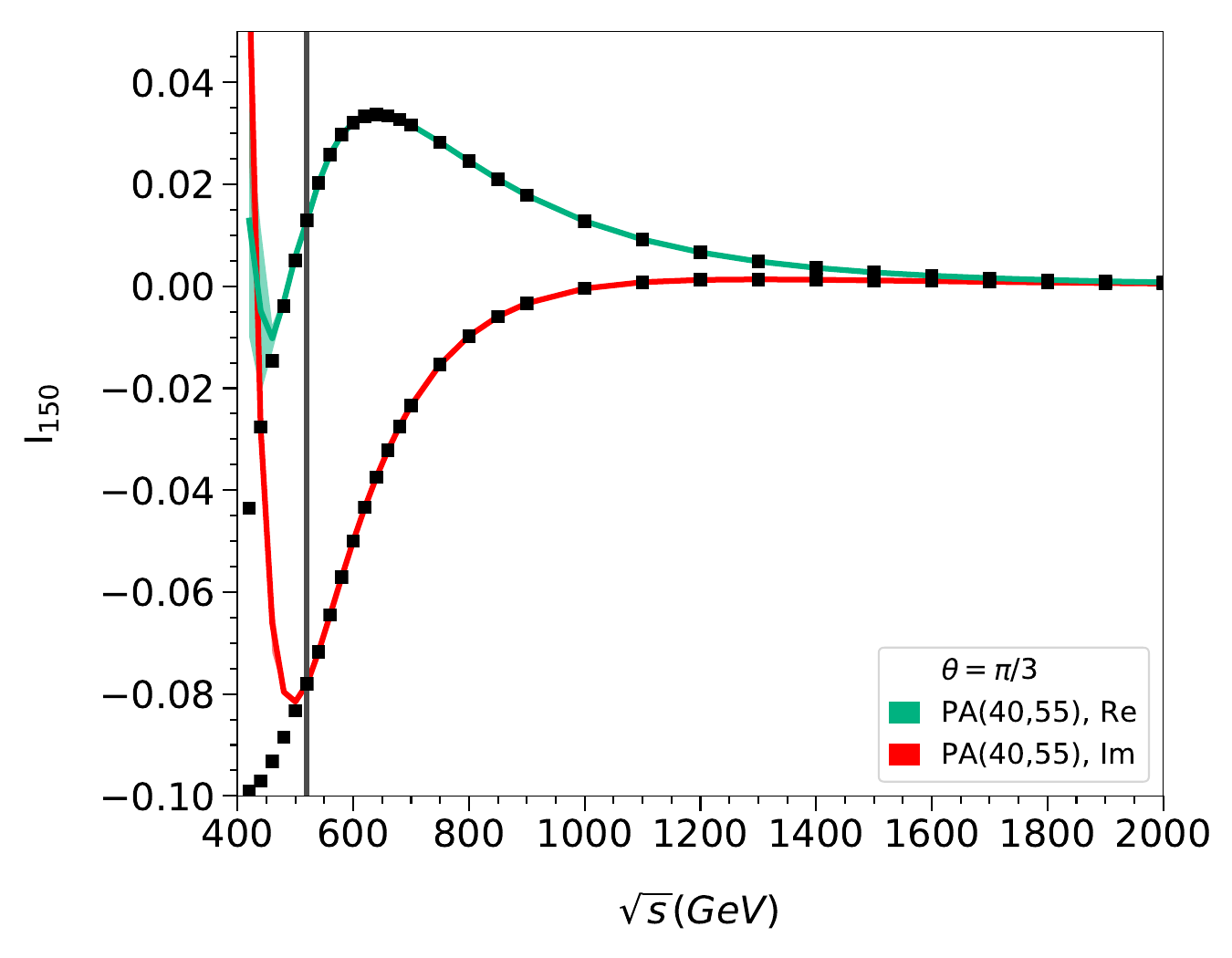}
    & 
    \includegraphics[width=0.30\textwidth]{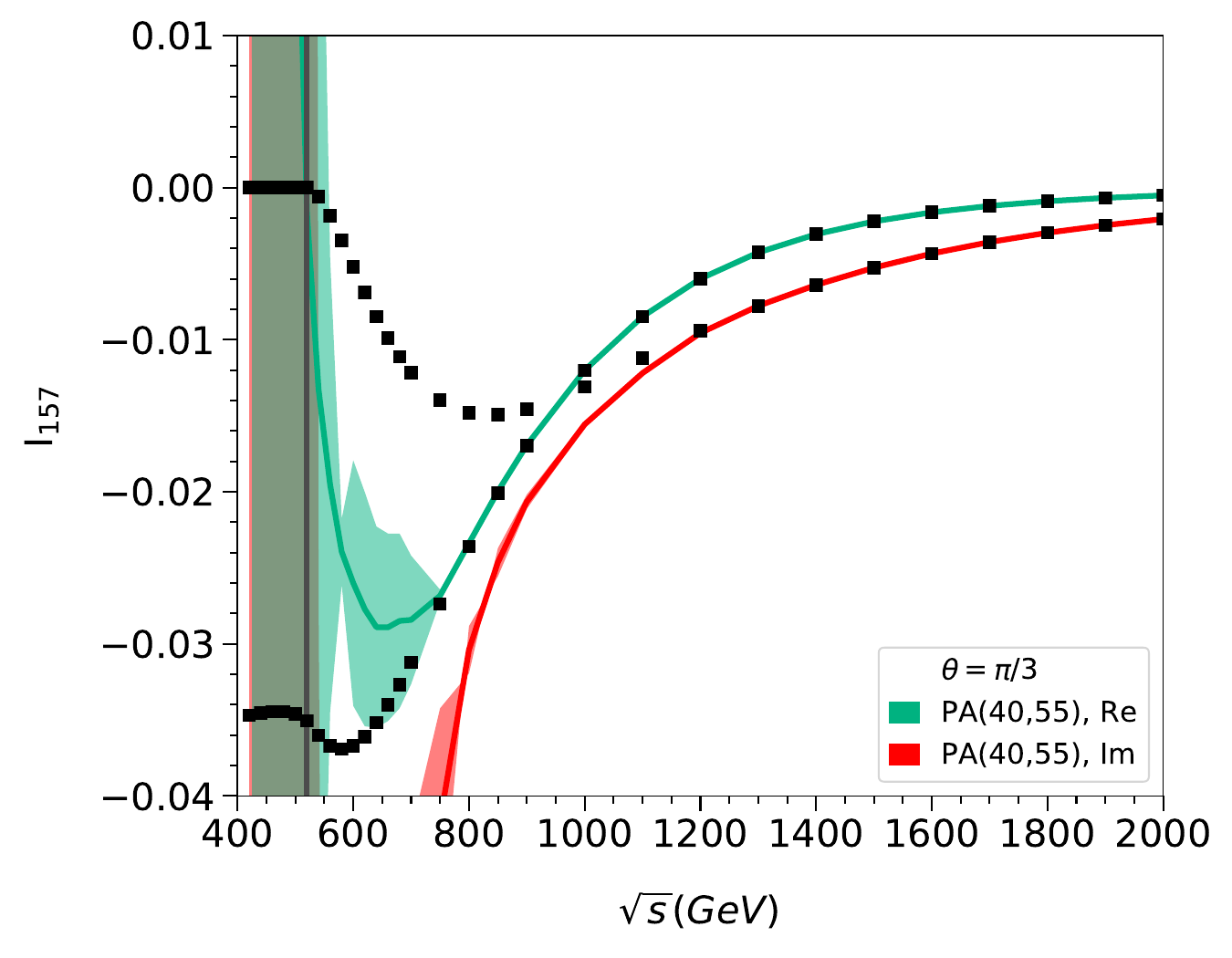}
    & 
    \includegraphics[width=0.30\textwidth]{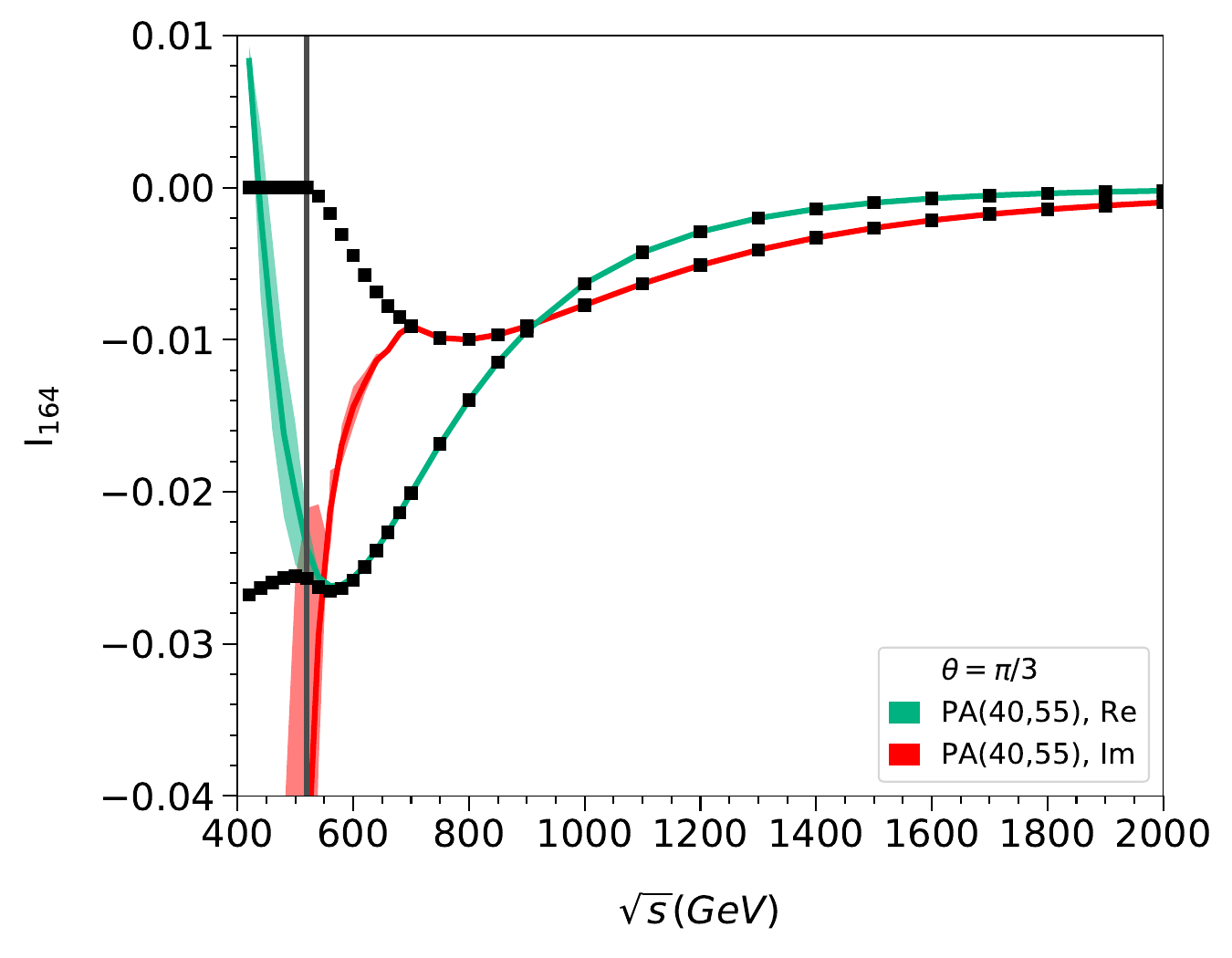}
    \\
    \includegraphics[width=0.30\textwidth]{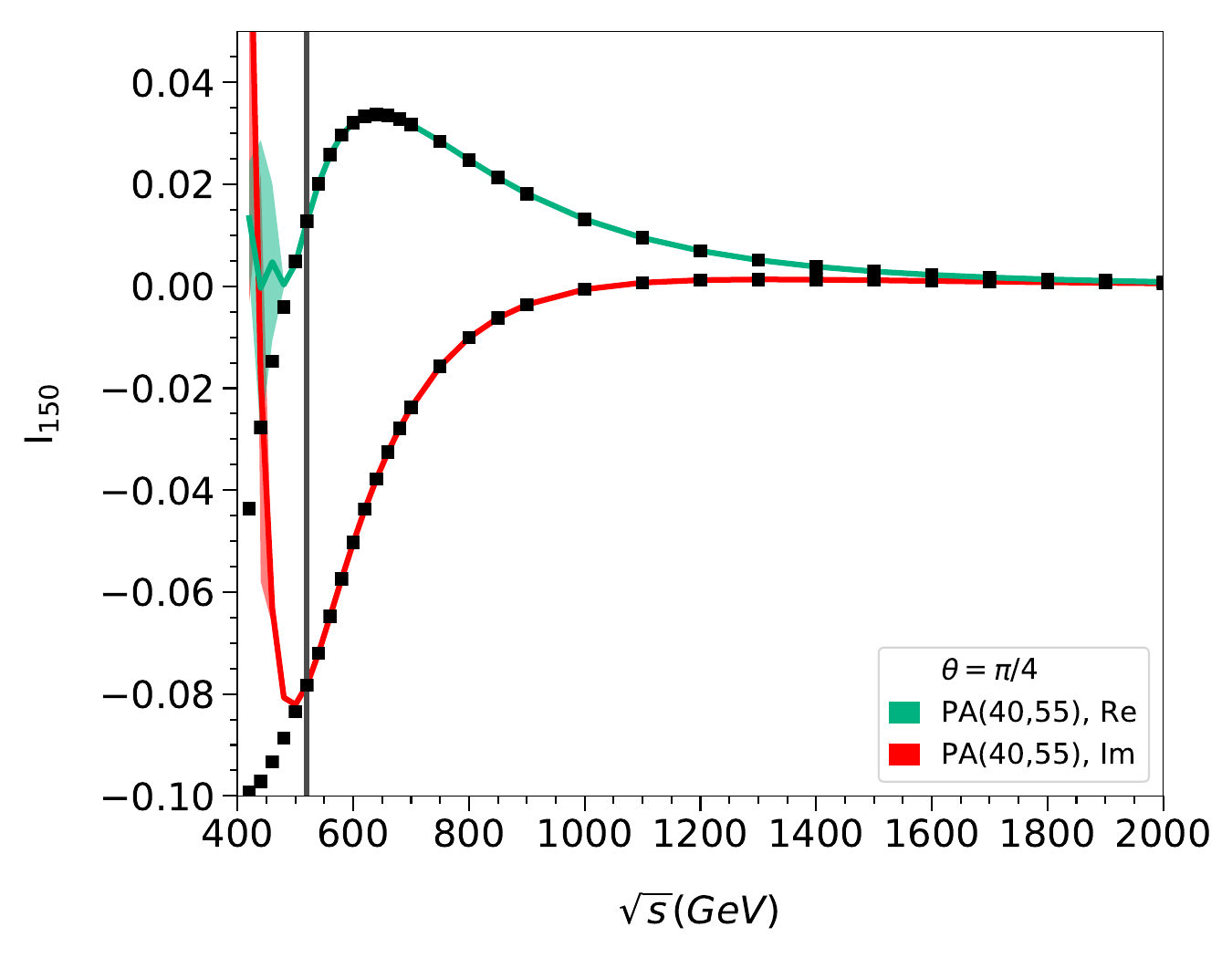}
    &
    \includegraphics[width=0.30\textwidth]{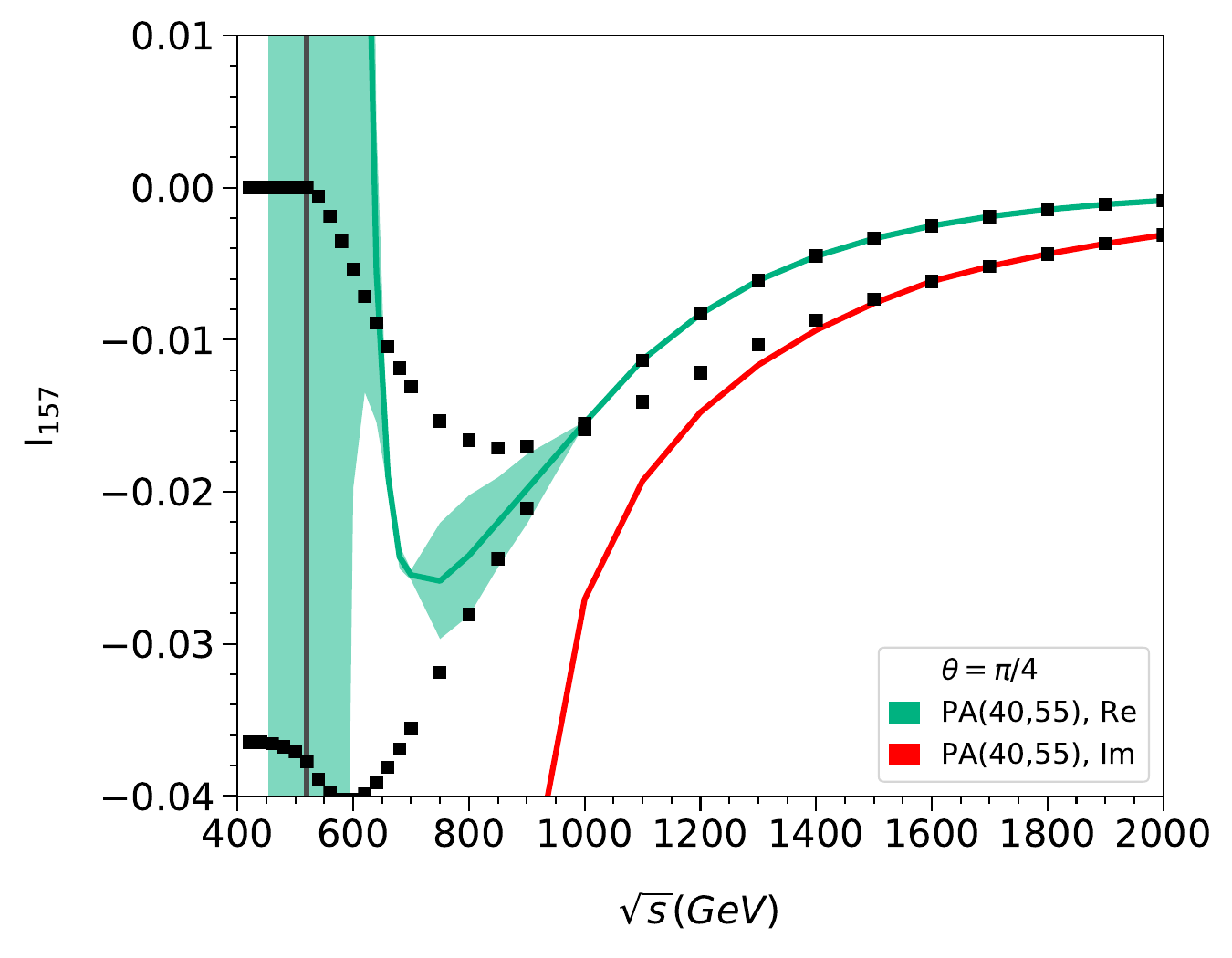} 
    & 
    \includegraphics[width=0.30\textwidth]{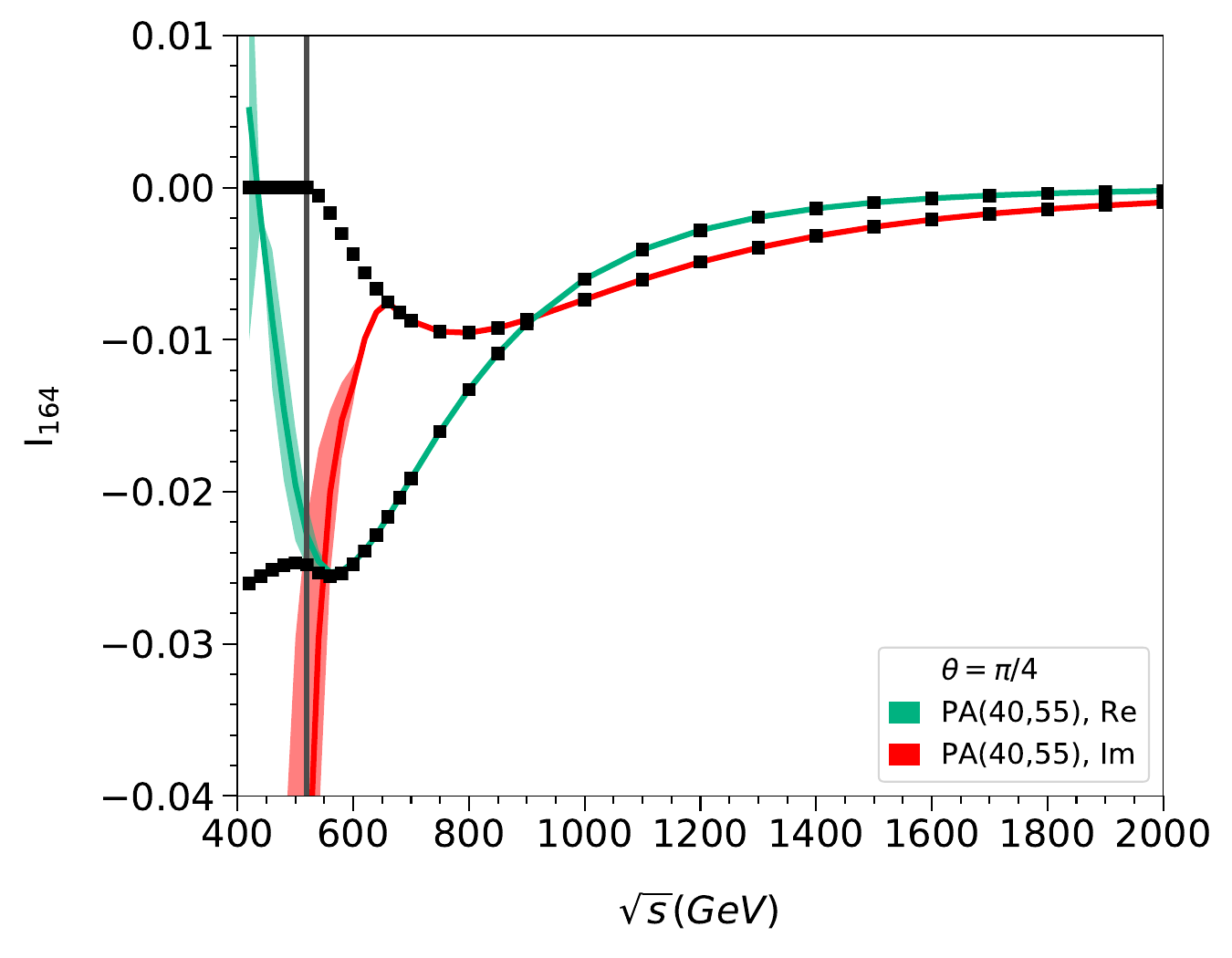}
    \end{tabular}
    \caption{\label{fig::masters_theta} Dependence on $\sqrt{s}$ for finite
      real and imaginary parts for the three top-level master integrals shown
      in Fig.~\ref{fig::masters} for scattering angles $\theta=\pi/2$ (top
      row) $\theta=\pi/3$ (middle row) and $\theta=\pi/4$ (bottom row).  We
      multiply the integrals by $m_t^{6}$ such they are dimensionless.  The
      coloured lines and bands correspond to the results based on the
      Pad\'e-improved high-energy expansion and the black dots to the
      numerical results obtained with {\tt AMFlow}. The gray vertical line
      indicates the threshold at $3m_t=519$~GeV. }
\end{figure}

In the phase-space region covered in Fig.~\ref{fig::masters_res} the master
integral $I_{157}$ shows bad convergence properties. In order to show that in
the high-energy region it reproduces the numerical results from {\tt AMFlow}
we show in Fig.~\ref{fig::masters_theta} the results of the three top-level
master integrals for fixed scattering angles $\theta$, which is obtained from
Eq.~(\ref{eq::pT_costhe}).  We choose three scattering angles, $\pi/2$,
$\pi/3$ and $\pi/4$.  For $I_{150}$ we observe good agreement for all angles
up to the threshold for $\sqrt{s}=3m_t$. This is also true for the real part
of $I_{164}$ whereas the imaginary part starts to deviate from the exact
result for $\sqrt{s}\lesssim 700-800$~GeV.  In the high-energy region good
results are also obtained for $I_{157}$ which slightly deteriorate for smaller
scattering angles.

In summary, the comparison of the Pad\'e-improved results for the master
integrals to numerical results obtained with {\tt AMFlow} provides confidence
for the correctness of the analytic high-energy expansions of the master
integrals.

Note that out of our four two-loop form factors introduced in
Eq.~(\ref{eq::F}), only $F_{\rm box}^{(0,y_t^3g_3)}$ receives contributions
from the families $G_{53}$ and $G_{54}$. Thus, we anticipate worse convergence
properties for (the imaginary part of) $F_{\rm box}^{(0,y_t^3g_3)}$
compared to the remaining three form factors. In general, we expect precise
predictions for all form factors in the region $p_T\gtrsim 300$~GeV.

\begin{table}[t]
\centering
  \begin{tabular}{c|c|c|c}
     $p_T$  & $\sqrt{s}$  & Pad\'e-improved results  & {\tt AMFlow} for $I_{150}$ \\
     (GeV) & (GeV) &  for $I_{150}$ up to $\mathcal{O}(m_t^{110})$ &  \\[2pt]
    \hline
     100 & 520 & $0.012231 (840) $ & $0.012616$ \\
     && $-0.07818 ( 670) \, \rm i $ & $-0.07848 \, \rm i$ \\[2pt] \hline
     100 & 1000 & $0.013853	(180) $ & $0.013856$\\
     && $-0.000986 (146)\, \rm i $ & $-0.000990 \, \rm i$ \\[2pt] \hline
     200 & 520 & $0.01290367 (594)$ & $0.01292119$ \\
     && $-0.07801176 (26) \, \rm i$ & $-0.07801270 \, \rm i $ \\[2pt] \hline
     200 & 1000 & $0.01363961528613 (121) $ & $0.01363961528606$ \\
     && $-0.00086534770114	(198) \,\rm i$ & $-0.00086534770126 \,\rm i $\\[2pt] \hline
     300 & 660 & $0.0333941347347566 (83)$ & $0.0333941347347600$ \\
     && $-0.0317960435292528 (113) \,\rm i$ & $-0.0317960435292533 \,\rm i$ \\[2pt] \hline
     300 & 1000 & $0.013302917959524752$ &  $0.013302917959524752 $ \\
     && $-0.0006827034768774084 \,\rm i$ & $-0.0006827034768774084 \, \rm i$\\[2pt] 
  \end{tabular}
  \caption{\label{tab::num_val} Pad\'e-improved and numerical results for
    selected values of $p_T$ and $\sqrt{s}$ for the $\epsilon^0$ term of the
    master integral $I_{150}$. For the Pad\'e approximation, we choose
    $\{N_{\rm low},N_{\rm high}\}=\{40,55\}$.  The shown uncertainties are
    from the Pad\'e procedure. We multiply this integral by $m_t^{6}$ such
    that it is dimensionless. }
\end{table}

To demonstrate the quality of the Pad\'e predictions within the radius of
convergence we compare in Tab.~\ref{tab::num_val} results for the master
integral $I_{150}$ for selected values of $p_T$ and $\sqrt{s}$ to the results
obtains with {\tt AMFlow}. It is impressive that for large values of
$\sqrt{s}$ and $p_T$ the Pad\'e method can reproduce more than 15 significant
digits of the exact result. Even for $p_T=200$~GeV and $\sqrt{s}=520$, which is
very close to the $3m_t=519$~GeV threshold, we obtain three significant
digits, and for $p_T=100$~GeV still two. In all cases the uncertainty estimate
covers the exact result.

\begin{figure}[tb]
  \centering
    \includegraphics[width=0.35\textwidth]{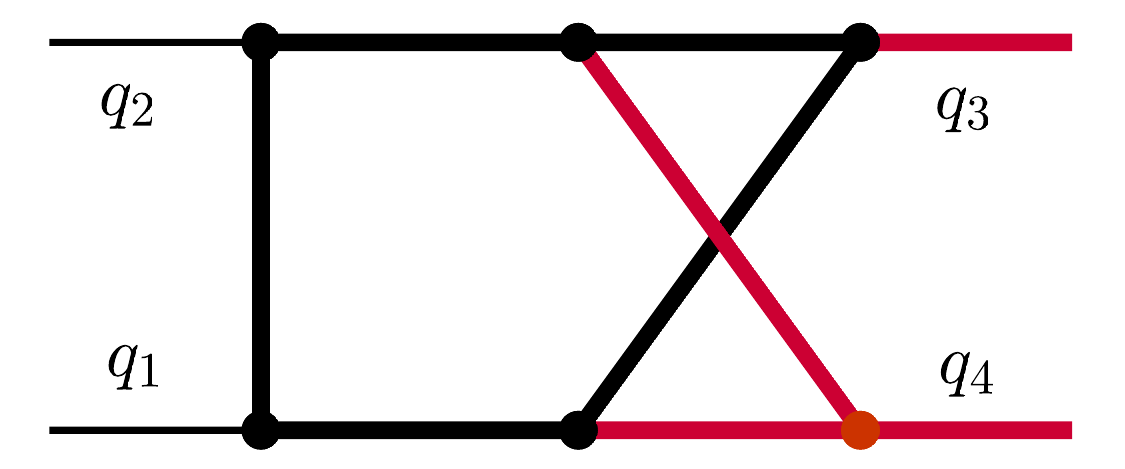}
    \caption{\label{fig::scalar} Scalar Feynman integral
        $I_{150}^{(y_t^3 \, g_3)}$ of family $G_{52}$ where thick black lines
        have mass $m_t$ and thick red lines have mass $m_H$. The external
        lines with momenta $q_1$ and $q_2$ are massless.}
\end{figure}

In a next step we study the quality of the expansion in the external Higgs
boson mass and in $\delta$. In Tab.~\ref{tab::num_val2} we show results for
the scalar non-planar integral $I_{150}^{(y_t^3 \, g_3)}$ depicted in
Fig.~\ref{fig::scalar} for selected values of $p_T$ and $\sqrt{s}$. We provide
both the central values and the Pad\'e uncertainties and compare to the
numerical results from {\tt AMFlow} (which always has an negligible
uncertainty).  For reference we also provide results for the expansion in the
external Higgs boson mass and $\delta^\prime$.

For higher values of $p_T$ and/or $\sqrt{s}$ the Pad\'e uncertainty is
negligible and we find agreement at the percent level or better. This
remaining difference can be attributed to the neglected terms in $\delta$ and
$m_H$. For smaller values of $p_T$ or values of $\sqrt{s}$ close to the
production threshold ($3m_t$) the Pad\'e uncertainty becomes larger; it covers
well the exact result. It is impressive that even for $p_T=100$~GeV and
$\sqrt{s}=520$~GeV we can reproduce one digit in the exact result.  In all
cases the expansion in $\delta$ is slightly better than the expansion
in~$\delta^\prime$.

\begin{table}[b]
\centering
  \begin{tabular}{c|c|c|c|c}
     $p_T$  & $\sqrt{s}$  &  $I_{150}^{(y_t^3  g_3)}$ &   $I_{150}^{(y_t^3  g_3)}$ & {\tt AMFlow} for $I_{150}^{(y_t^3 g_3)}$ \\
     (GeV) & (GeV) & up to $\mathcal{O}(m_H^4 \, \delta^{'3})$ & up to $\mathcal{O}(m_H^4\, \delta^{3})$ &  \\[4pt]
    \hline
    100 & 520 & $0.0640 (35)  $ & $ 0.0667 (37)  $ & $ 0.06982  $ \\
    & & $-0.1457 (124)\,\rm i$ & $-0.1472 (119) \,\rm i$ & $-0.14878 \, i$ \\[4pt] \hline
    100 & 1000 & $0.02084 ( 50 ) $ & $ 0.02105 ( 52 ) $ & $0.02115 $ \\
    && $+ 0.00240 (21) \,\rm i$ & $+ 0.00254 (21)\,\rm i$ & $+ 0.00265 \, i$ \\[4pt] \hline
    200 & 520 & $0.06502 (9) $ & $ 0.06826 (67) $ & $0.07034 $ \\
    && $-0.14501 (8) \,\rm i$ & $- 0.14648 (1) \,\rm i$ & $-0.14728 \,\rm i$ \\[4pt] \hline
    200 & 1000 & $0.02038+ 0.00254 \,\rm i$ & $0.02058 +0.00268 \,\rm i$ & $0.02066 + 0.00279\,\rm i$ \\[4pt] \hline
    300 & 660 & $0.068325 - 0.038190 \,\rm i$ & $ 0.069487 -0.038045 \,\rm i$ & $0.070304 -0.037667 \,\rm i$ \\[4pt] \hline
    300 & 1000 & $0.019667 +0.002729 \,\rm i$ & $ 0.019841 + 0.002864 \,\rm i$ & $0.019915 + 0.002973\,\rm i$ \\[4pt] \hline
     400 & 840 & $0.034359 -0.002381\,\rm i $ & $0.034717 -0.002176 \, i$ & $0.034897 - 0.001972 \,\rm i $ \\[4pt] \hline
     400 & 1200 & $0.009551 + 0.003347 \,\rm i$ & $0.009616 + 0.003420 \,\rm i$ & $0.009640 + 0.003468 \,\rm i$
       \end{tabular}
       \caption{\label{tab::num_val2} Results for the $\epsilon^0$ term of the
         scalar non-planar integral $I_{150}^{(y_t^3 \, g_3)}$ (see
         Fig.~\ref{fig::scalar}) which enters $F_{\rm
           box}^{(0,y_t^3\,g_3)}$. The expansion in $m_H$ and $\delta^\prime$
         respectively $m_H$ and $\delta$ is compared to the numerical result
         form {\tt AMFlow} for various values of $p_T$ and $\sqrt{s}$.  The
         uncertainties shown are from the Pad\'e procedure.}
\end{table}

To conclude this subsection we want to stress that: (1) the Pad\'e method as
applied within the region of convergence is a precision tool and (2) the
uncertainty estimate is reliable.  (3) For the form factors we expect
agreement with the exact result within 1\% or better if $p_T$ is sufficiently
large, say above 200 to 300~GeV, with the exception of
$\mbox{Im}(F_{\rm box}^{(0,y_t^3 g_3)})$, where $p_T$ needs to be
above about $500$~GeV.

We provide results for all 168 master integrals expanded up to order
$m_t^{100}$ in the supplementary material to the paper which can be downloaded
from~\cite{progdata}.


\subsection{Form factors}
\label{sec::FFs}

This subsection is devoted to analytic and numerical results of the form
factors. We study in particular the quality of the various expansions.  Let us
start with the analytic structure of our result.  As an illustration we show in
the following the two leading terms in the high-energy expansion of the form
factor $F_{\rm box1}^{(0,y_t^2 g_3^2)}$. They are of order $m_t^2$ and
$m_t^3$, respectively. We have
\begin{eqnarray}
   F_{\rm box1}^{(0,y_t^2 g_3^2)} &=& -
   m_H^2 
   \Biggl\{
        m_t^2 \biggl[
                 \frac{18 L_m^2}{s^2}
                +L_m \biggl(
                        -\frac{9 \pi ^2}{s^2}
                        -\frac{18 t }{s^2 (s+t)} H_{0,0}
                        -\frac{18 }{s^2} H_{0,1}
                        -\frac{18 }{s^2} H_{1,0}
                        \nonumber \\ &&
                        -\frac{18 (s+t) }{s^2 t} H_{1,1}
                \biggr)
                +\pi ^2 \biggl(
                        -\frac{27}{2 s^2}
                        +\frac{9 t }{2 s^2 (s+t)} H_0
                        -\frac{9 (s+t) }{2 s^2 t} H_1
                \biggr)
                \nonumber \\ &&
                -\frac{18 t }{s^2 (s+t)} H_{0,0}
                -\frac{18 }{s^2} H_{0,1}
                -\frac{18 }{s^2} H_{1,0}
                -\frac{18 (s+t) }{s^2 t} H_{1,1}
                +\frac{18 t }{s^2 (s+t)} H_{0,0,0}
                \nonumber \\ &&
                +\frac{9 }{s^2} H_{0,0,1}
                +\frac{9 }{s^2} H_{0,1,0}
                -\frac{9 }{s t} H_{0,1,1}
                -\frac{9 }{s (s+t)} H_{1,0,0}
                -\frac{9 }{s^2} H_{1,0,1}
                -\frac{9 }{s^2} H_{1,1,0}
                \nonumber \\ &&
                -\frac{18 (s+t) }{s^2 t} H_{1,1,1}
                -\frac{9 t }{s^2 (s+t)} \zeta_{3}
                + {\rm i} \pi \Biggl(
                        L_m \biggl(
                                \frac{18}{s^2}
                                +\frac{18 }{s (s+t)} H_0
                                +\frac{18 }{s t} H_1
                        \biggr)
                        \nonumber \\ &&
                        \frac{18}{s (s+t)}  H_0
                        +\frac{18 }{s t} H_1
                        -\frac{9 }{s (s+t)} H_{0,0}
                        +\frac{9 }{s t} H_{0,1}
                        -\frac{9 }{s (s+t)} H_{1,0}
                        +\frac{9 }{s t} H_{1,1}
                        \nonumber \\ &&
                        -\frac{3 \pi ^2}{2 s (s+t)} 
                \Biggr)
        \biggr]
        + {\rm i} m_t^3 \sqrt{\frac{-t}{s+t}} \frac{9 \pi ^2 (5 c_Z+16 c_{Z_2}) }{8 t s^{3/2}}
    + {\cal O}(m_t^4)
    \Biggr\} 
    \nonumber \\ && 
    + {\cal O}(m_H^4) + {\cal O}(\delta)
   \,,
\end{eqnarray}
with $L_m = \log(m_t/\sqrt{s})$ and $H_{\vec{w}} \equiv H_{\vec{w}}(-t/s)$ are
harmonic polylogarithms. Note that the overall factor $m_H^2$ arises due to
the way we write the prefactor in Eq.~(\ref{eq::F}).  It is not counted in our
expansion in $m_H^{\rm ext}$.  We see the general feature that the even powers
in $m_t$ can be expressed in terms of harmonic polylogarithms and rather well
known polylogarithmic constants, while the odd powers also contain the
elliptic constants $c_Z$ and $c_{Z_2}$.  Analytic results for all form factors
introduced in Eq.~(\ref{eq::F}) can be found in the supplementary
material~\cite{progdata}.

Next we turn to the numerical analysis.  For the input parameters we choose
$m_t=173$~GeV, $m_H^{\rm int}=m_H^{\rm ext}=m_H=125$~GeV and set $\mu^2=s$.
Furthermore, we fix $p_T$ and then show the dependence on $\sqrt{s}$.  As
discussed for the individual master integrals in the previous subsection we
construct, for each phase-space point, the Pad\'e-improved result which consists
of a central value and an uncertainty.  If not stated otherwise we include
expansion terms between $m_t^{80}$ and $m_t^{100}$ to construct the Pad\'e
approximants.  A detailed analysis of the one-loop results of the form factors
can be found in Ref.~\cite{Davies:2022ram}. In the following we will thus
concentrate on the bare two-loop expressions. Furthermore, we will restrict
the discussion to $F_{\rm box1}$; we also have results for $F_{\rm box2}$ (see
Ref.~\cite{progdata}).  Note that results for $F_{\rm box1}^{(0,y_t^4)}$ can
also be found in Ref.~\cite{Davies:2022ram}, although only for the
renormalized form factor.

\begin{figure}[t]
  \centering
  \begin{tabular}{cc}
    \includegraphics[width=0.45\textwidth]{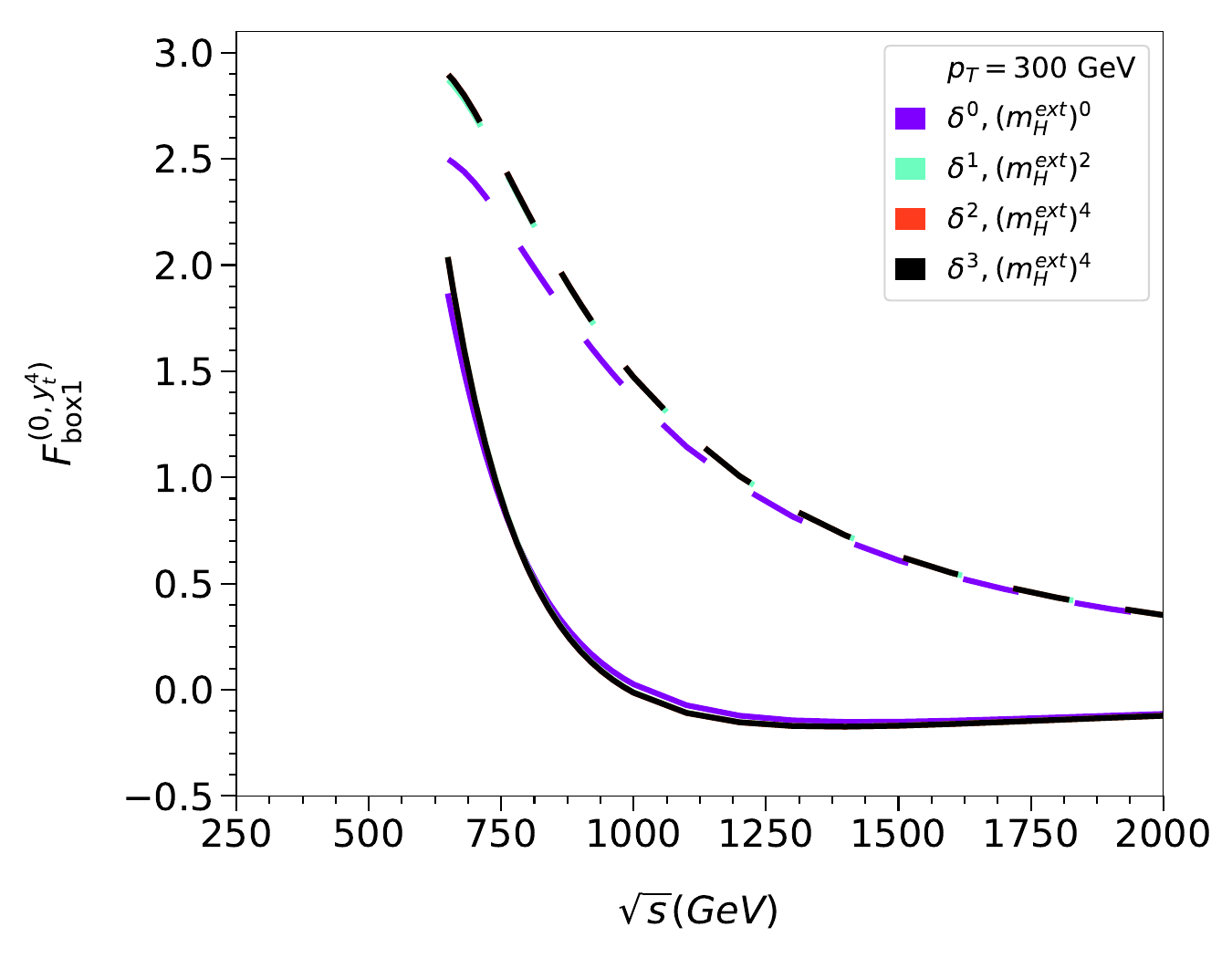} &
    \includegraphics[width=0.45\textwidth]{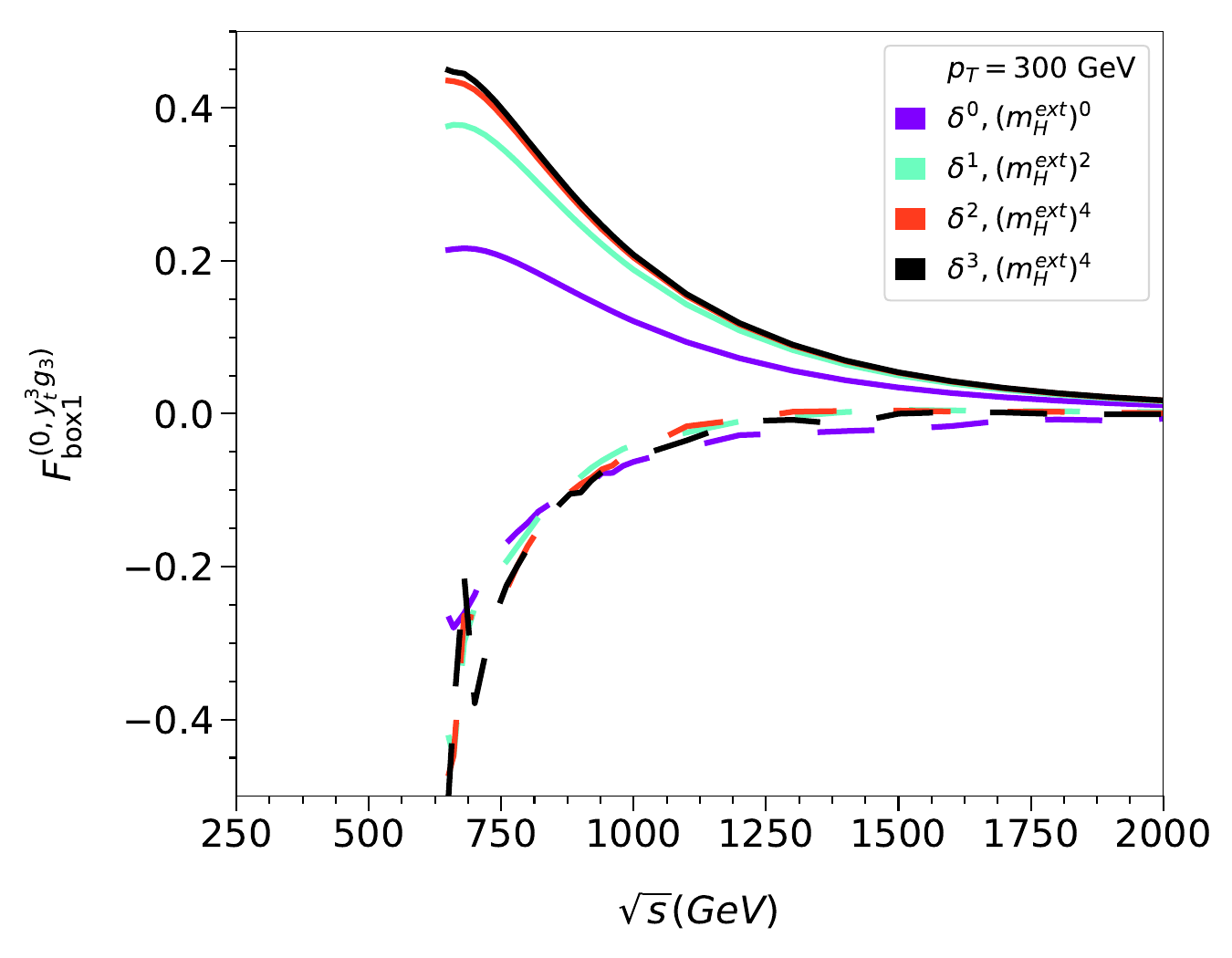}
    \\
    \includegraphics[width=0.45\textwidth]{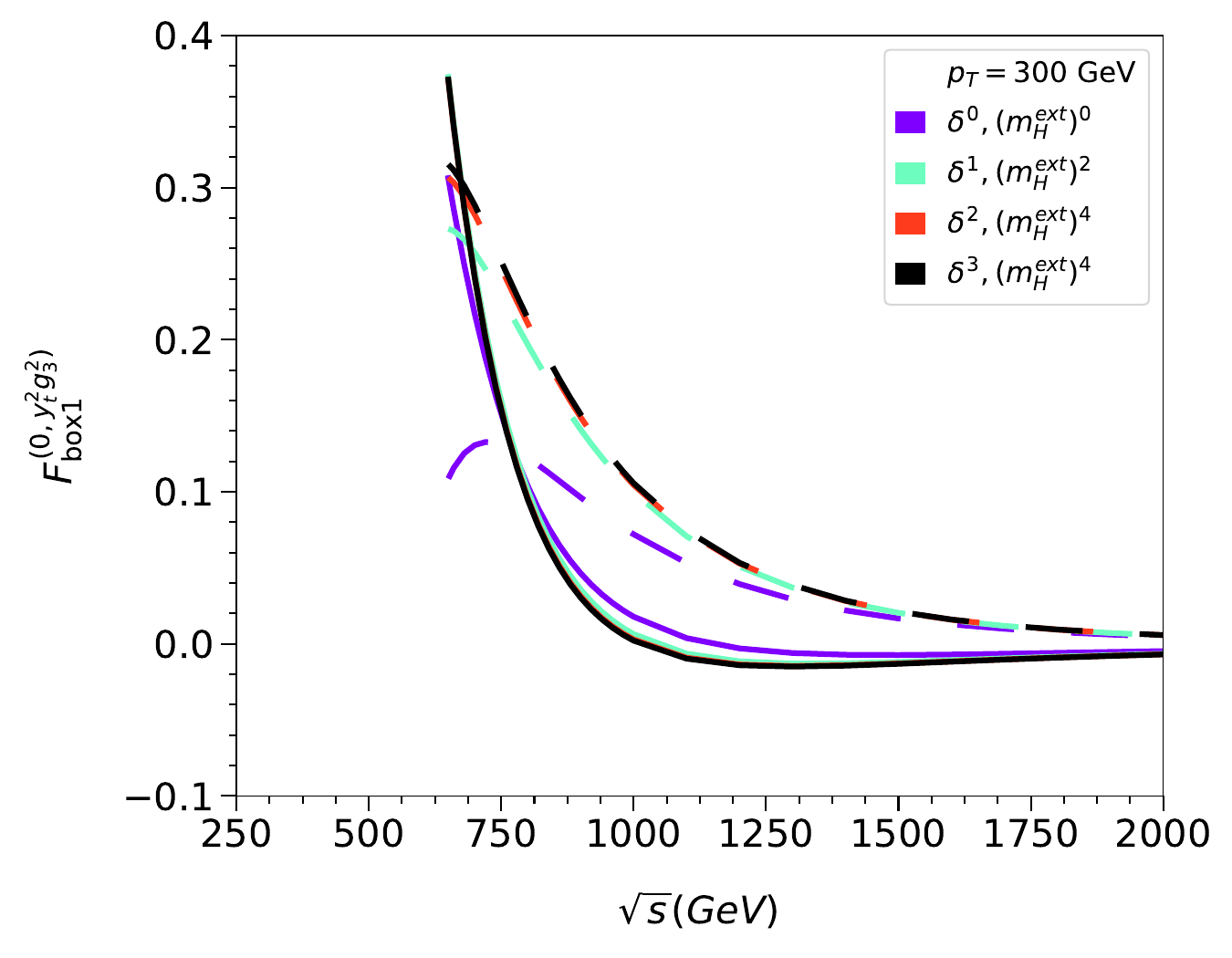} &
    \includegraphics[width=0.45\textwidth]{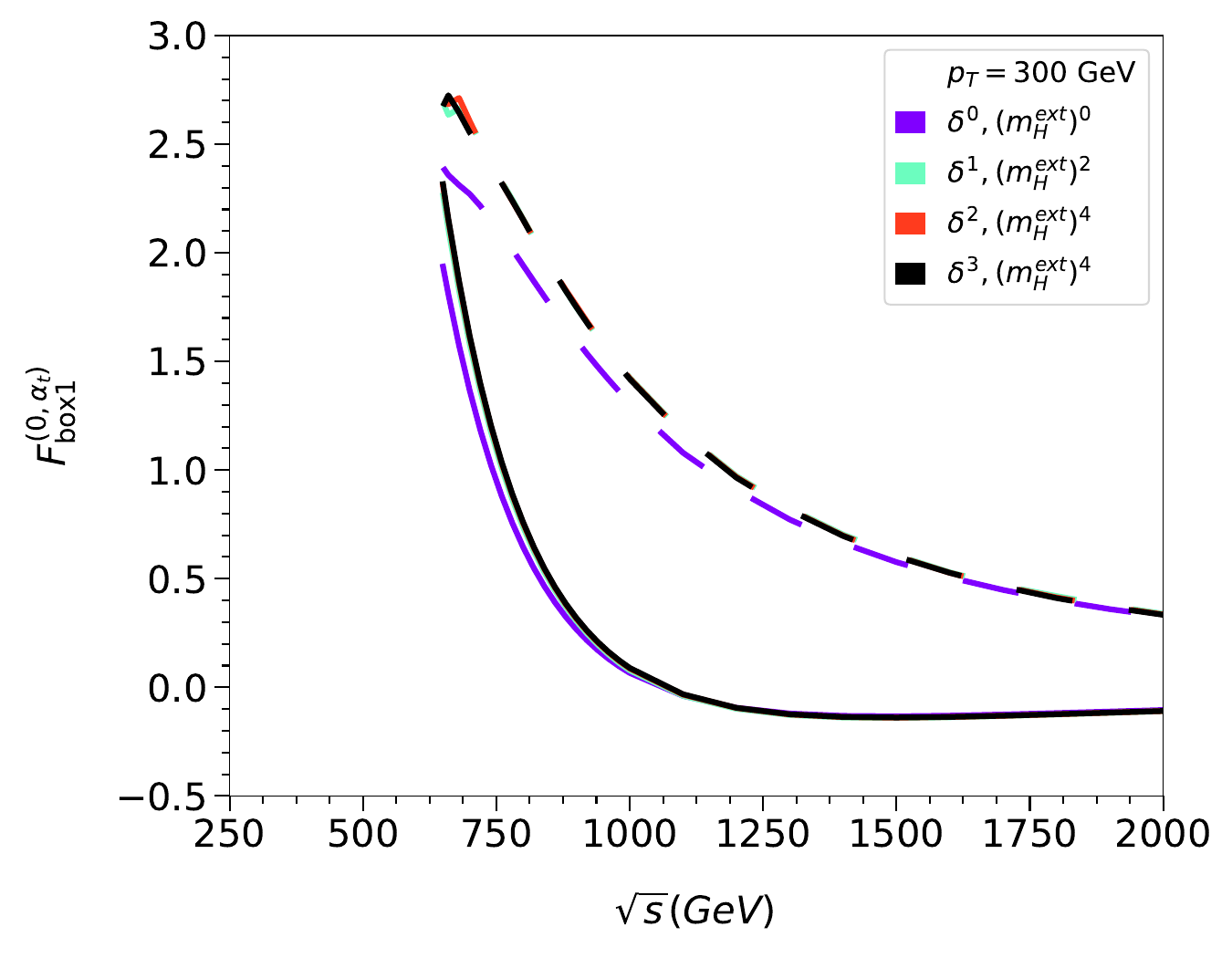}
  \end{tabular}
  \caption{\label{fig::F1yt2g23_del_mhs_300} Form factors as a function of
    $\sqrt{s}$ for $p_T=300$~GeV. Real and imaginary parts are shown as solid
    and dashed curves.
   }
\end{figure}

\begin{figure}[t]
  \centering
   \begin{tabular}{cc}
    \includegraphics[width=0.45\textwidth]{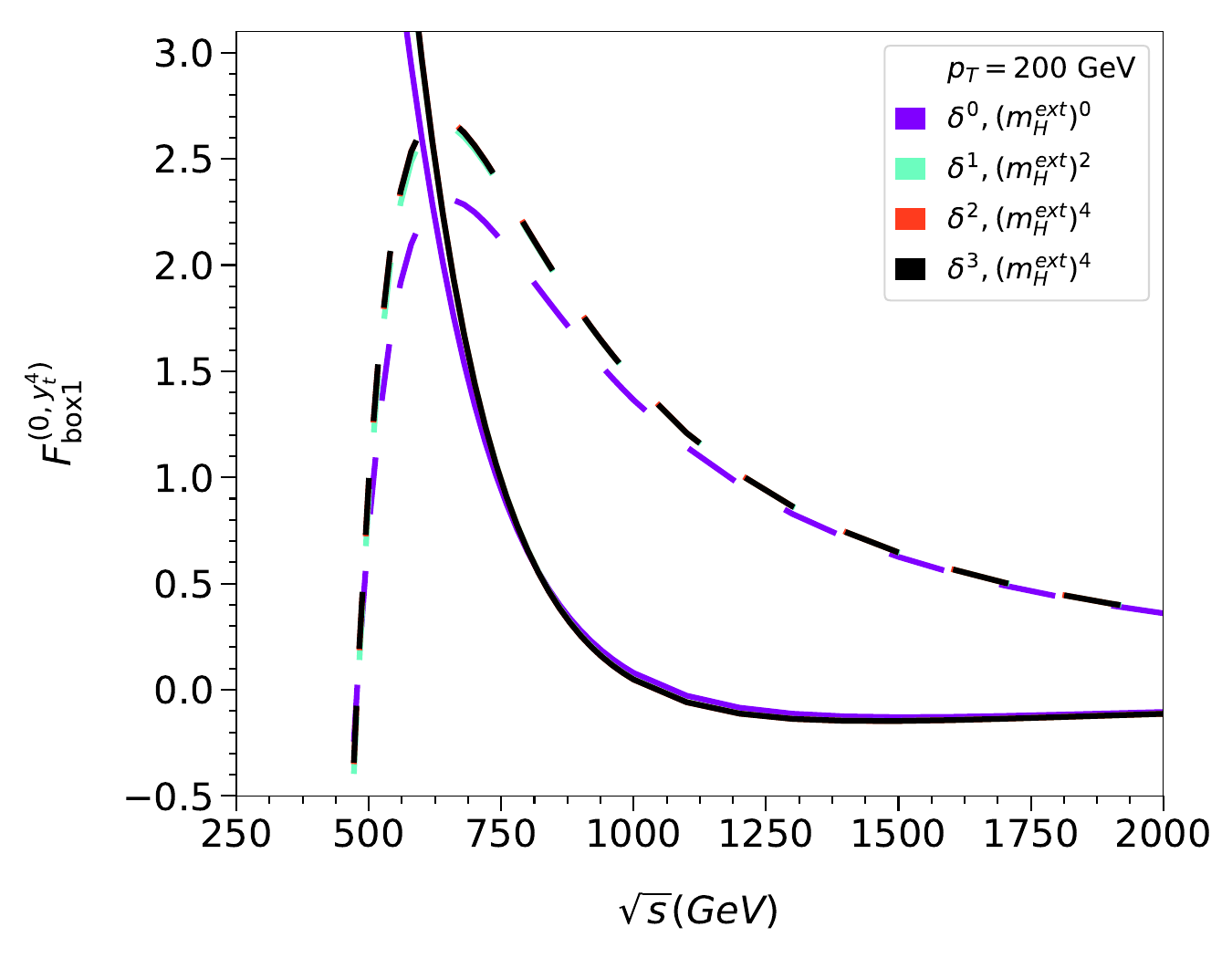} &
    \includegraphics[width=0.45\textwidth]{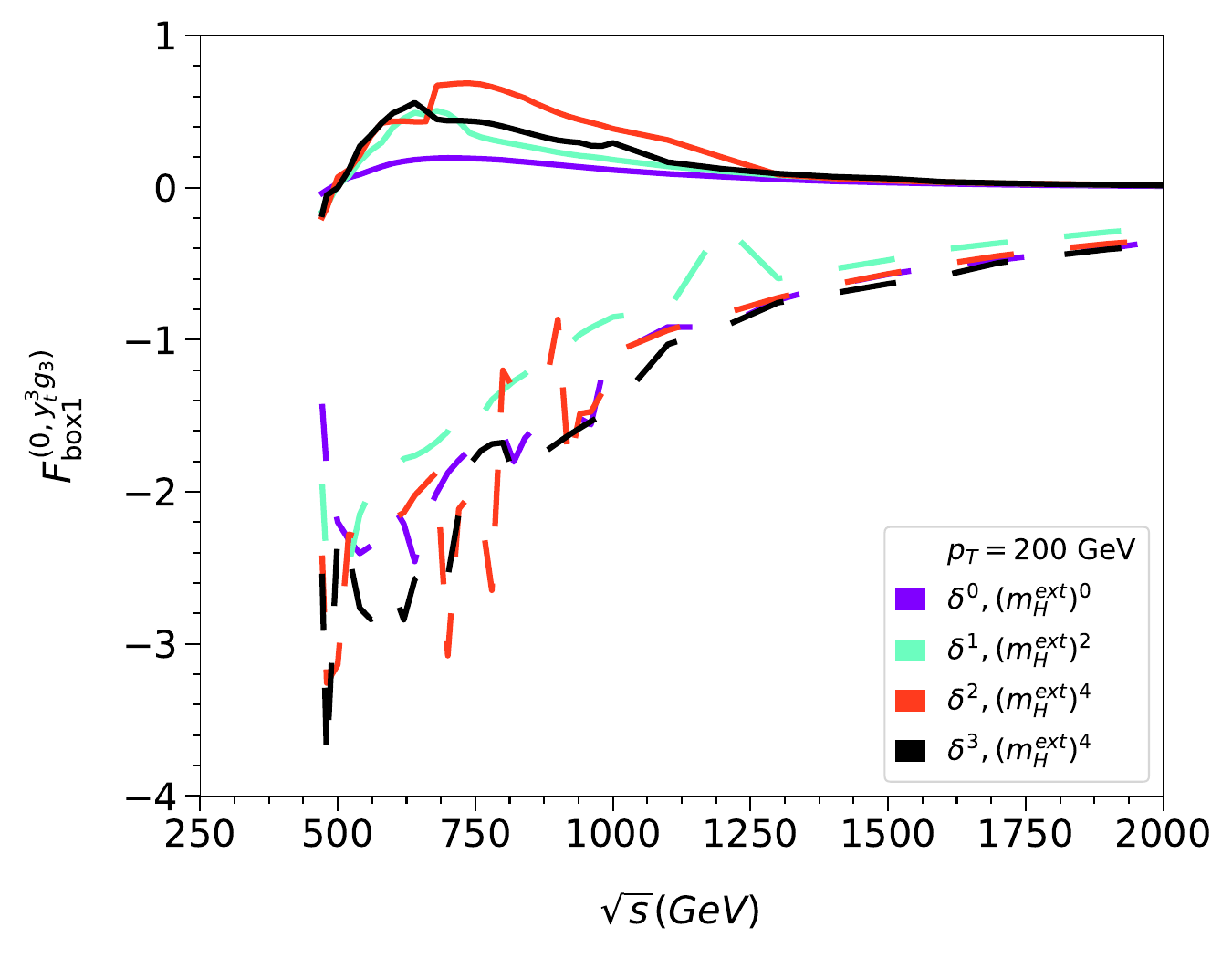}\\
    \includegraphics[width=0.45\textwidth]{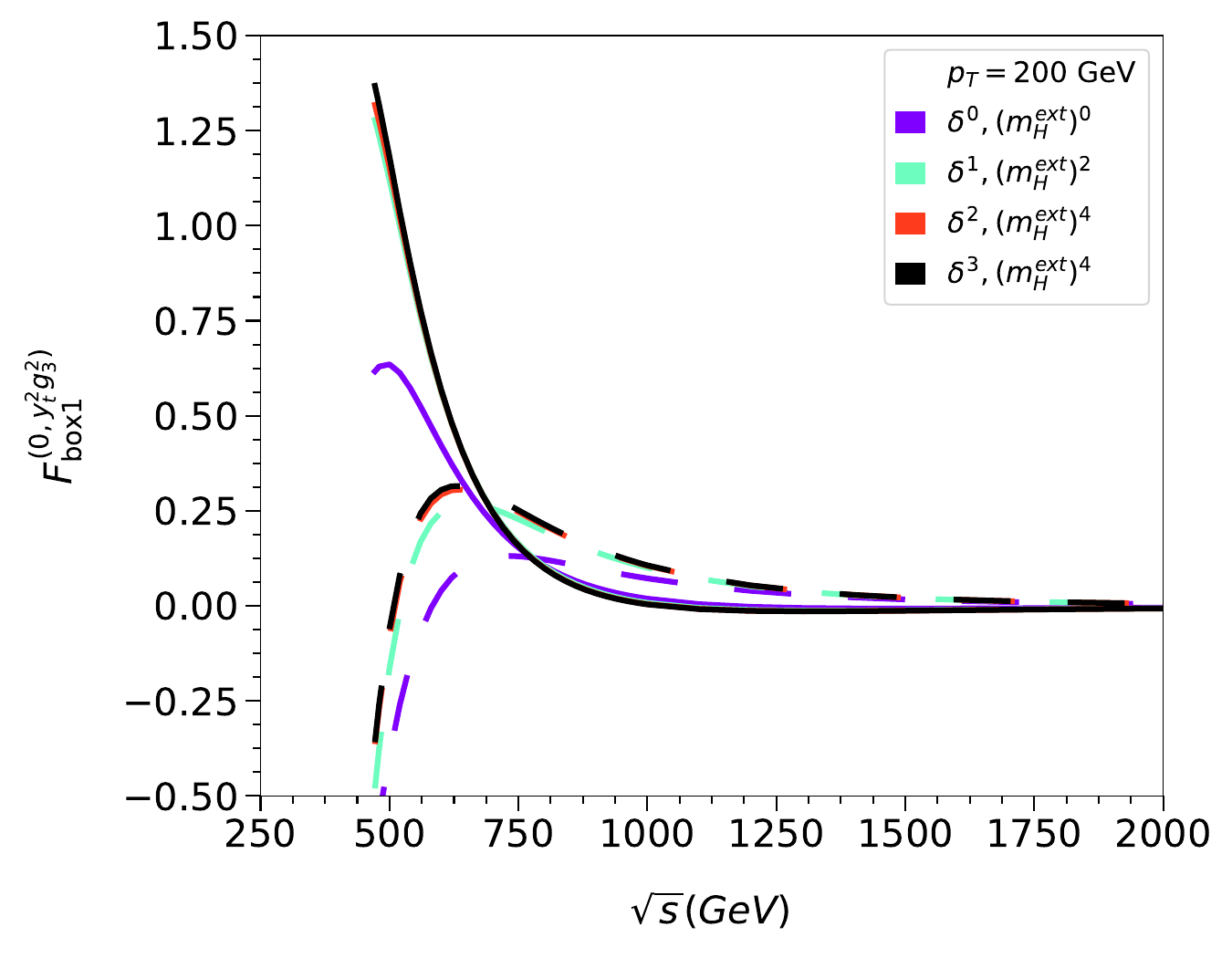} &
   \end{tabular}
  \caption{\label{fig::F1yt2g23_del_mhs_200}Same as
    Fig.~\ref{fig::F1yt2g23_del_mhs_300} but for $p_T=200$~GeV.
    }
\end{figure}

In Figs.~\ref{fig::F1yt2g23_del_mhs_300} and~\ref{fig::F1yt2g23_del_mhs_200}
we study the expansions in $m_H^{\rm exp}$ and $\delta$ for $p_T=300$~GeV and
$p_T=200$~GeV, respectively.  We show results for the real and imaginary parts
for the individual form factors as solid and dashed lines, respectively.  In
Fig.~\ref{fig::F1yt2g23_del_mhs_300} also results for the sum as given in
Eq.~(\ref{eq::F}) are shown.
Note that in the approximations labelled ``$\delta^n$'' in the plot legend,
all higher order $\delta$ terms are set to zero for each of the
expansion terms in $m_H^{\rm exp}$.

For $p_T=300$~GeV we observe a rapid convergence in both expansion parameters
for all form factors. In all cases the red and black
curves more-or-less lie on top of each other.

For $p_T=200$~GeV (see Fig.~\ref{fig::F1yt2g23_del_mhs_200}) the form factors
$F_{\rm box1}^{(0,y_t^4)}$ and $F_{\rm box1}^{(0,y_t^2 g_3^2)}$ behave very
similarly as for $p_T=300$~GeV.  However, for $F_{\rm box1}^{(0,y_t^3g_3)}$ we
observe an unstable behaviour, in particular for the imaginary part. This is
most likely due to the families $G_{53}$ and $G_{54}$ (see the discussion in
Sec.~\ref{sub::MIs}) which do not contribute to $F_{\rm box1}^{(0,y_t^4)}$
and $F_{\rm box1}^{(0,y_t^2 g_3^2)}$. We refrain from showing the total
contribution in this case, which inherits the instability from $F_{\rm box1}^{(0,y_t^3g_3)}$.
Note that instabilities come together with large uncertainty bands (which, for clarity,
are not shown in Fig.~\ref{fig::F1yt2g23_del_mhs_200}) from the Pad\'e method.

\begin{figure}[t]
  \centering
  \begin{tabular}{cc}
    \includegraphics[width=0.45\textwidth]{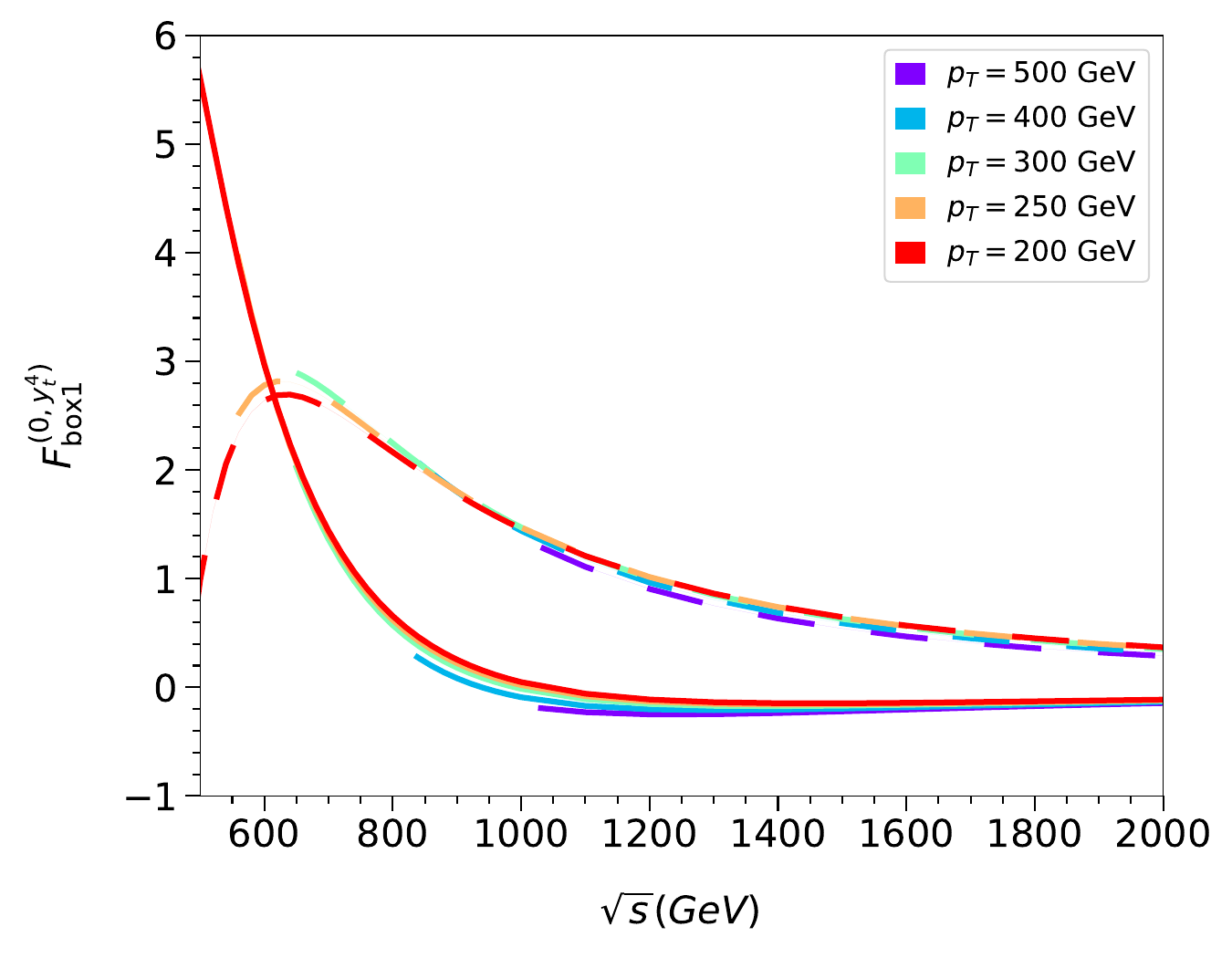} &
    \includegraphics[width=0.45\textwidth]{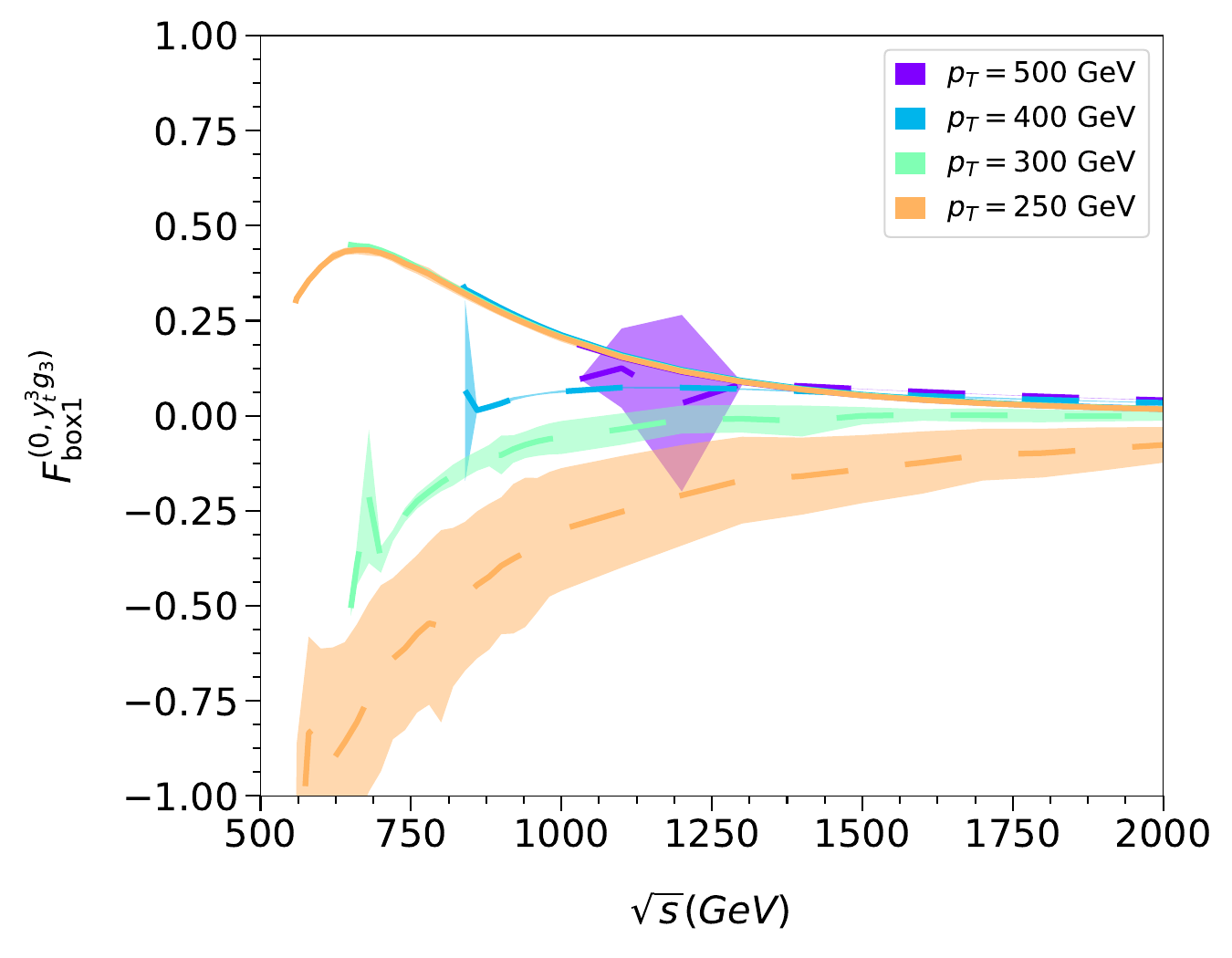}\\
    \includegraphics[width=0.45\textwidth]{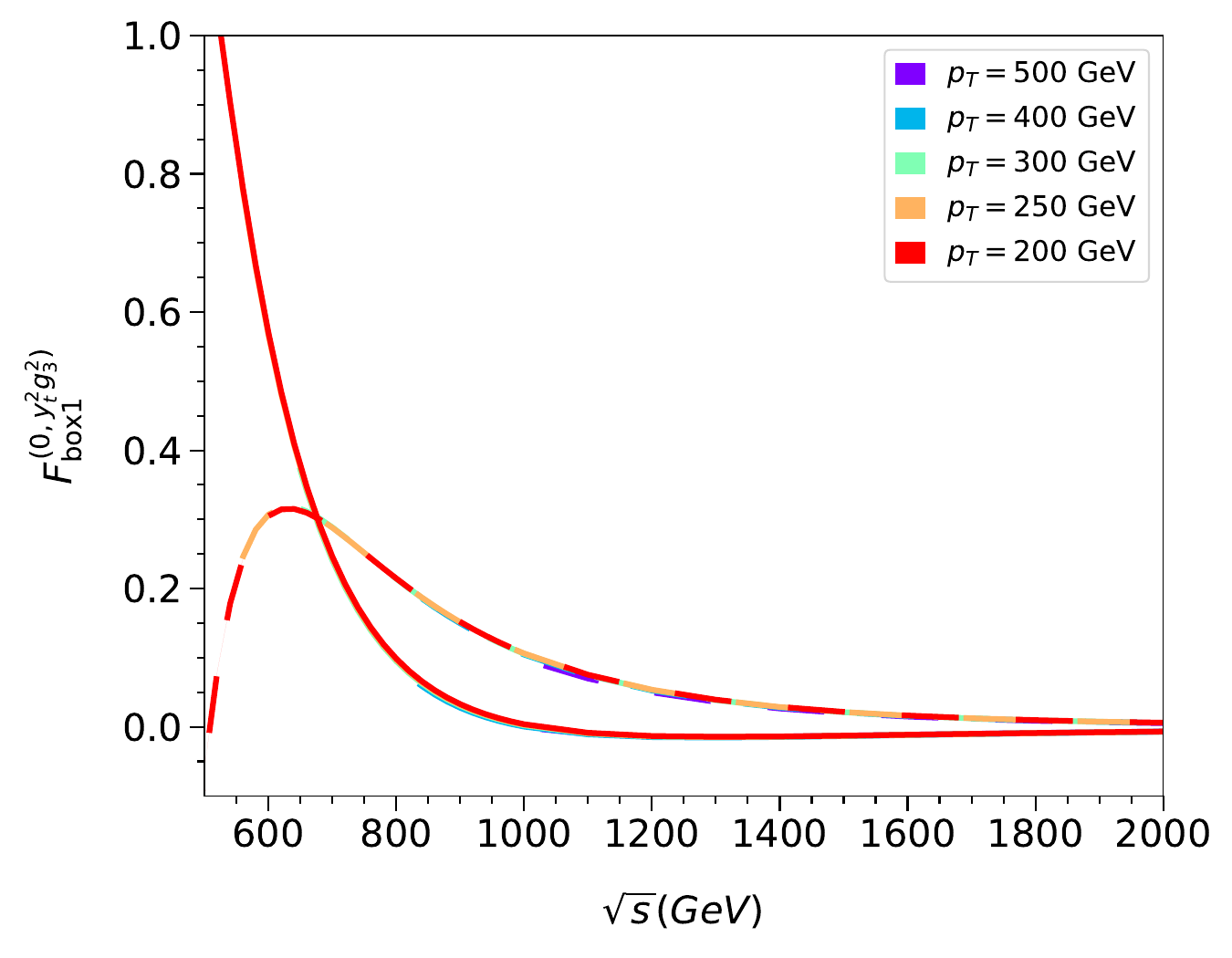}&
    \includegraphics[width=0.45\textwidth]{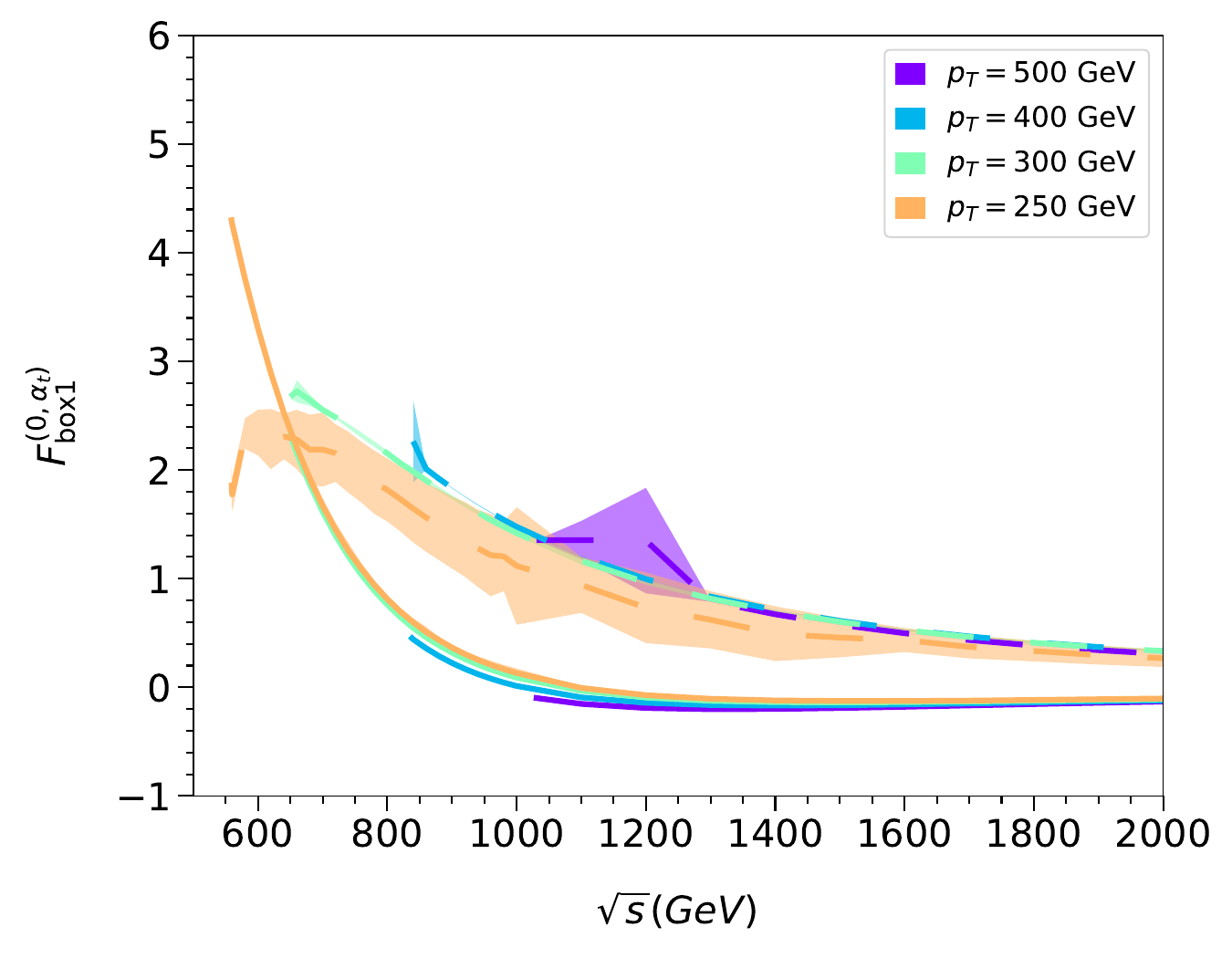}
  \end{tabular}
  \caption{\label{fig::F1_pT} Form factors as a function of
    $\sqrt{s}$ for different value of $p_T$. Solid and dashed lines
    refer to the real and imaginary parts.
    }
   
\end{figure}

In Fig.~\ref{fig::F1_pT} we show the real and imaginary parts of the form
factors $F_{\rm box1}$ for values of $p_T$ between 200~GeV and 500~GeV.
Note that in all cases we plot both the central value and the uncertainty bands
obtained by the Pad\'e approach.  Each curve starts at the lowest value for
$\sqrt{s}$ allowed by kinematic constraints. Stable results are obtained for
$F_{\rm box1}^{(0,y_t^4)}$ and $F_{\rm box1}^{(0,y_t^2 g_3^2)}$ where, even
for smaller values of $p_T$, the uncertainty bands are not visible. As
expected from the previous considerations, $F_{\rm box1}^{(0,y_t^3g_3)}$
becomes unstable below $p_T\approx 300$~GeV whereas for larger values of
$\sqrt{s}$ and $p_T$ precise results are obtained.

In our practical calculation we generate these deep analytic
expansions of the sum of all coupling structures split into the 12 pieces of
the $m_H$ and $\delta$ expansions. The gzipped version of each part is about
40~MB. Currently we perform the numerical evaluation in {\tt
  Mathematica}. Here inserting the kinematics takes of the order of a
minute. The subsequent construction of the Pad\'e approximants takes about a
second. We plan to implement our results into a {\tt C++} code and expect that
the total evaluation time will reduce to a few milliseconds as in the case of the QCD
corrections, see Ref.~\cite{Davies:2023vmj}.


\subsection{\label{sub::comp}Comparison to Ref.~\cite{Heinrich:2024dnz}}

In this section we compare our results for the form factors with the numerical evaluations
of Ref.~\cite{Heinrich:2024dnz}. By comparing Eqs.~(2.14)-(2.16) of Ref.~\cite{Heinrich:2024dnz}
with our definitions in Eqs.~(\ref{eq::F}) and (\ref{eq::Fsplit}), we can establish relations
between the two form factor definitions, which are given by
\begin{align}
  F^{(0)}_{g_t^2,{\rm box}} & = - \frac{s}{32 \, m_t^2 \, \pi^2}\, F_{\rm box}^{(0)}
                              \label{eq::conversionLO}
\end{align}
at one loop and
\begin{align}
  F^{(1)}_{g_t^4,{\rm box}} & = - \frac{s \, e^{-2\gamma_E \epsilon}}{64 \, m_t^2 \, \pi^4}  \Big(\frac{\mu^2}{4\pi \, m_t^2}\Big)^{-2\epsilon}  \, F_{\rm box}^{(0,y_t^4)} \,, \nonumber\\
  F^{(1)}_{g_t^2 \, g_3^2,{\rm box}} & = - \frac{s}{ 576 \, m_H^2 \, m_t^2 \, \pi^4} \, F_{\rm box}^{(0,y_t^2 \, g_3^2)}\,,\nonumber\\
  F^{(1)}_{g_t^3\, g_3,{\rm box}} & = - \frac{s}{192 \, m_H \, m_t^2 \, \pi^4} \, F_{\rm box}^{(0,y_t^3 \, g_3)} \,,\nonumber\\
  F^{(1)}_{g_t^2 \, g_4,{\rm box}} & = - \frac{s}{192 \, m_t^2 \, \pi^4} \,  F_{\rm box}^{(0,y_t^2 \, g_4)} \,
  \label{eq::conversionNLO}
\end{align}
at two loops. Here we have omitted the subscript $i=1,2$ which denotes the two
Lorentz projections of the structures in Eq.~(\ref{eq::LorStr}), and added the
subscript ``box'' to the form factors on the left-hand side since we do not
consider triangle diagrams in the comparison. We also do not compare
one-loop results since the quality of their high-energy expansions is discussed
in Ref.~\cite{Davies:2023vmj}. Note that $F_{\rm box}^{(0,y_t^4)}$ is
divergent, thus its conversion factor is $\epsilon$ dependent. Some conversion
factors are dimensionful because our form factors are dimensionless and some
of those of Ref.~\cite{Heinrich:2024dnz} are not.

While in Section \ref{sec::FFs} we have used $\mu^2 = s$, we compare with data
for the form factors of Ref.~\cite{Heinrich:2024dnz}\footnote{We would like to
  thank the authors of Ref.~\cite{Heinrich:2024dnz} for providing this data,
  by private communication.}  for $\mu^2 = m_t^2$ and thus adopt that choice
here. Similarly we use common mass values of $m_t=173.055$~GeV and
$m_H=125$~GeV.

\begin{figure}[b]
  \centering
  \begin{tabular}{ccc}
  \includegraphics[width=0.35\textwidth]{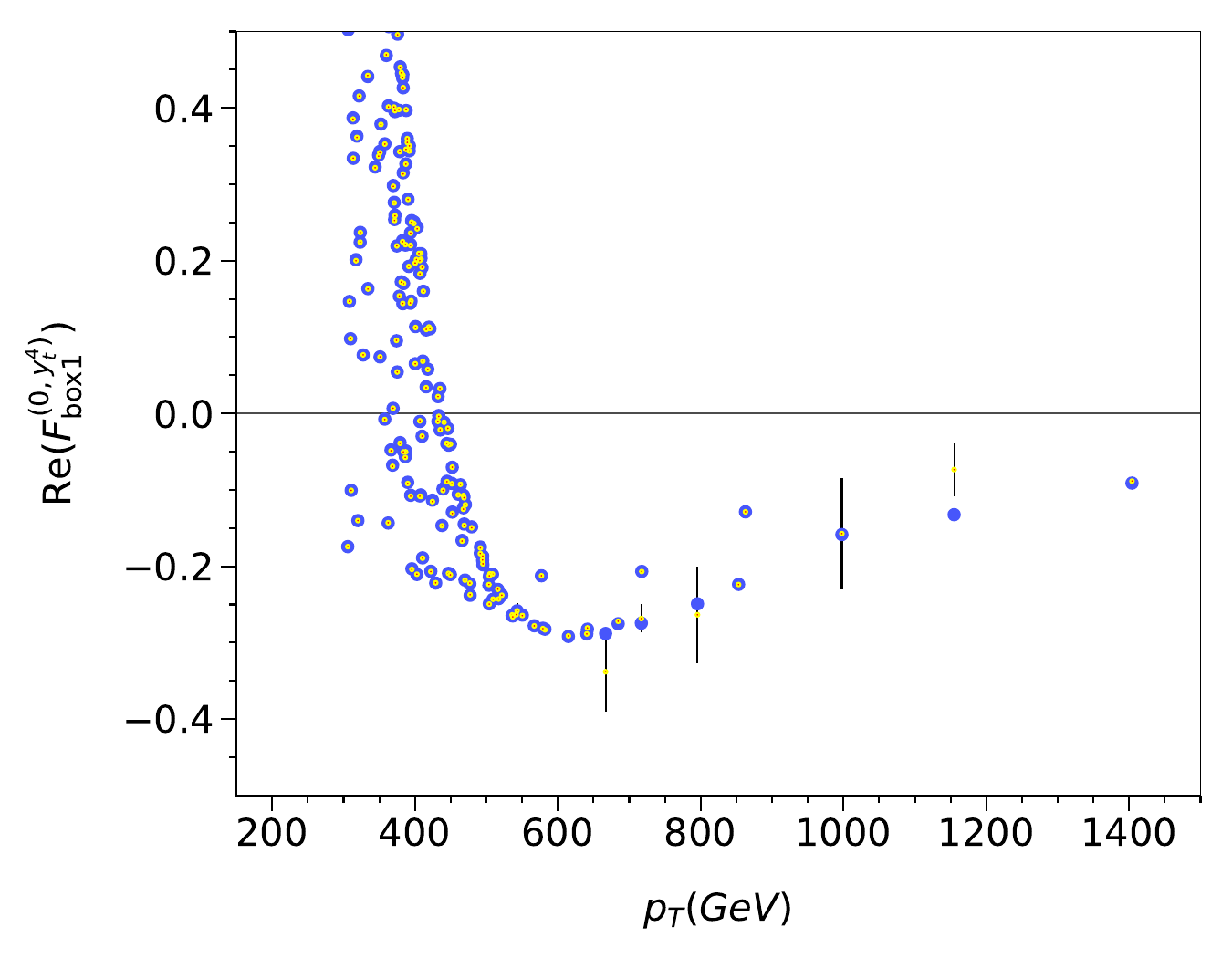}
  \includegraphics[width=0.35\textwidth]{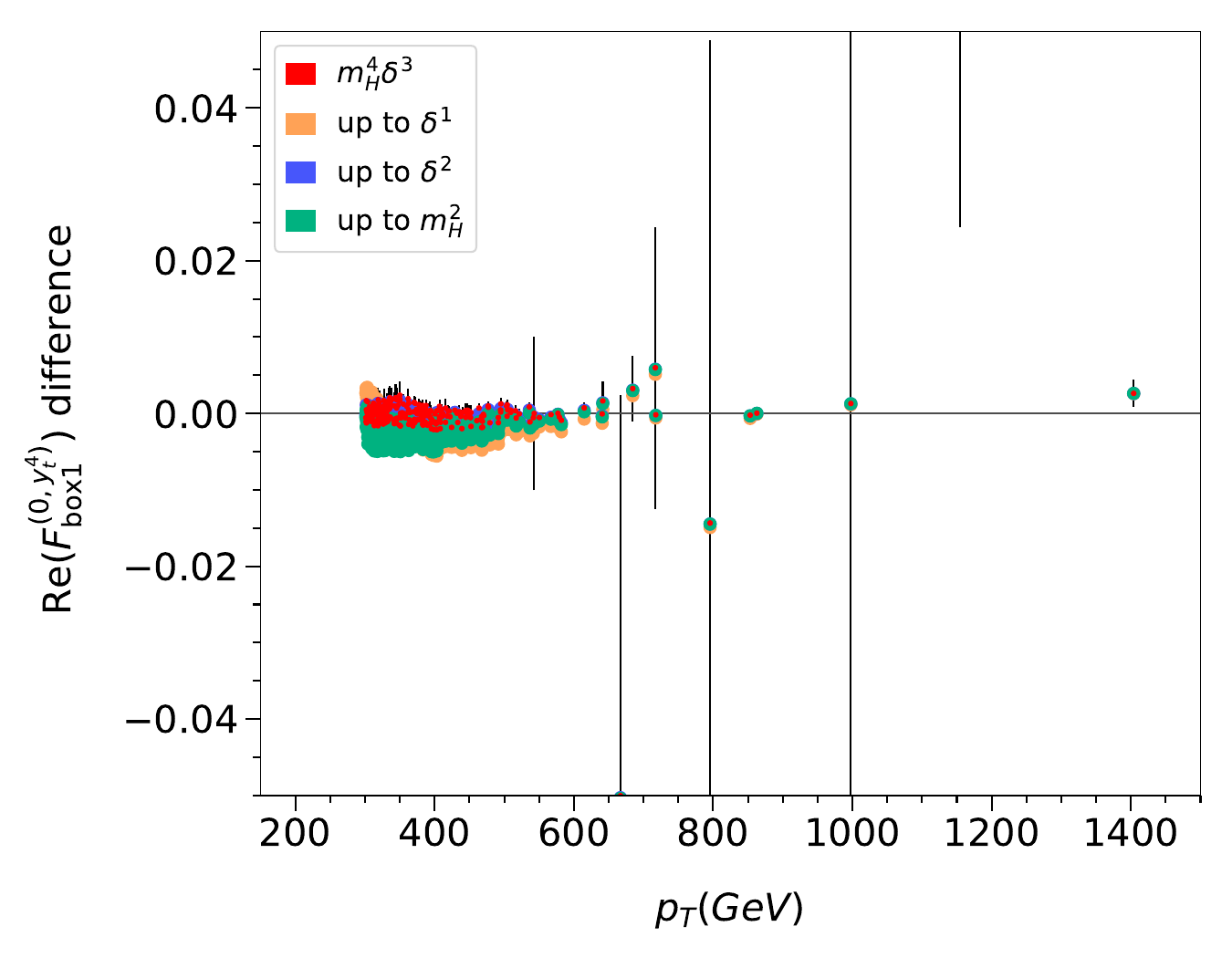}
  \includegraphics[width=0.35\textwidth]{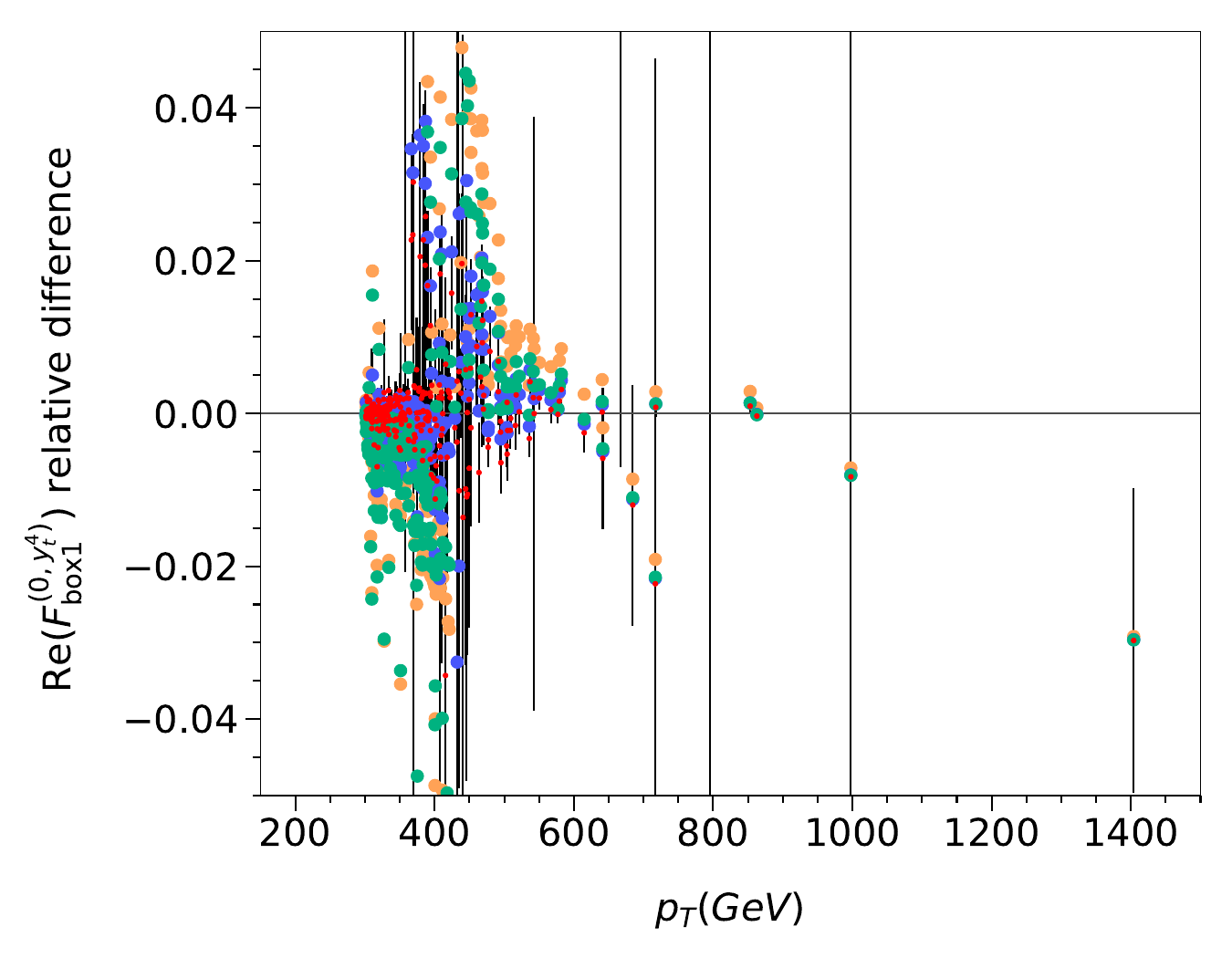}
\\
  \includegraphics[width=0.35\textwidth]{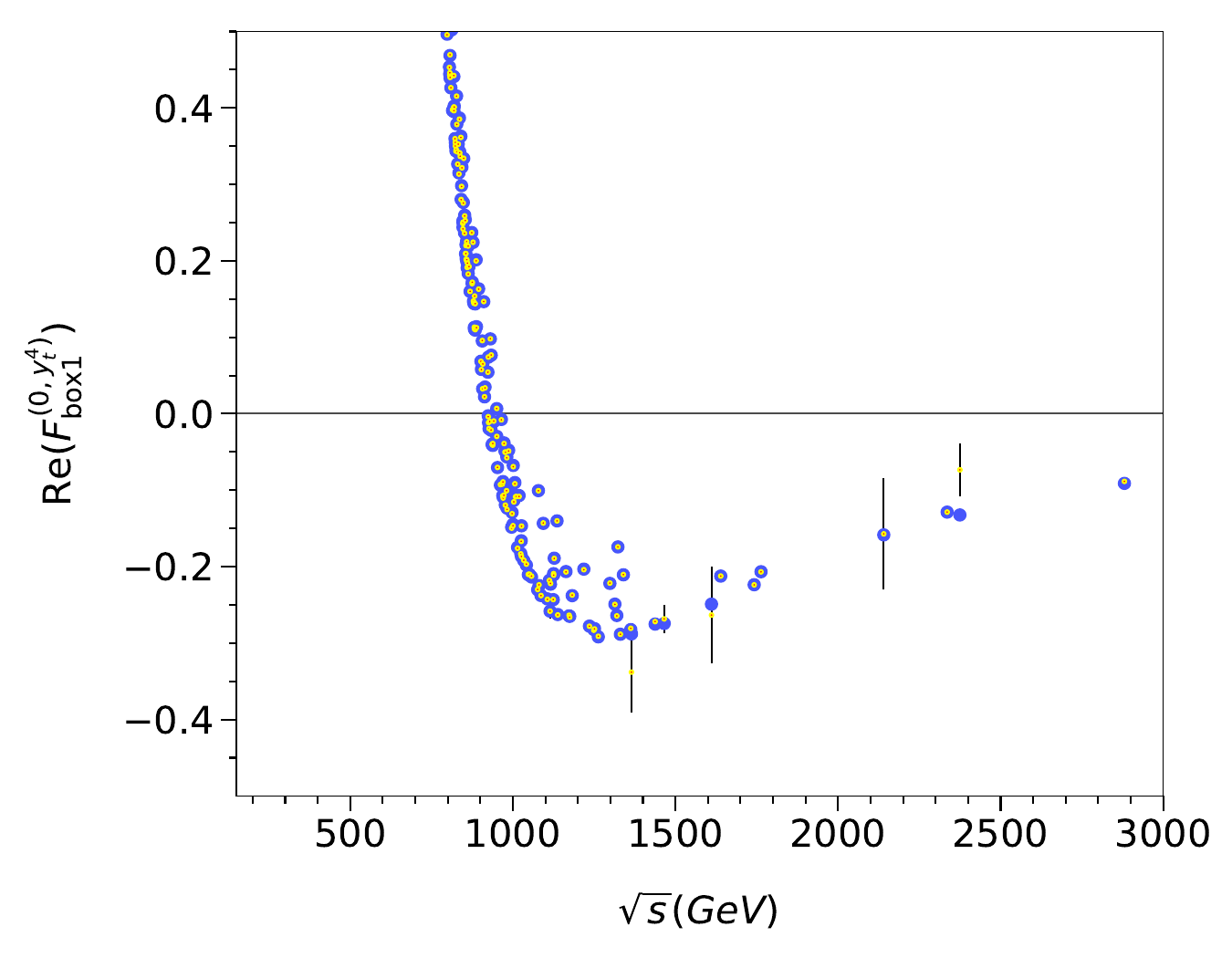}
  \includegraphics[width=0.35\textwidth]{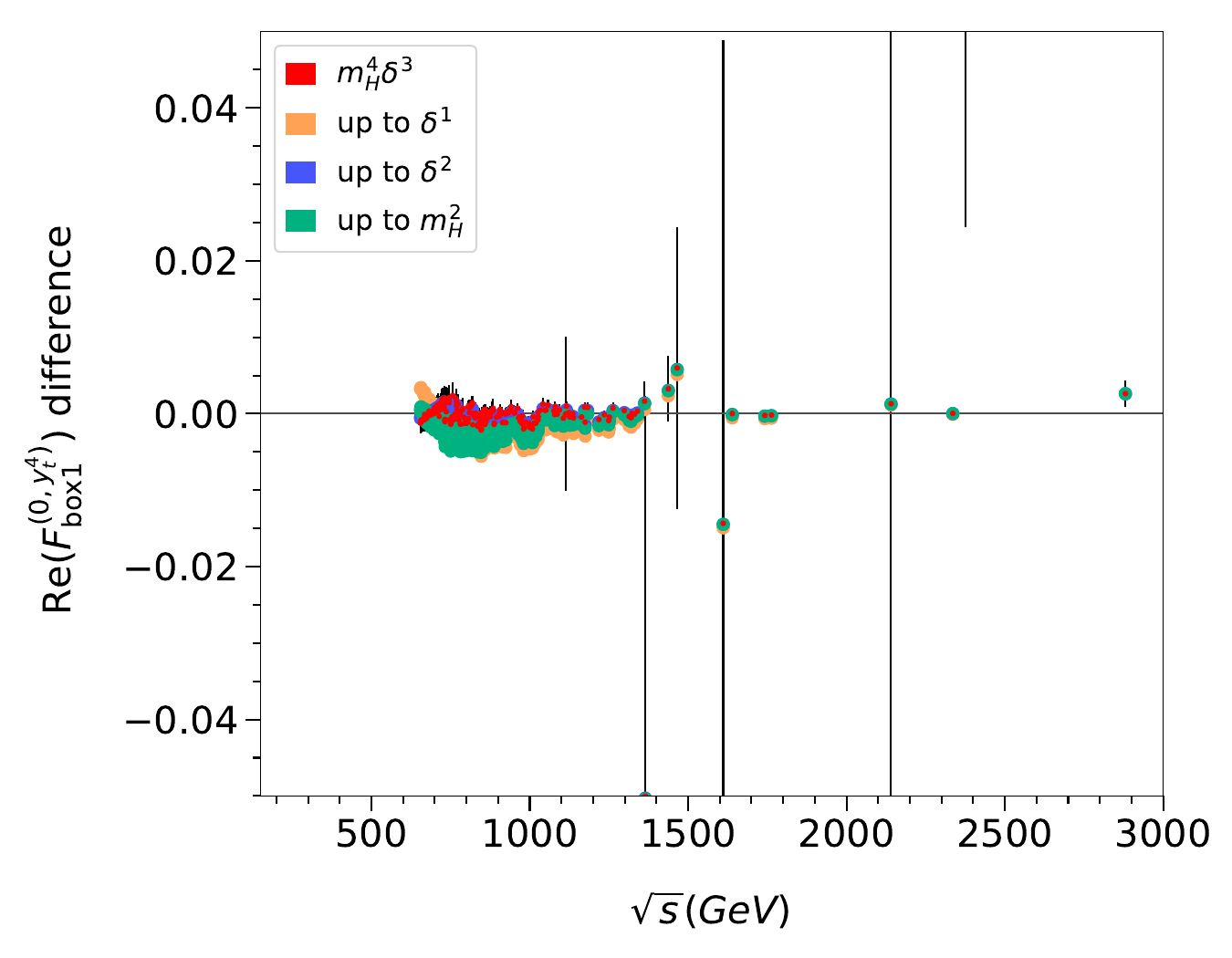}
  \includegraphics[width=0.35\textwidth]{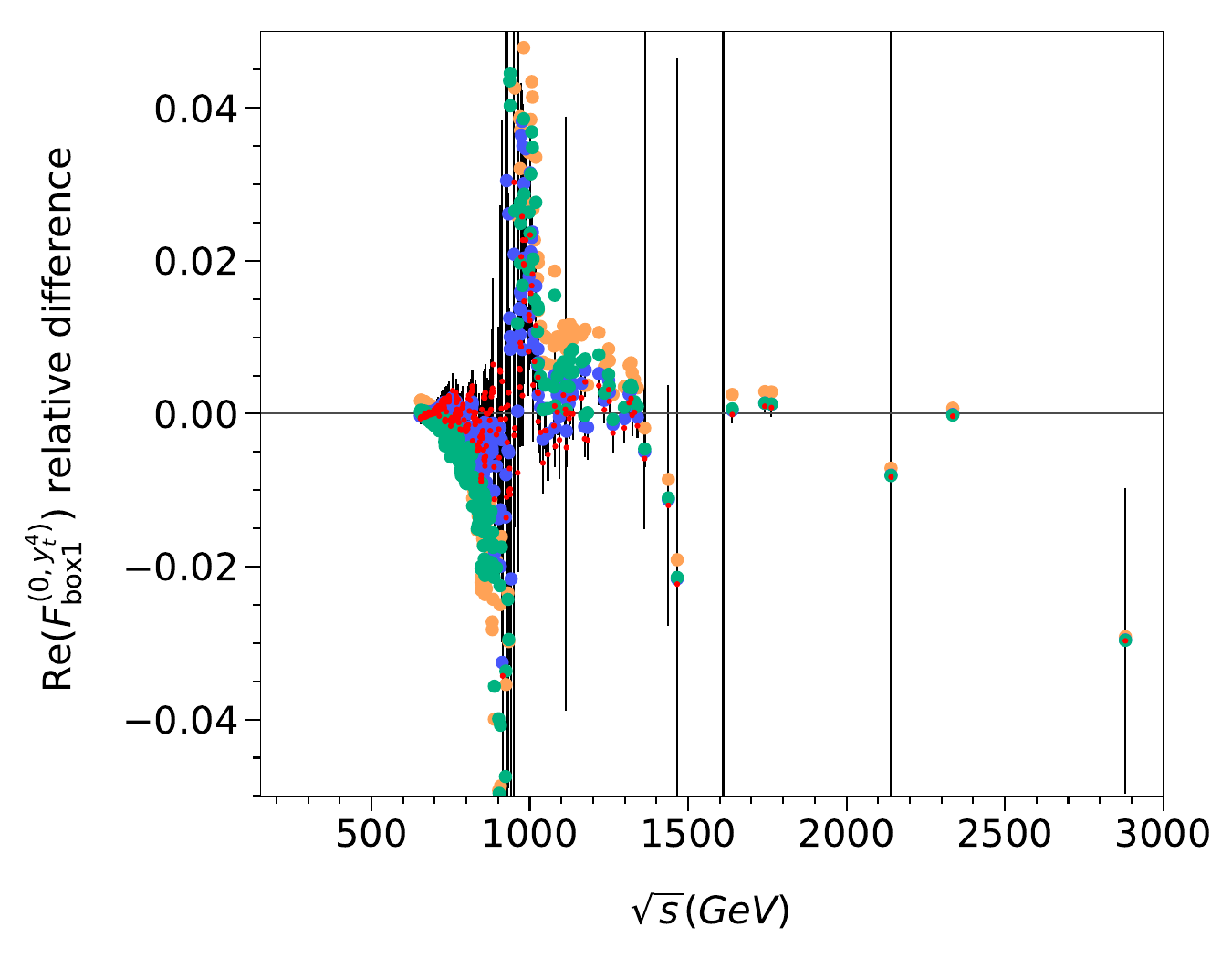}
\\
  \includegraphics[width=0.35\textwidth]{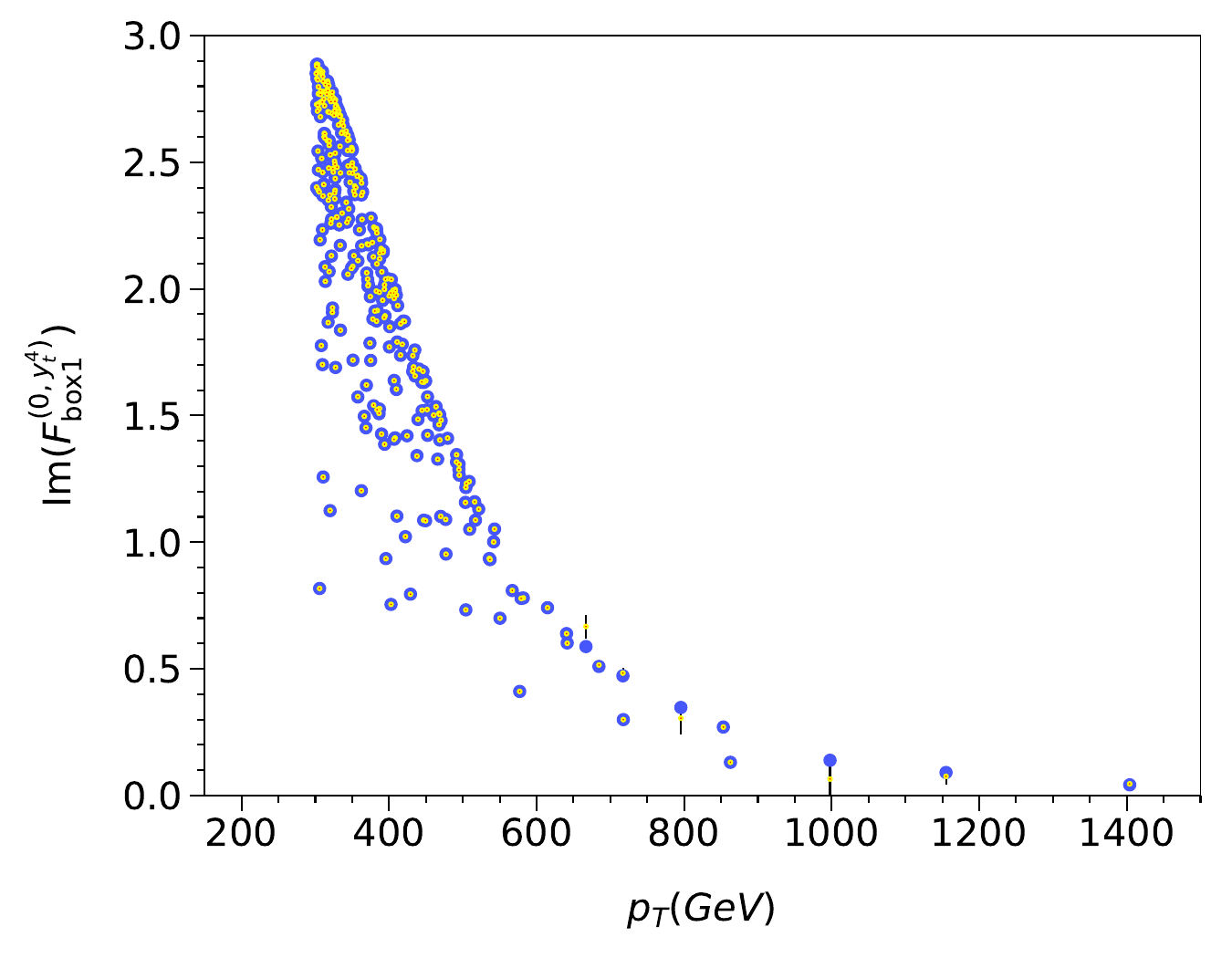}
  \includegraphics[width=0.35\textwidth]{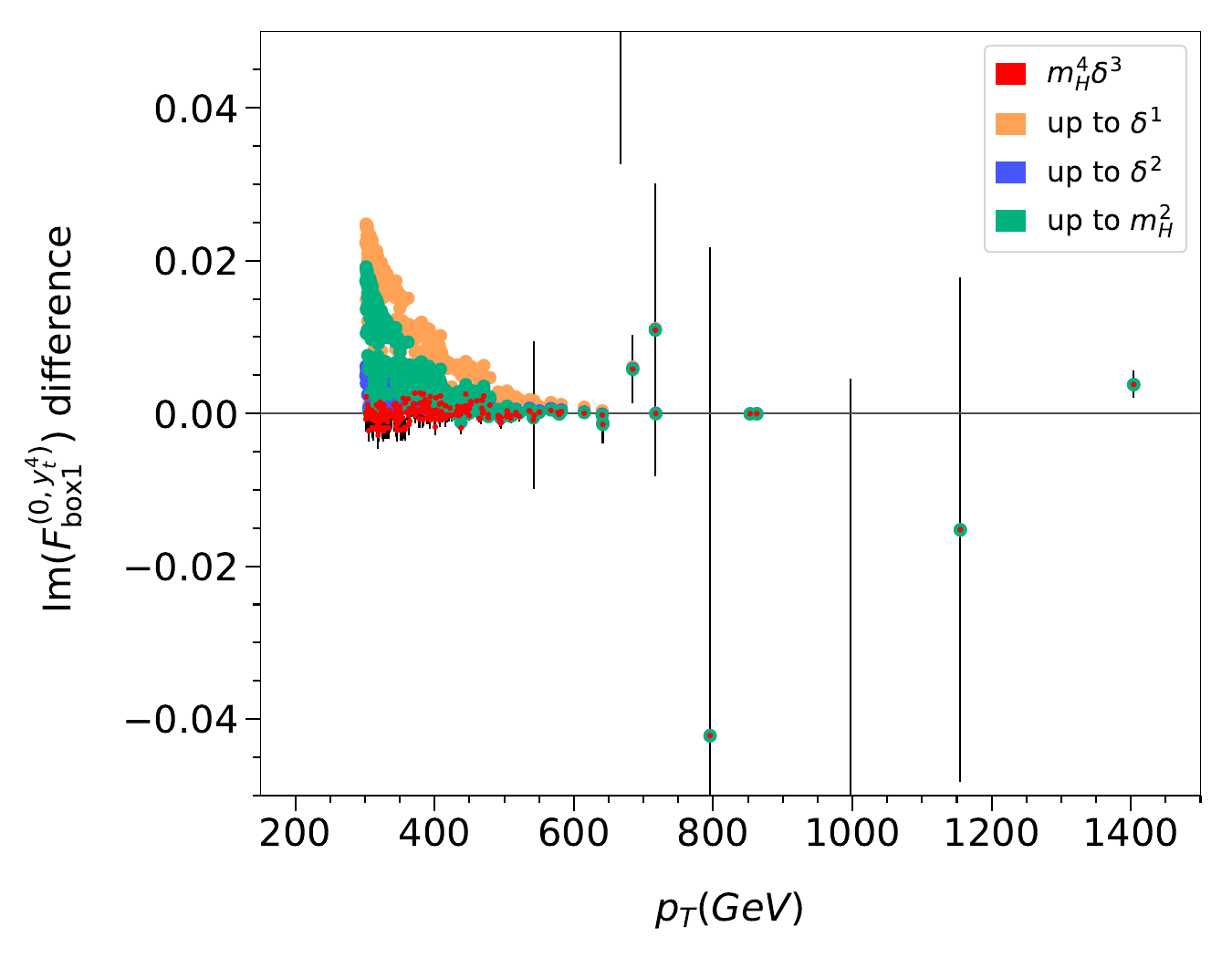}
  \includegraphics[width=0.35\textwidth]{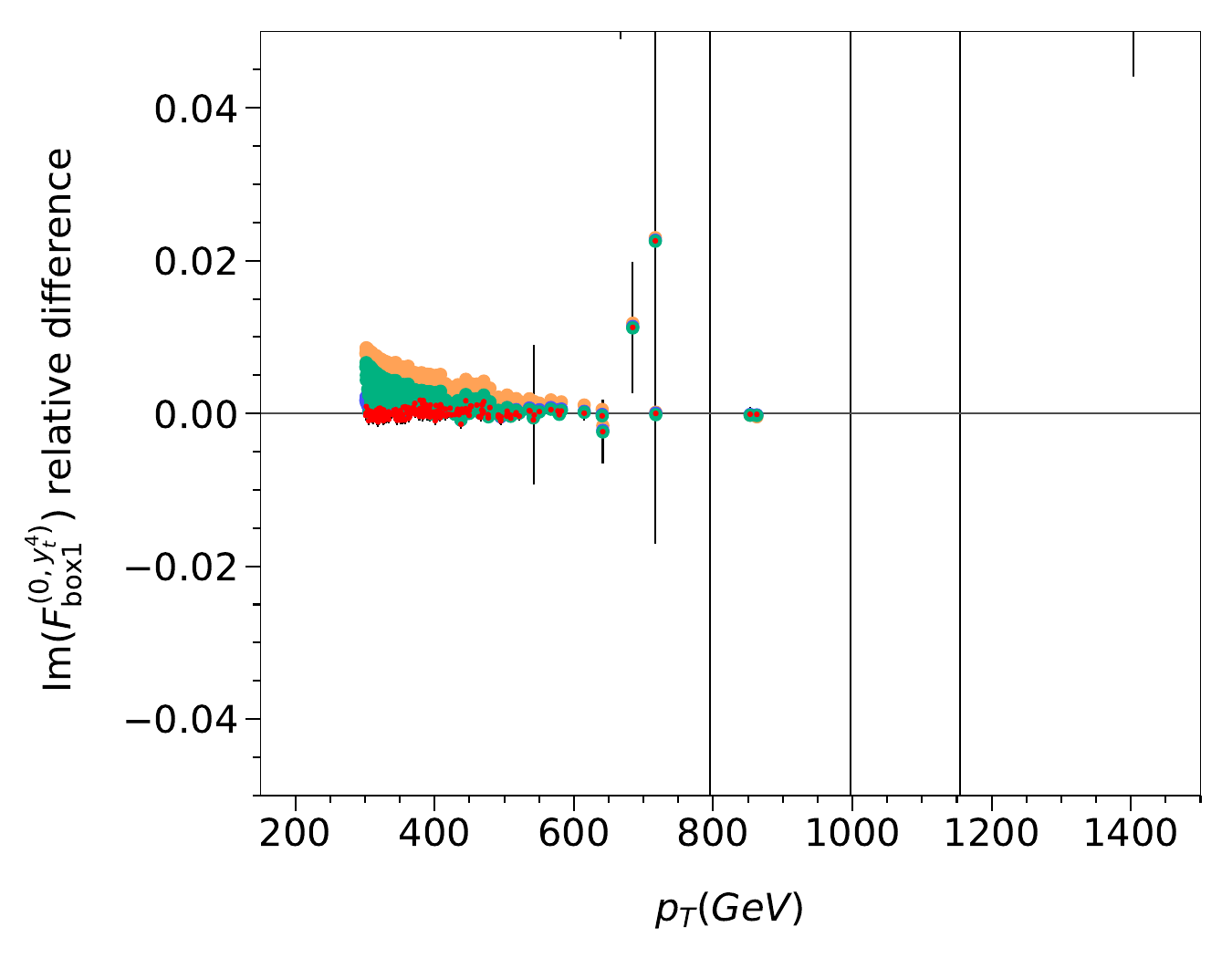}
\\
  \includegraphics[width=0.35\textwidth]{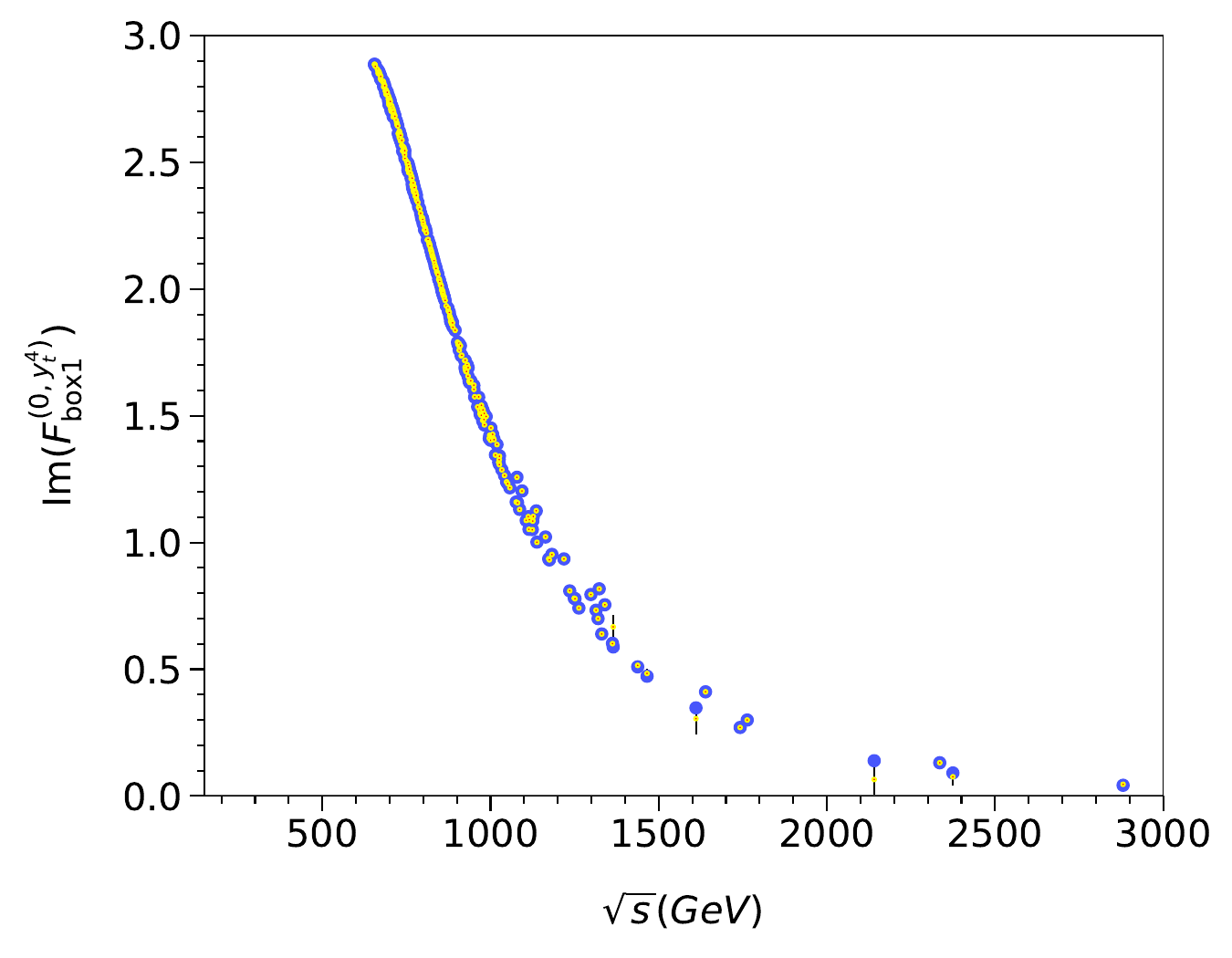}
  \includegraphics[width=0.35\textwidth]{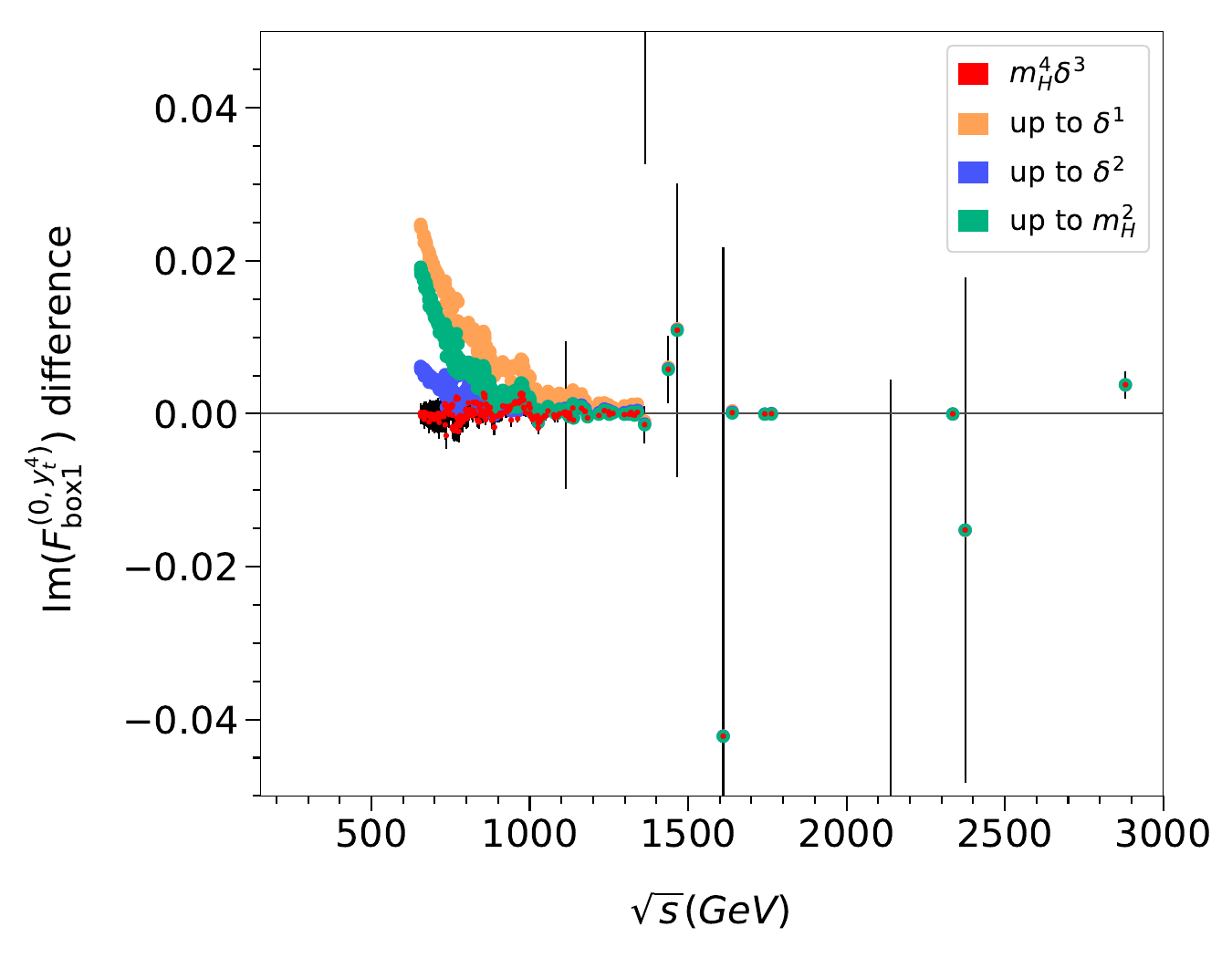}
  \includegraphics[width=0.35\textwidth]{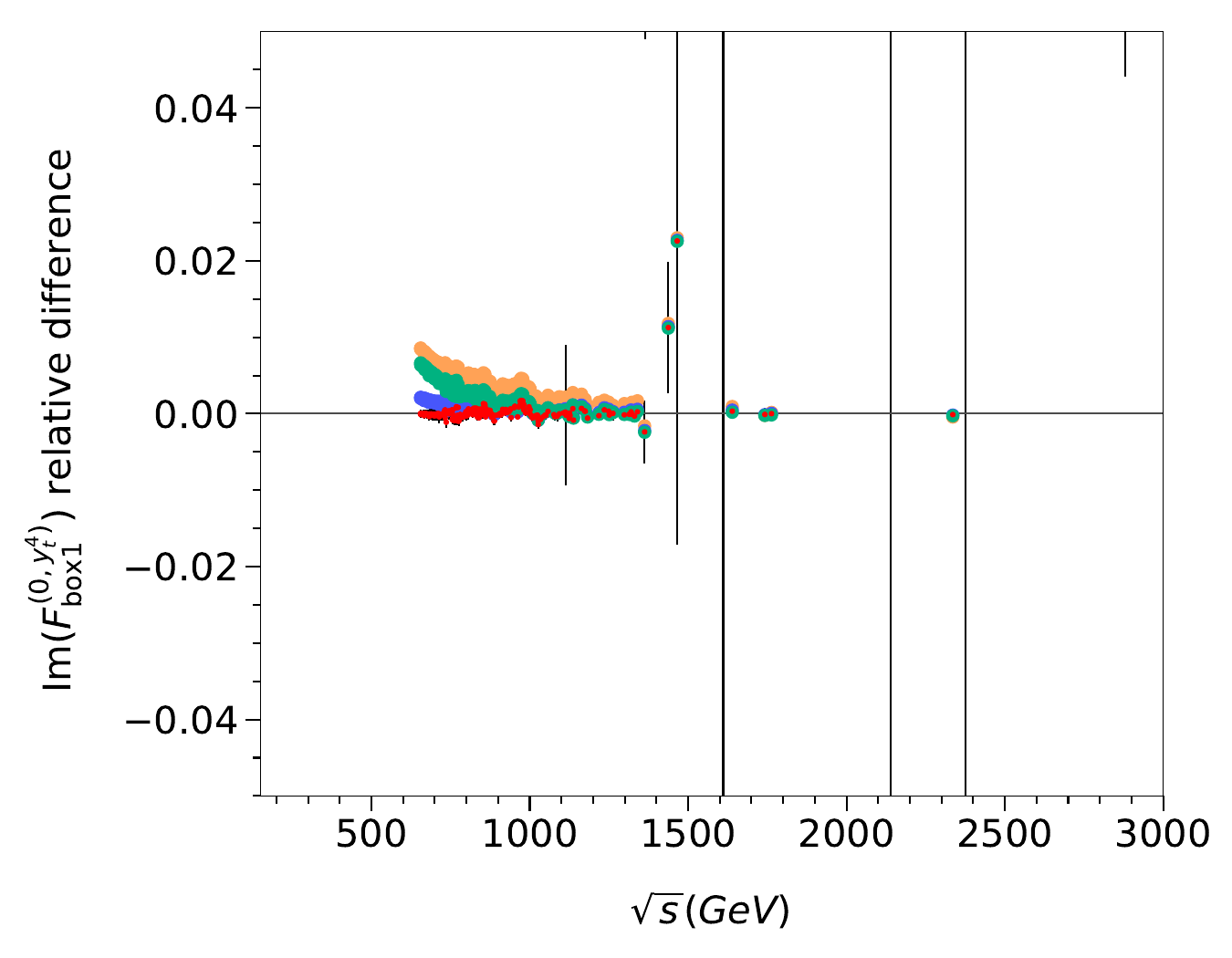}
    \end{tabular}
   \caption{\label{fig::diff_to_SD_yt4}
    $F_{\rm box1}^{(0,y_t^4)}$ form factor (left column), absolute (middle column) and relative (right column)
    difference compared to Ref.~\cite{Heinrich:2024dnz}. We show the real (upper two rows) and imaginary
    (lower two rows) parts as a function of $p_T$ and of $\sqrt{s}$. We show phase-space points with a
    $p_T$ value of at least $300$GeV. The data point colours are as defined in
    the text. }
\end{figure}

\begin{figure}[t]
  \centering
  \begin{tabular}{ccc}
  \includegraphics[width=0.35\textwidth]{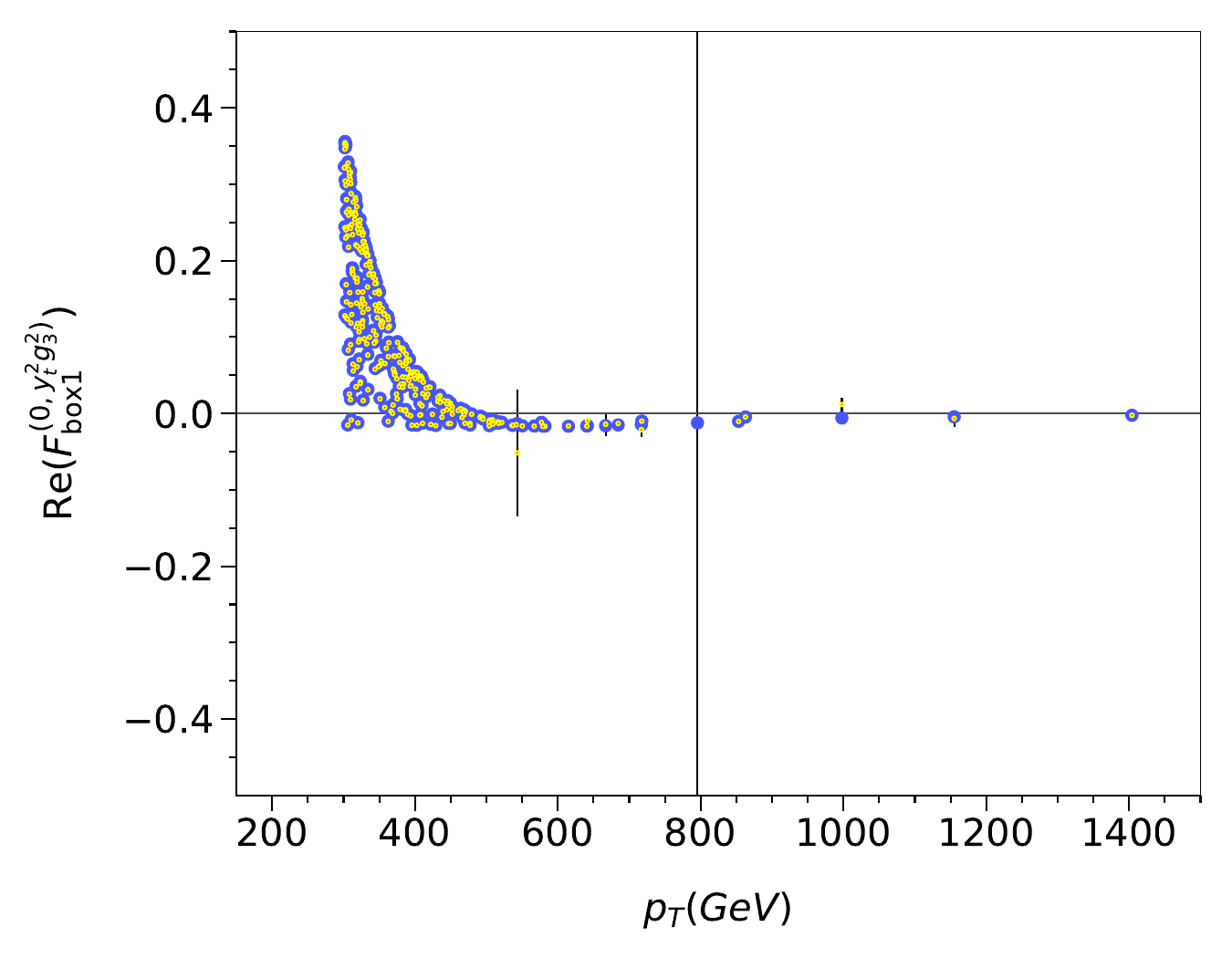}
  \includegraphics[width=0.35\textwidth]{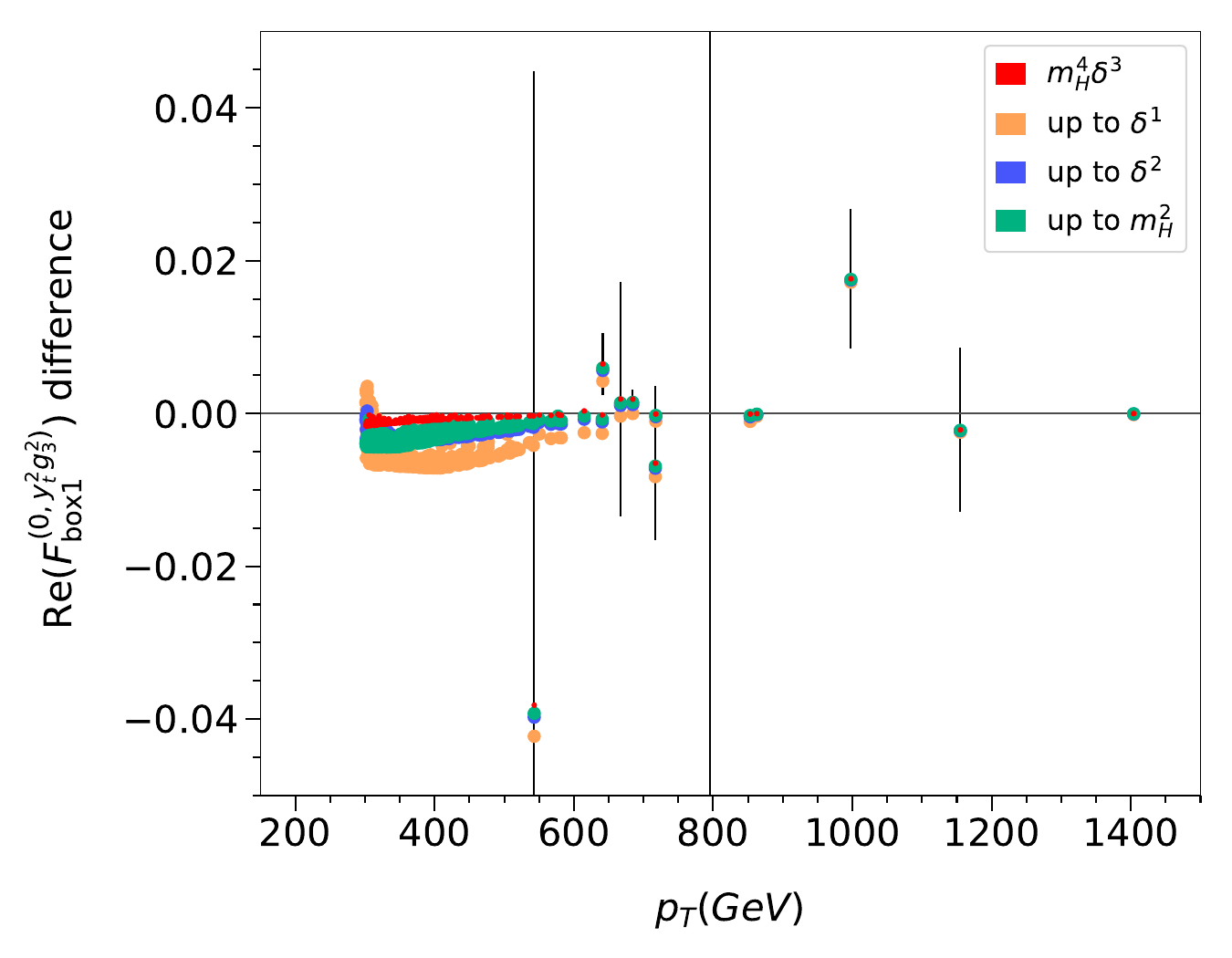}
  \includegraphics[width=0.35\textwidth]{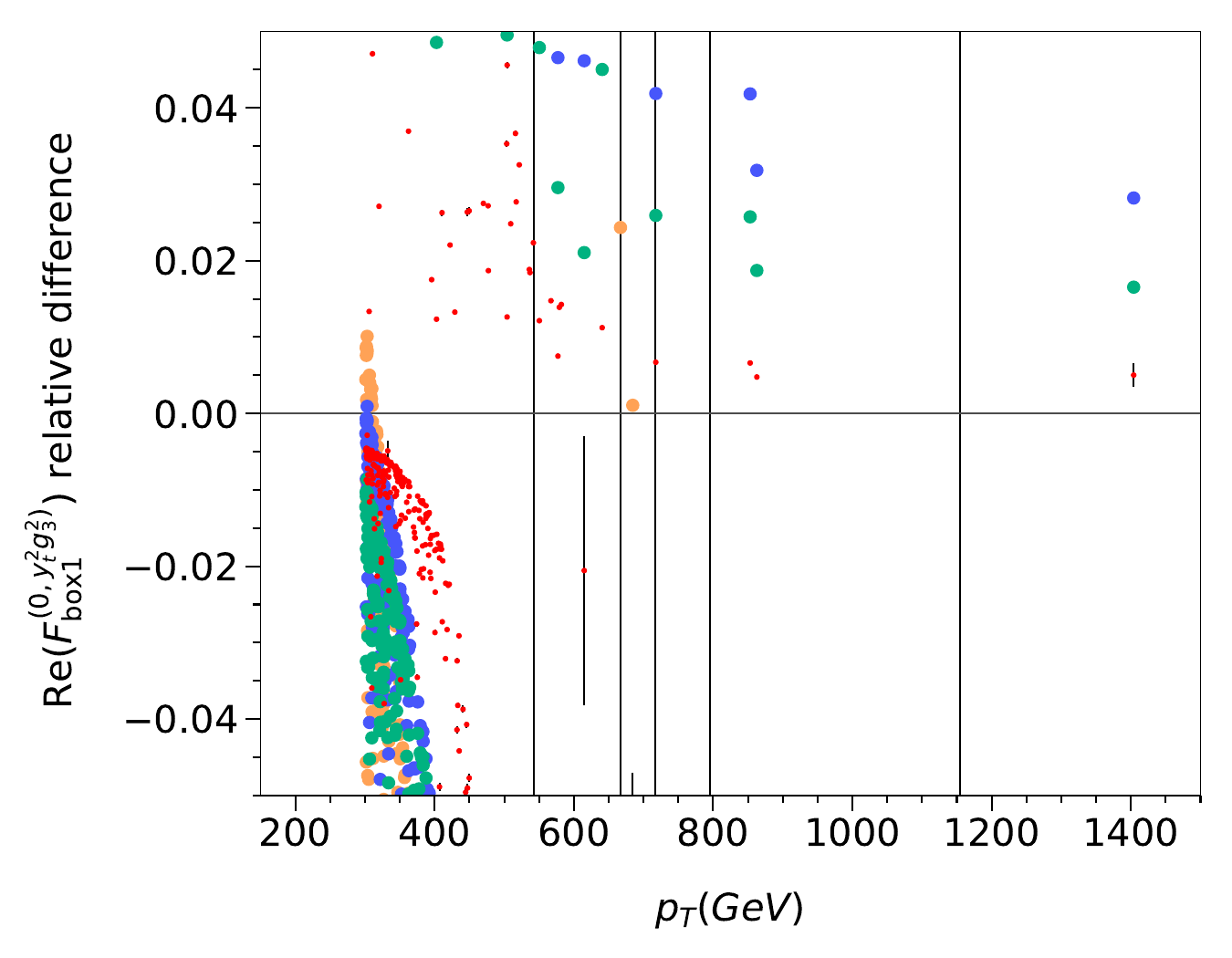}
\\
  \includegraphics[width=0.35\textwidth]{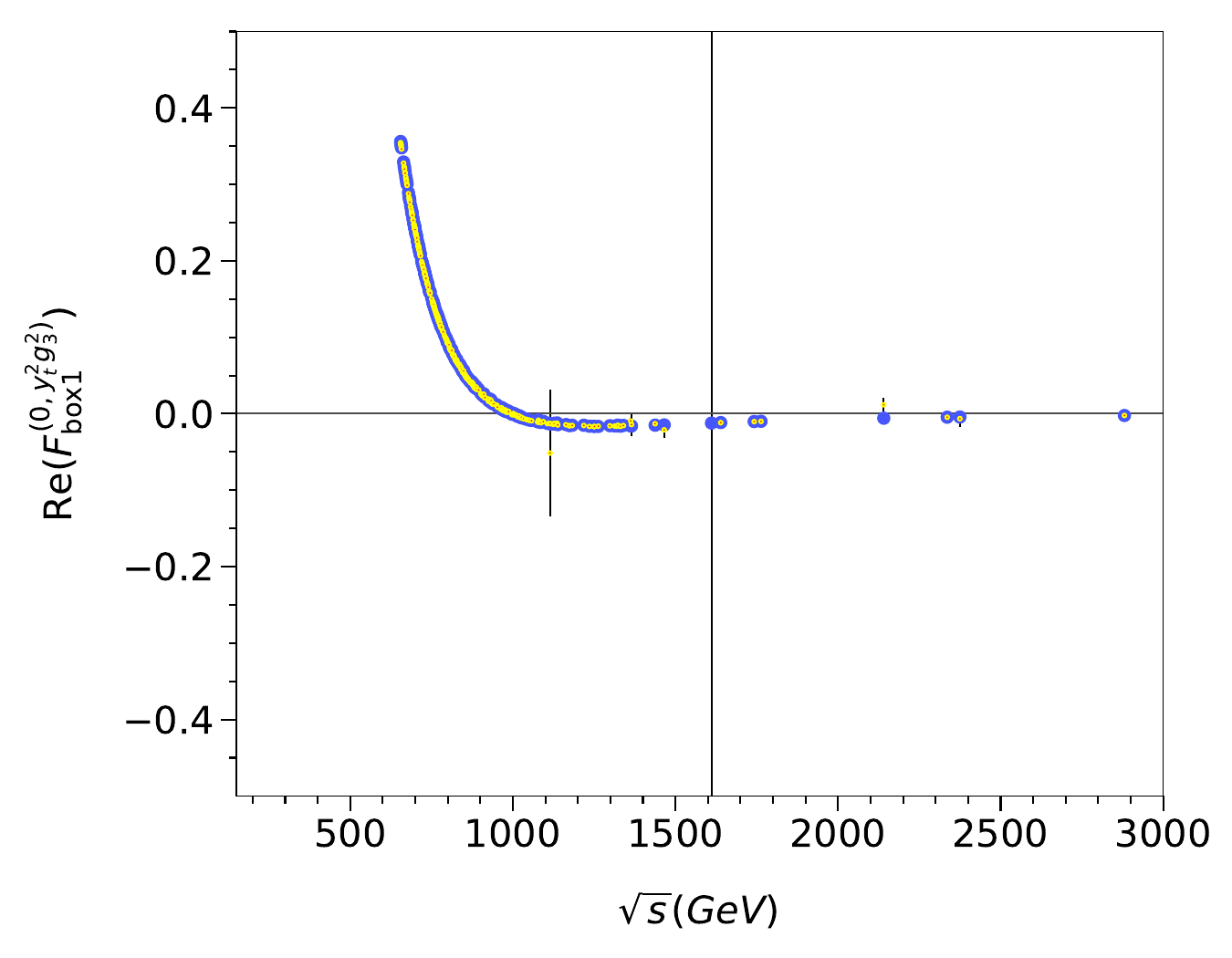}
  \includegraphics[width=0.35\textwidth]{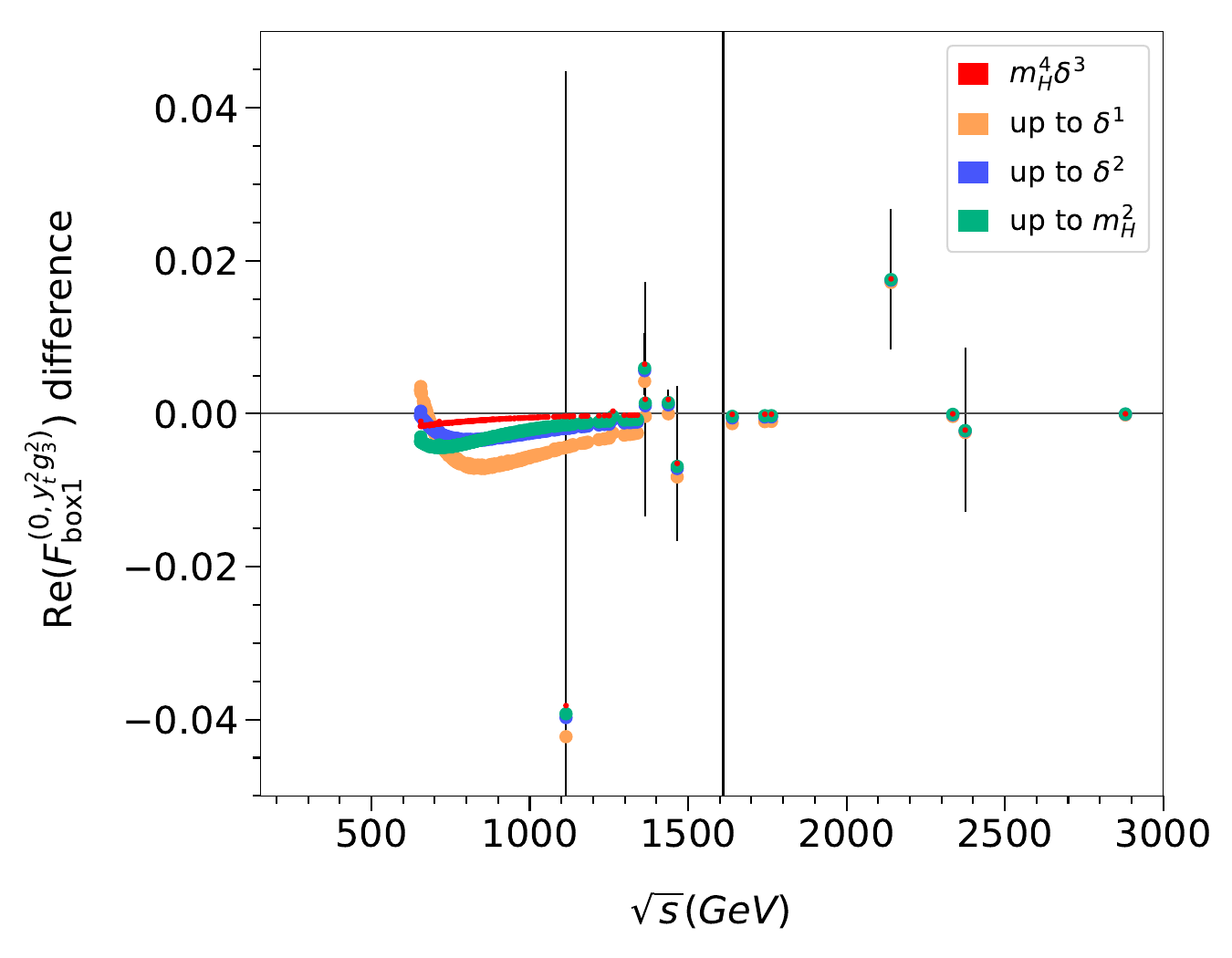}
  \includegraphics[width=0.35\textwidth]{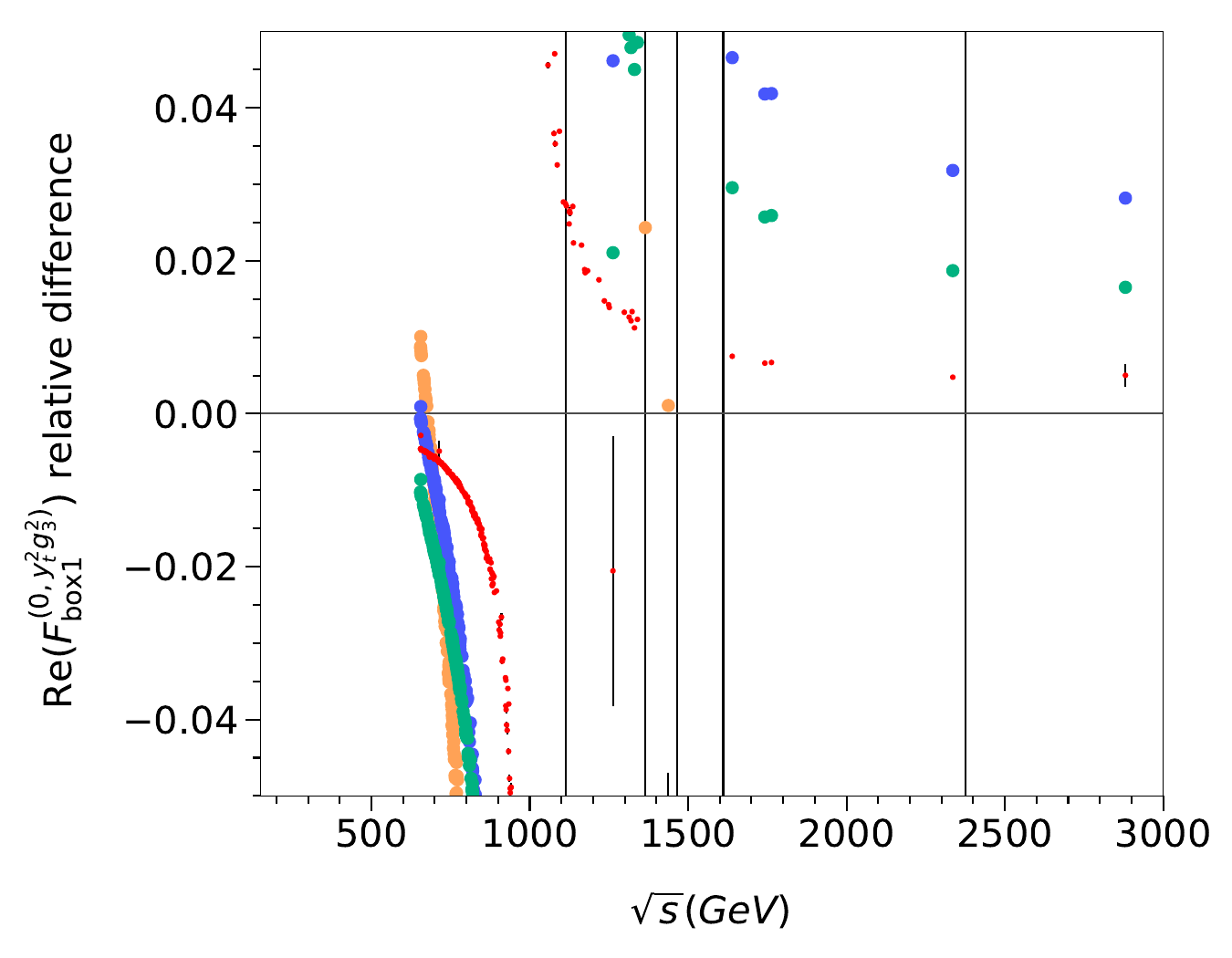}
\\
  \includegraphics[width=0.35\textwidth]{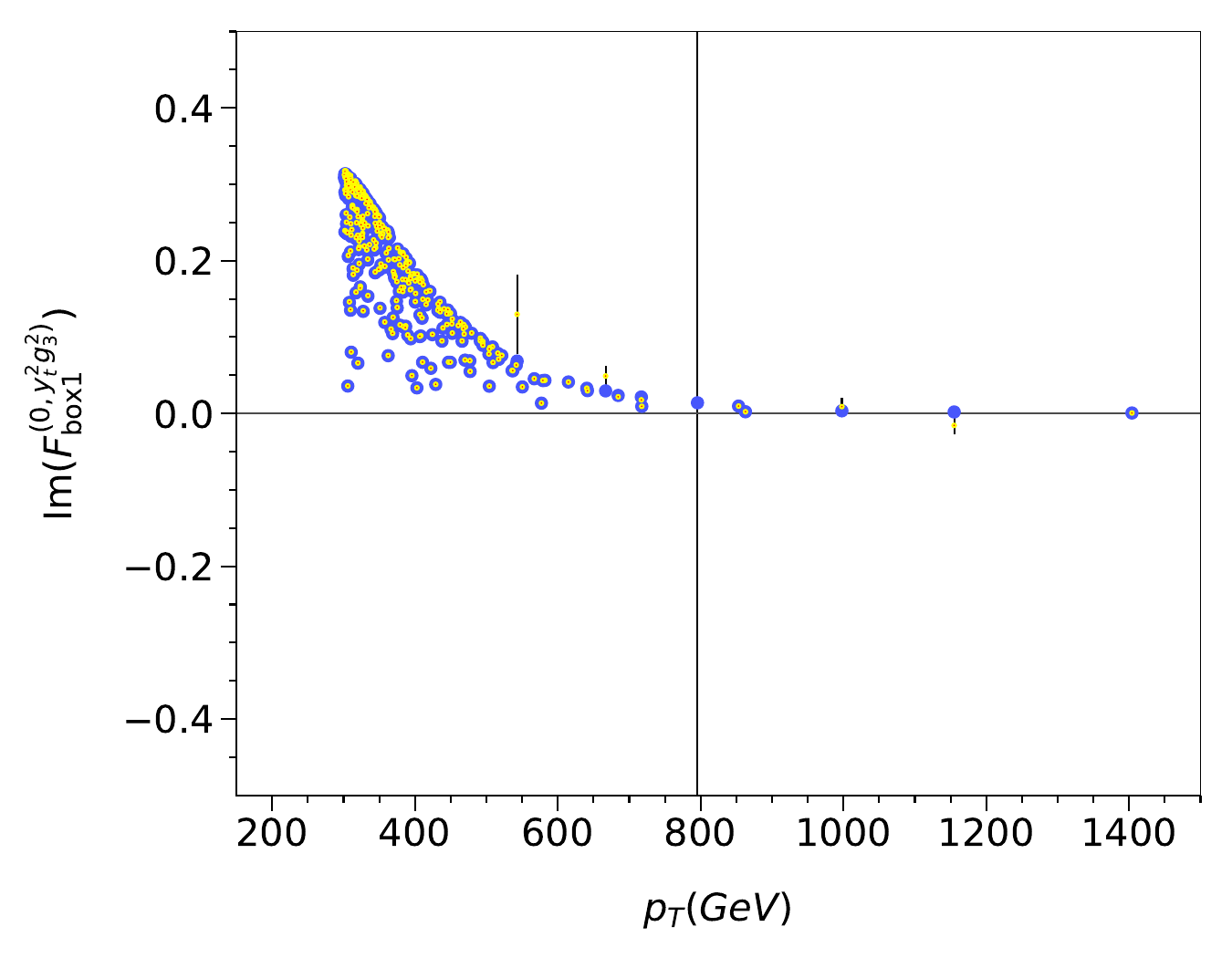}
  \includegraphics[width=0.35\textwidth]{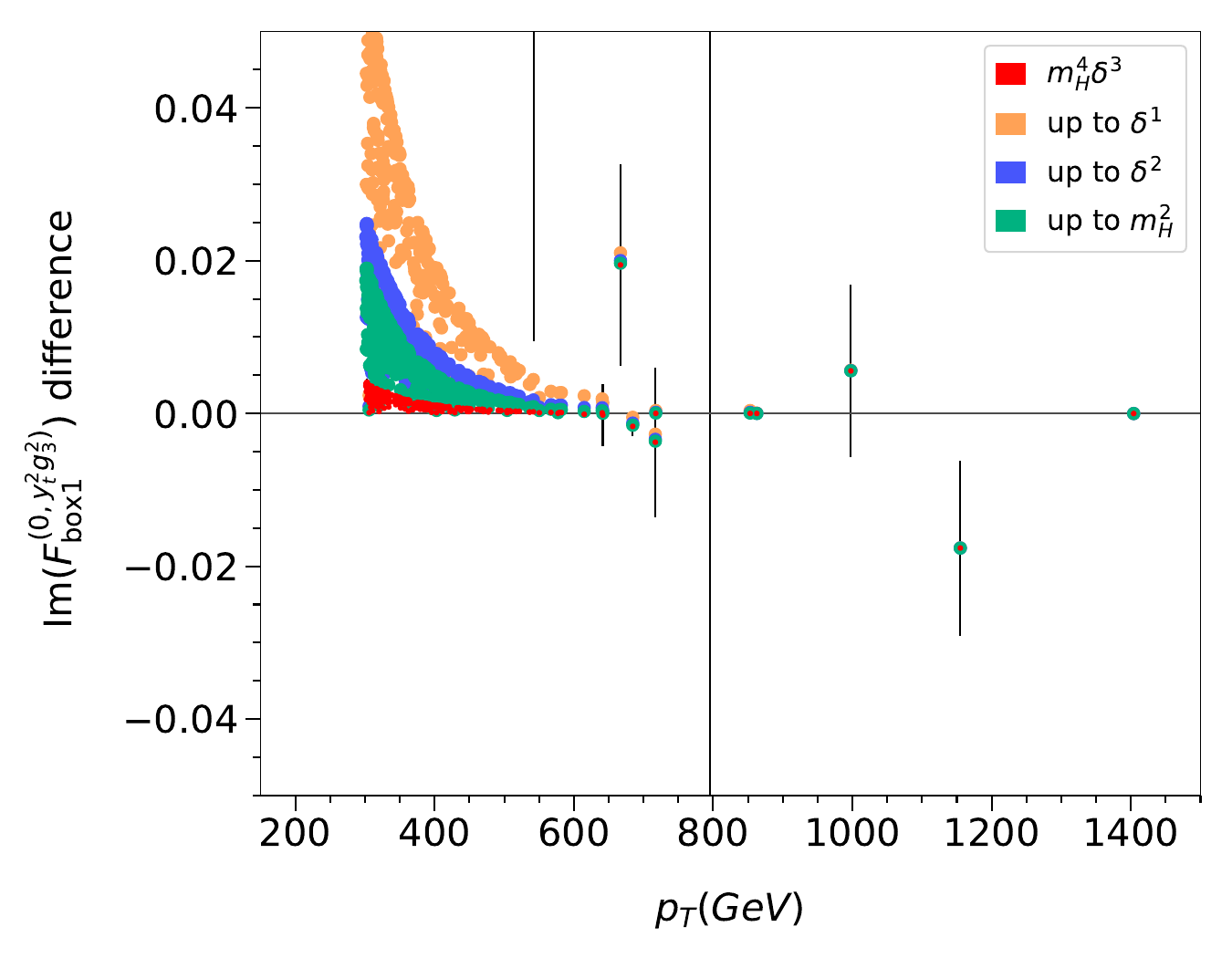}
  \includegraphics[width=0.35\textwidth]{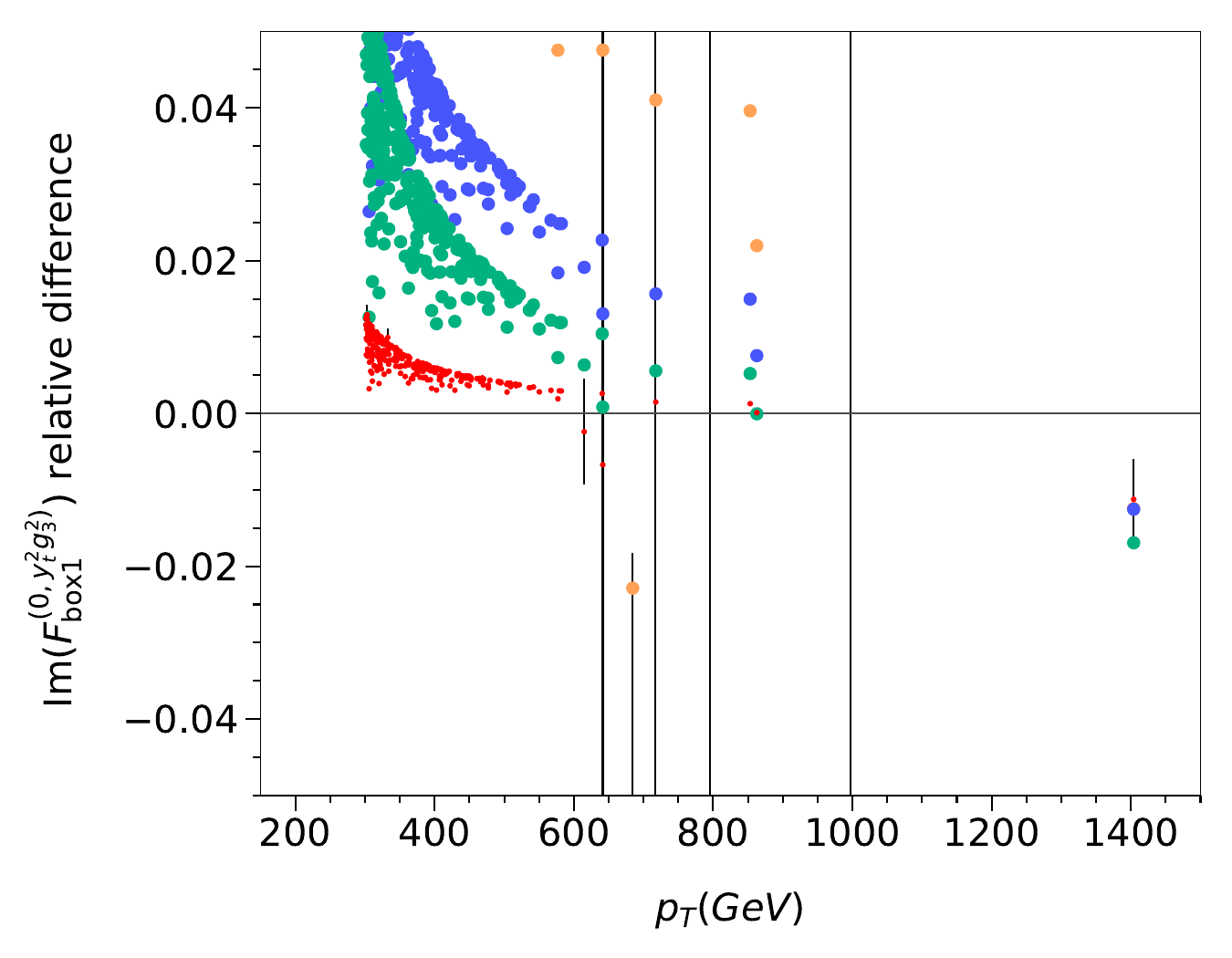}
\\
  \includegraphics[width=0.35\textwidth]{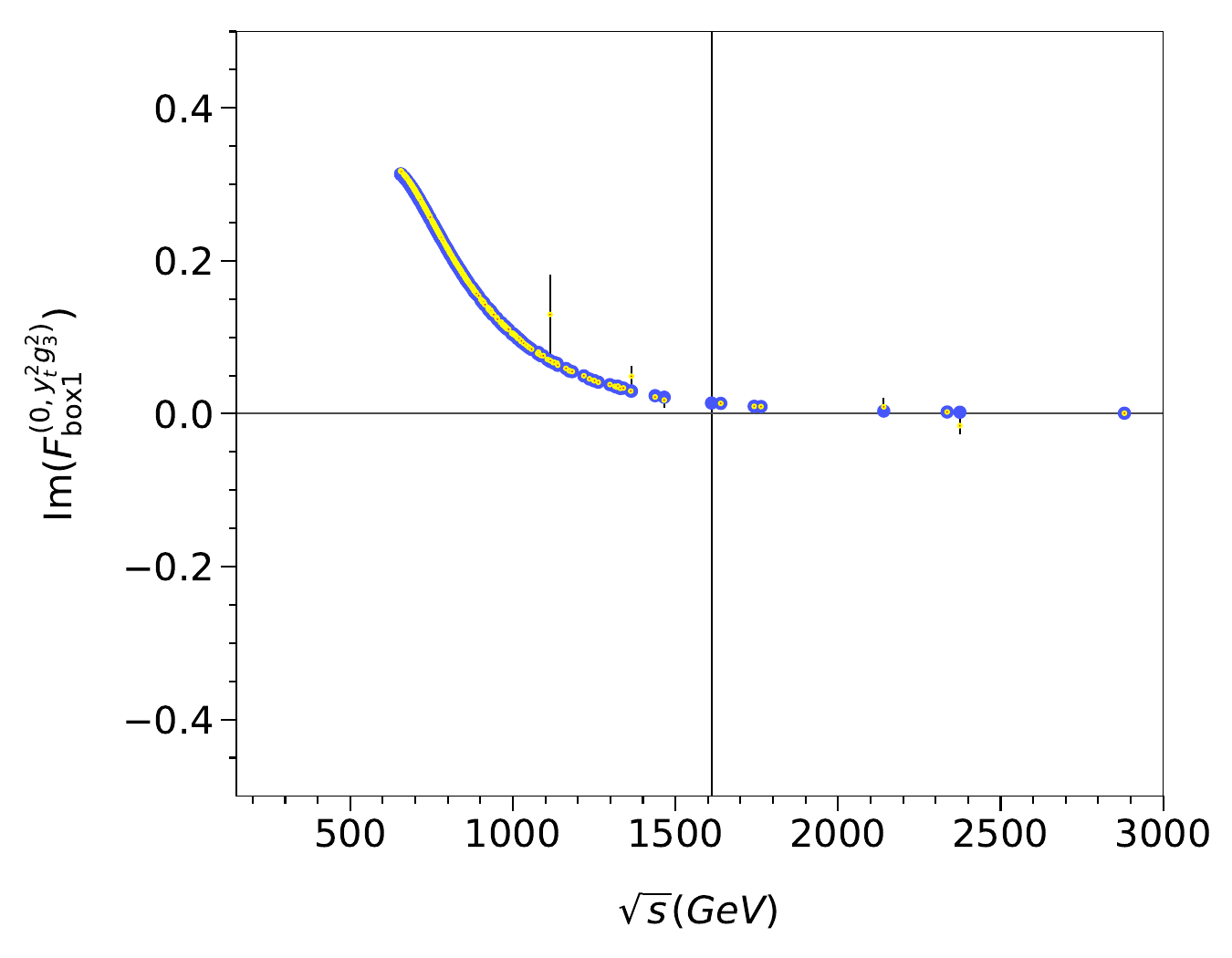}
  \includegraphics[width=0.35\textwidth]{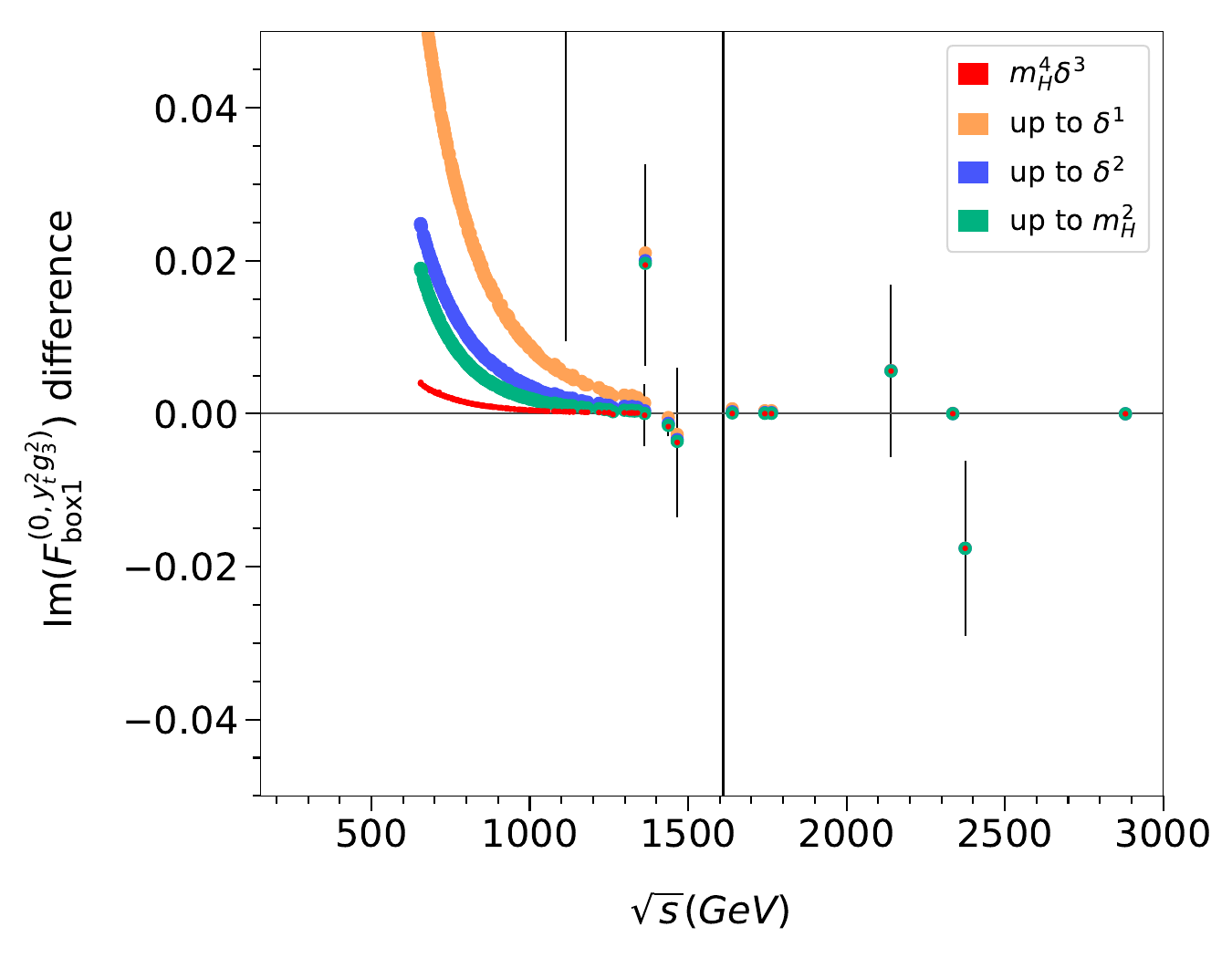}
  \includegraphics[width=0.35\textwidth]{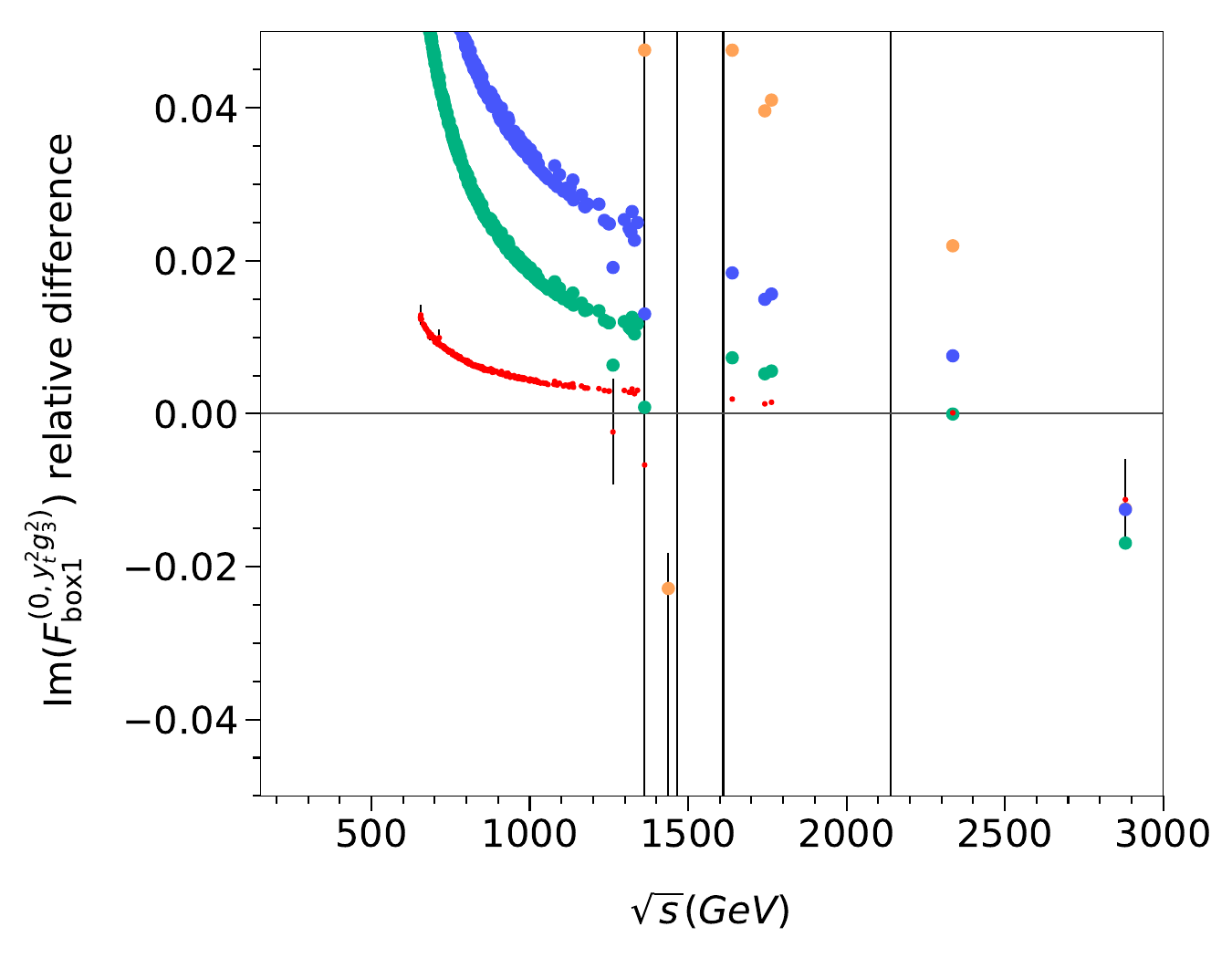}
    \end{tabular}
   \caption{\label{fig::diff_to_SD_yt2lam2}
   As Fig.~\ref{fig::diff_to_SD_yt4}, for $F_{\rm box1}^{(0,y_t^2 \, g_3^2)}$.}
\end{figure}

\begin{figure}[t]
  \centering
  \begin{tabular}{ccc}
  \includegraphics[width=0.35\textwidth]{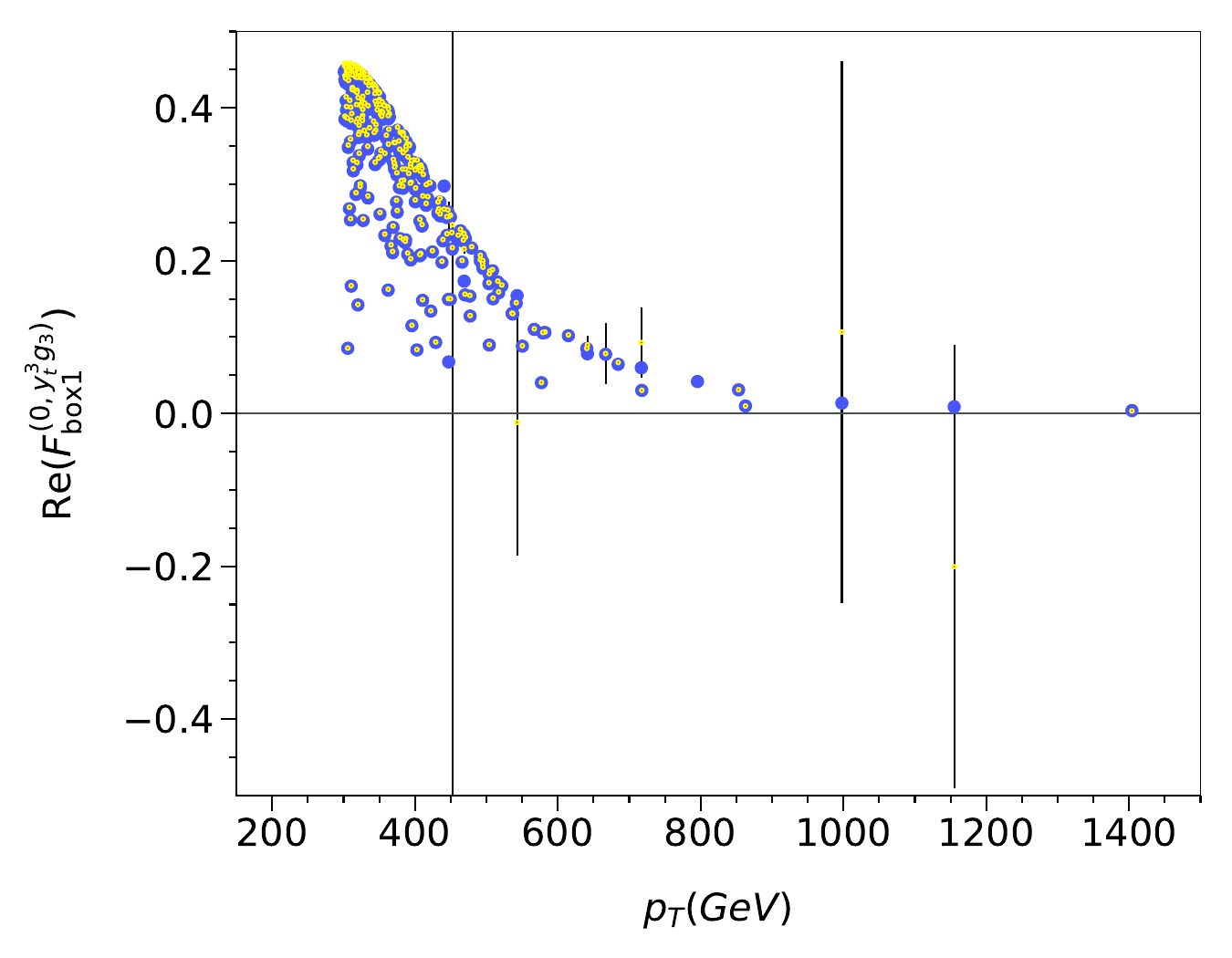}
  \includegraphics[width=0.35\textwidth]{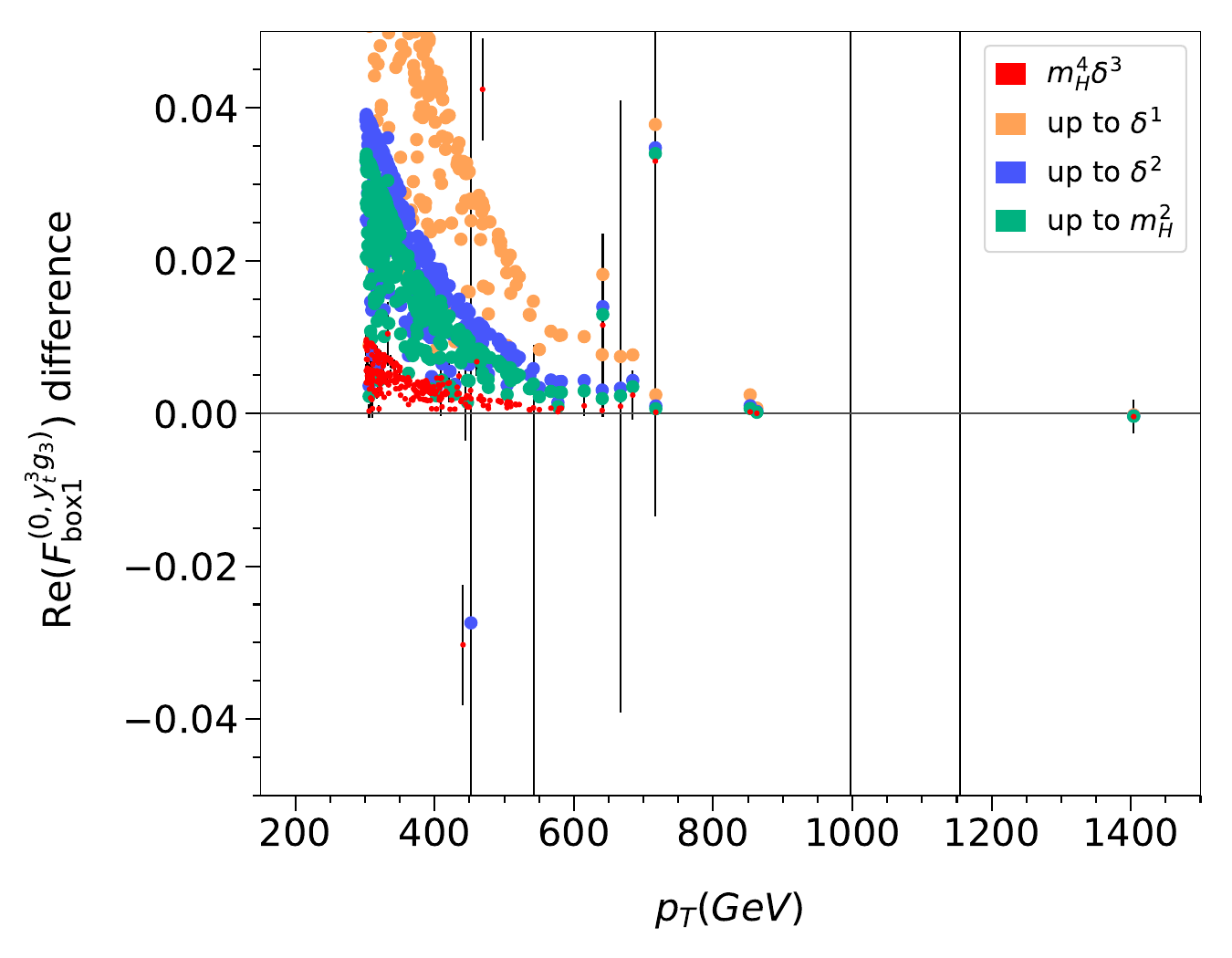}
  \includegraphics[width=0.35\textwidth]{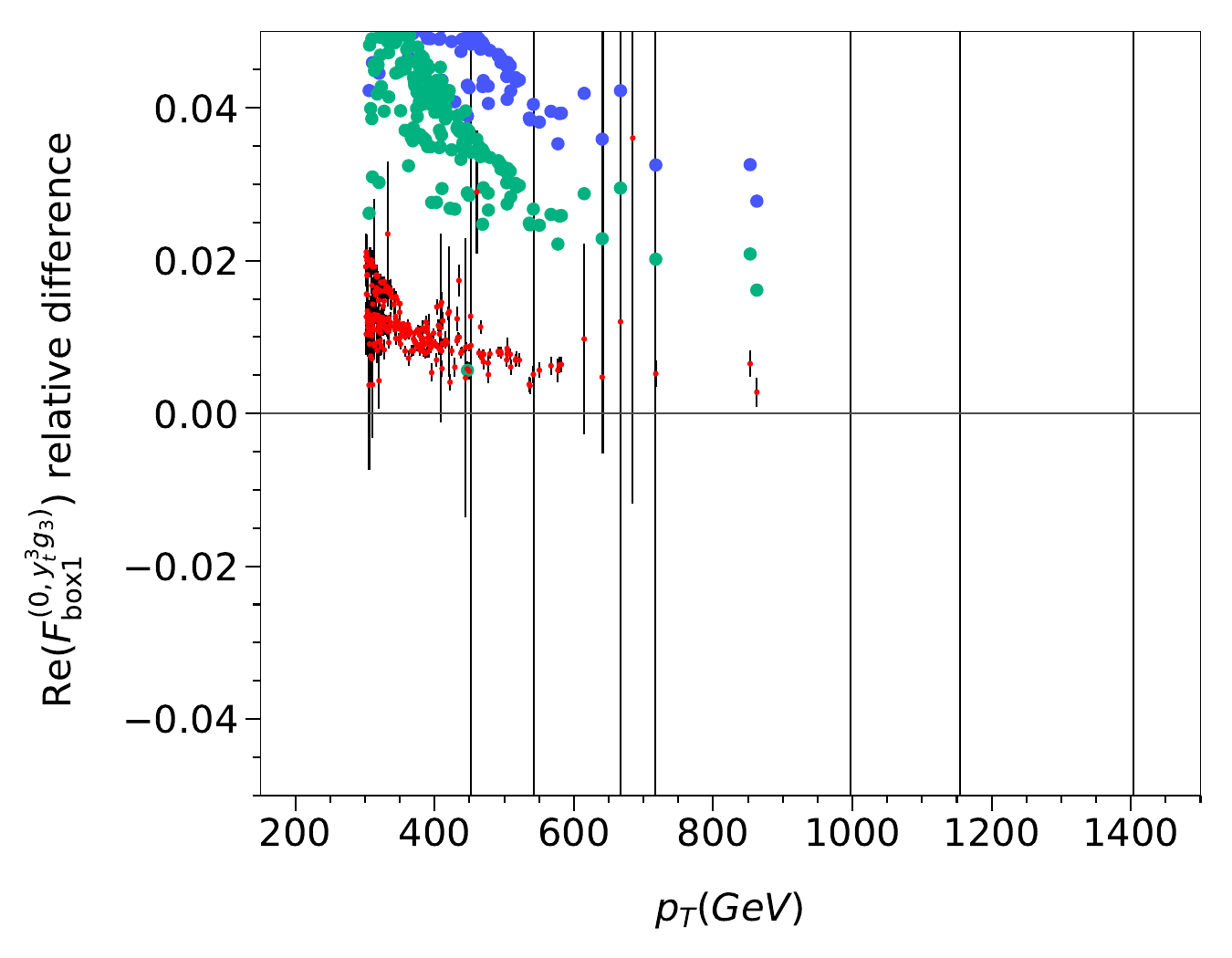}
\\
  \includegraphics[width=0.35\textwidth]{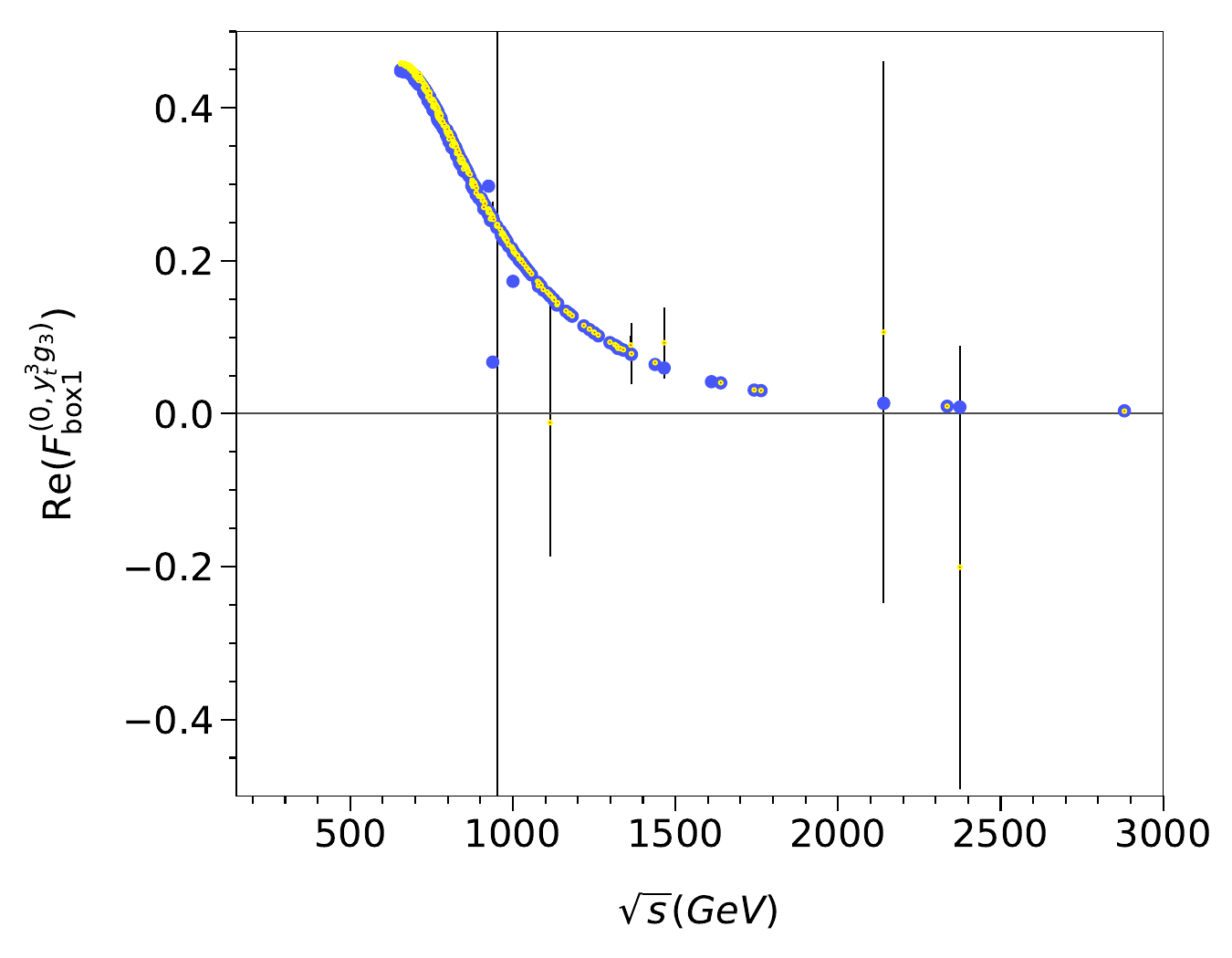}
  \includegraphics[width=0.35\textwidth]{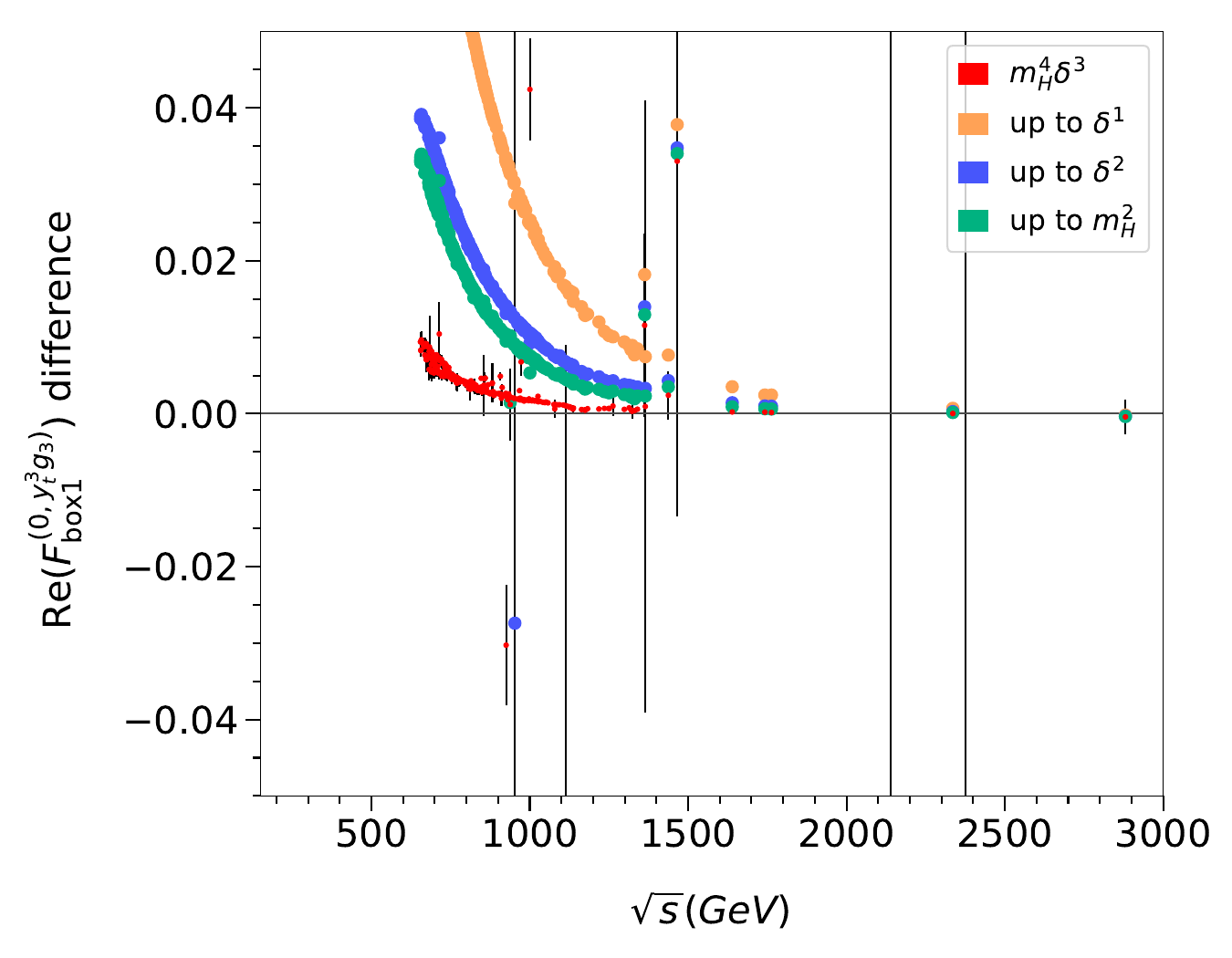}
  \includegraphics[width=0.35\textwidth]{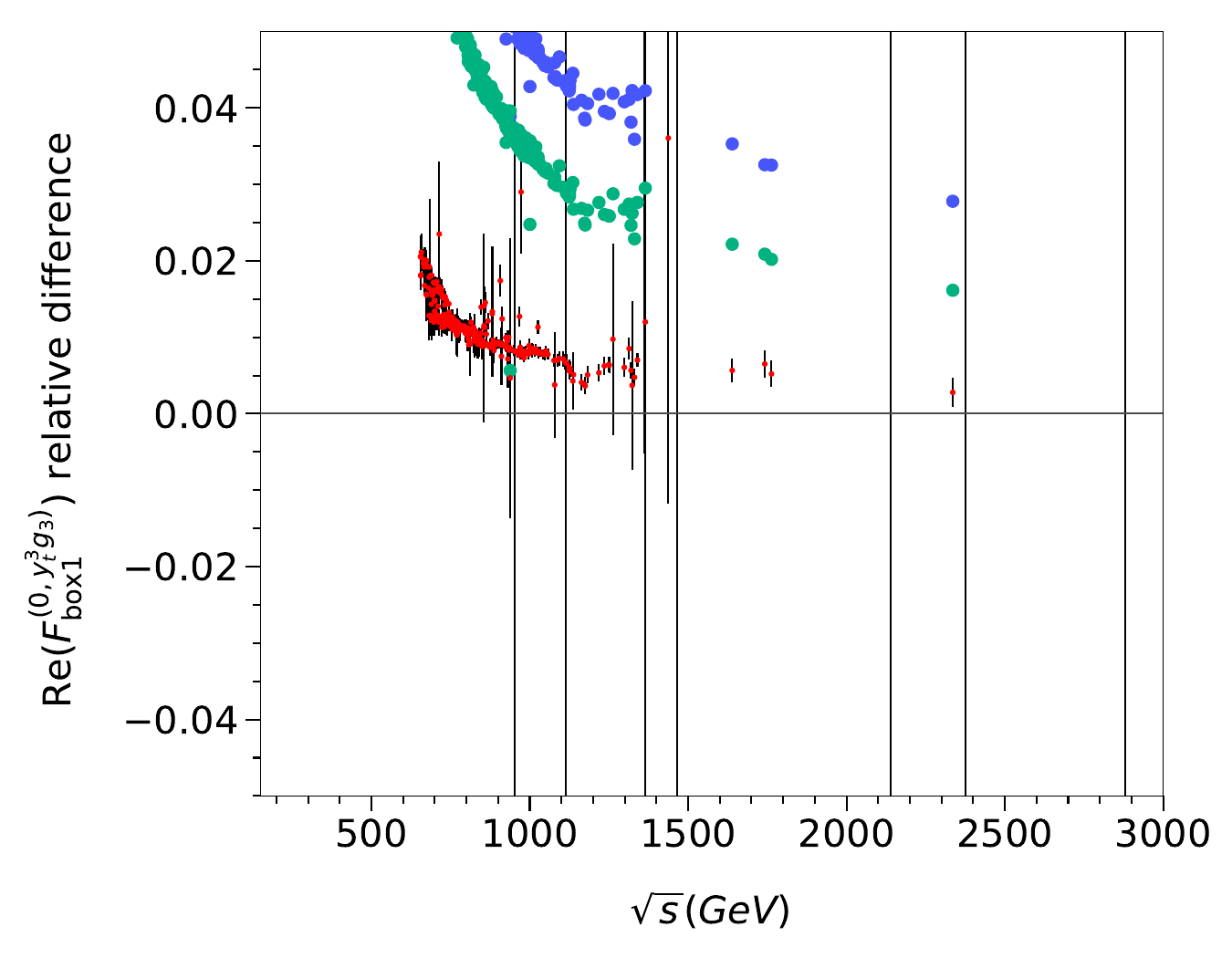}
\\
  \includegraphics[width=0.35\textwidth]{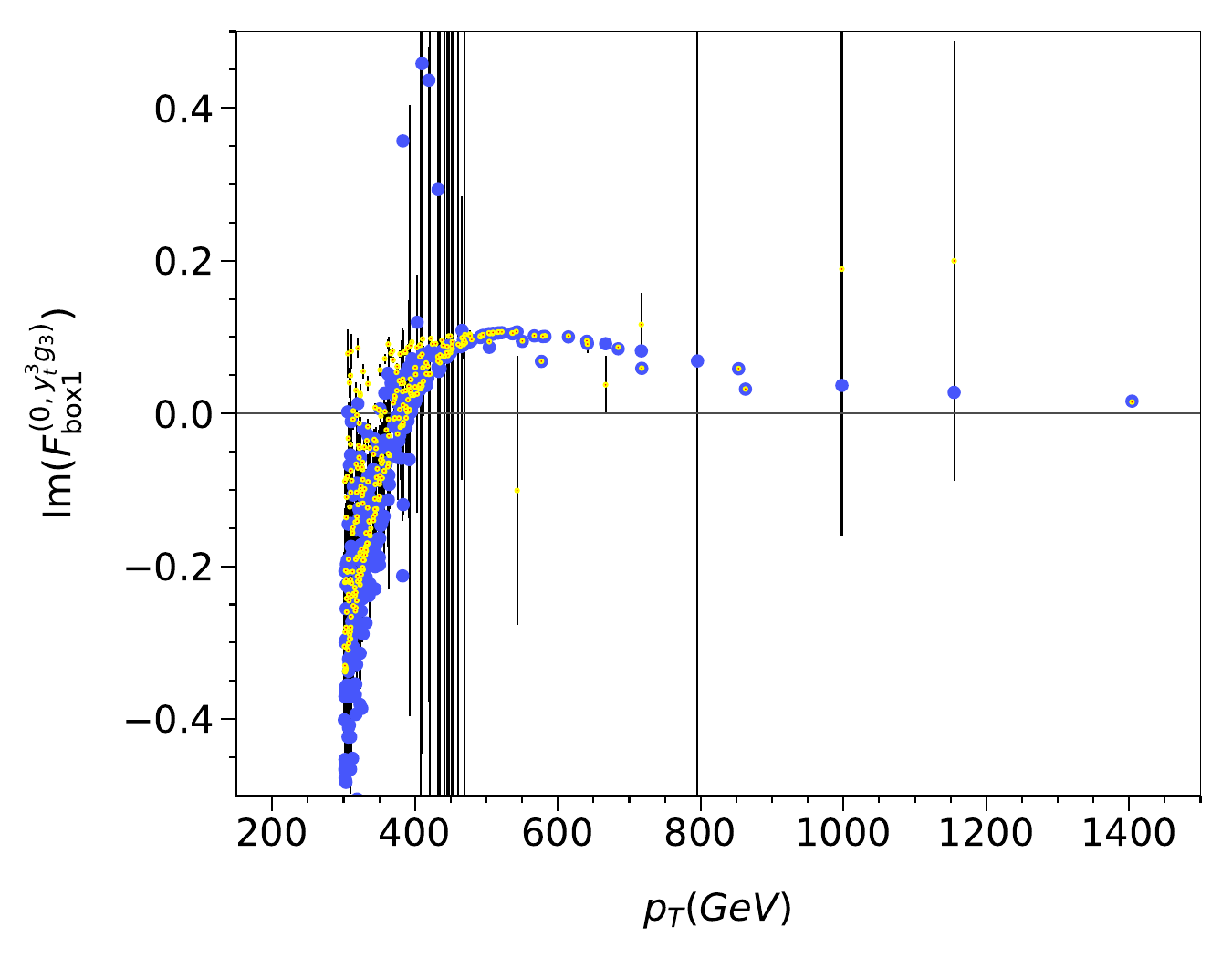}
  \includegraphics[width=0.35\textwidth]{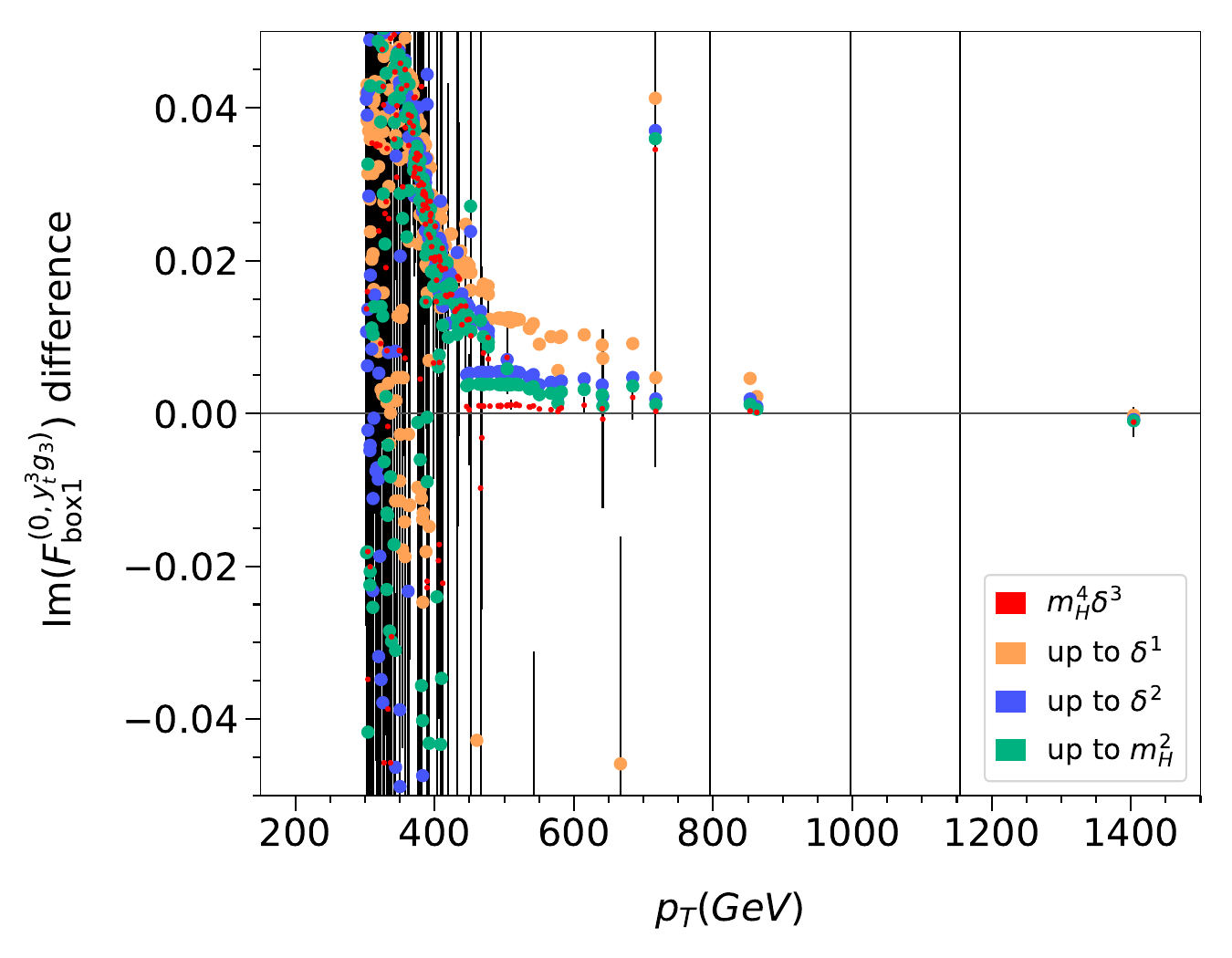}
  \includegraphics[width=0.35\textwidth]{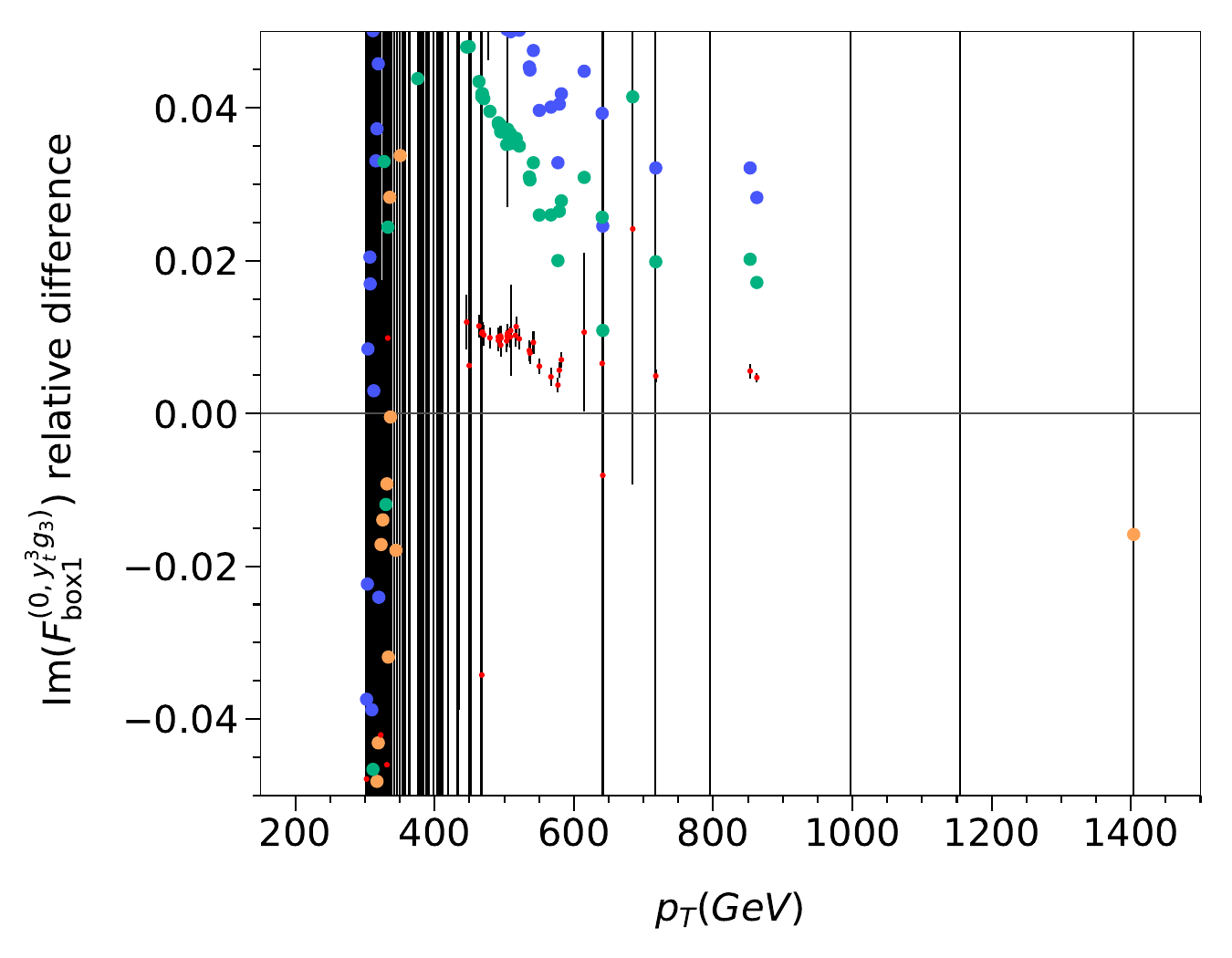}
\\
  \includegraphics[width=0.35\textwidth]{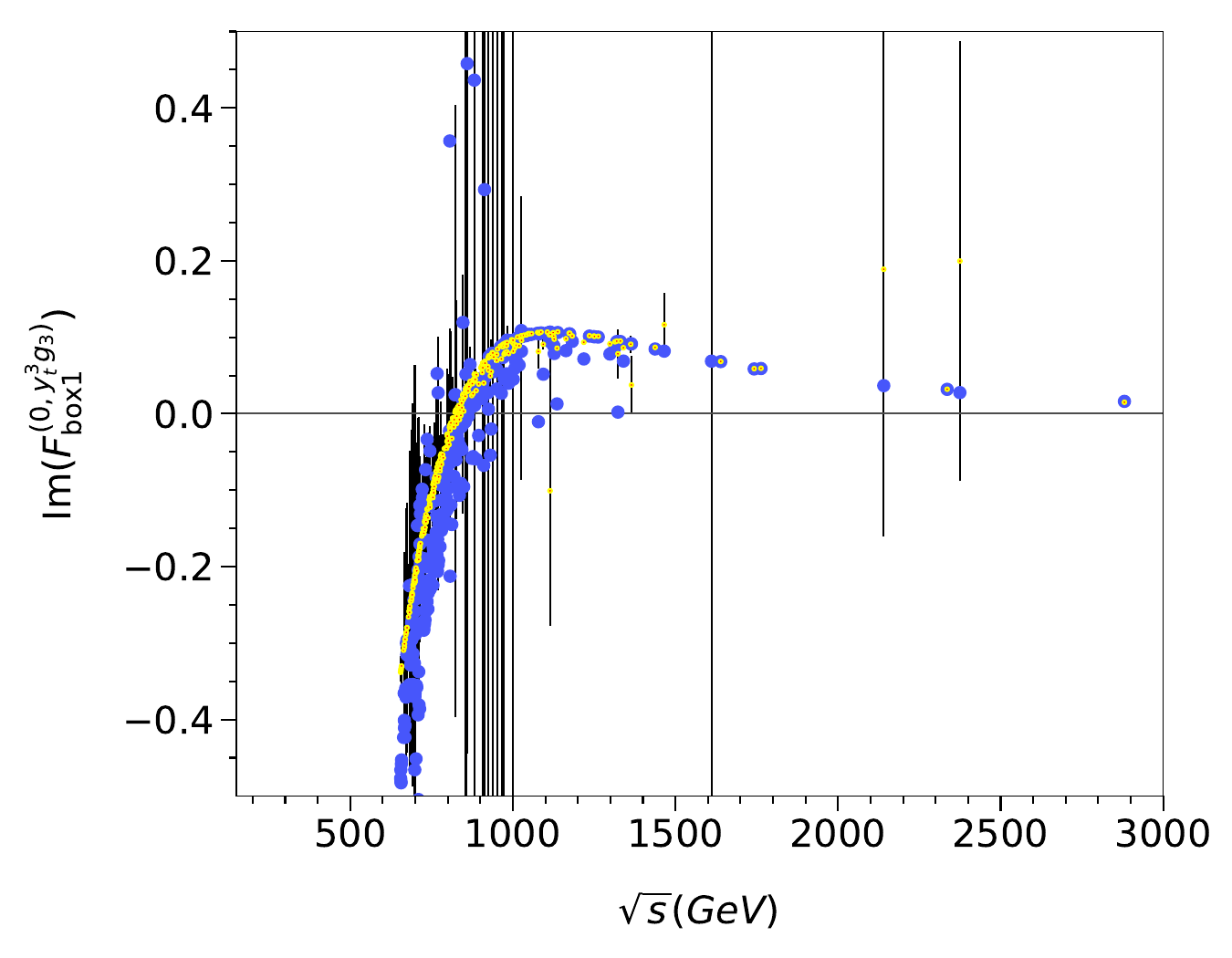}
  \includegraphics[width=0.35\textwidth]{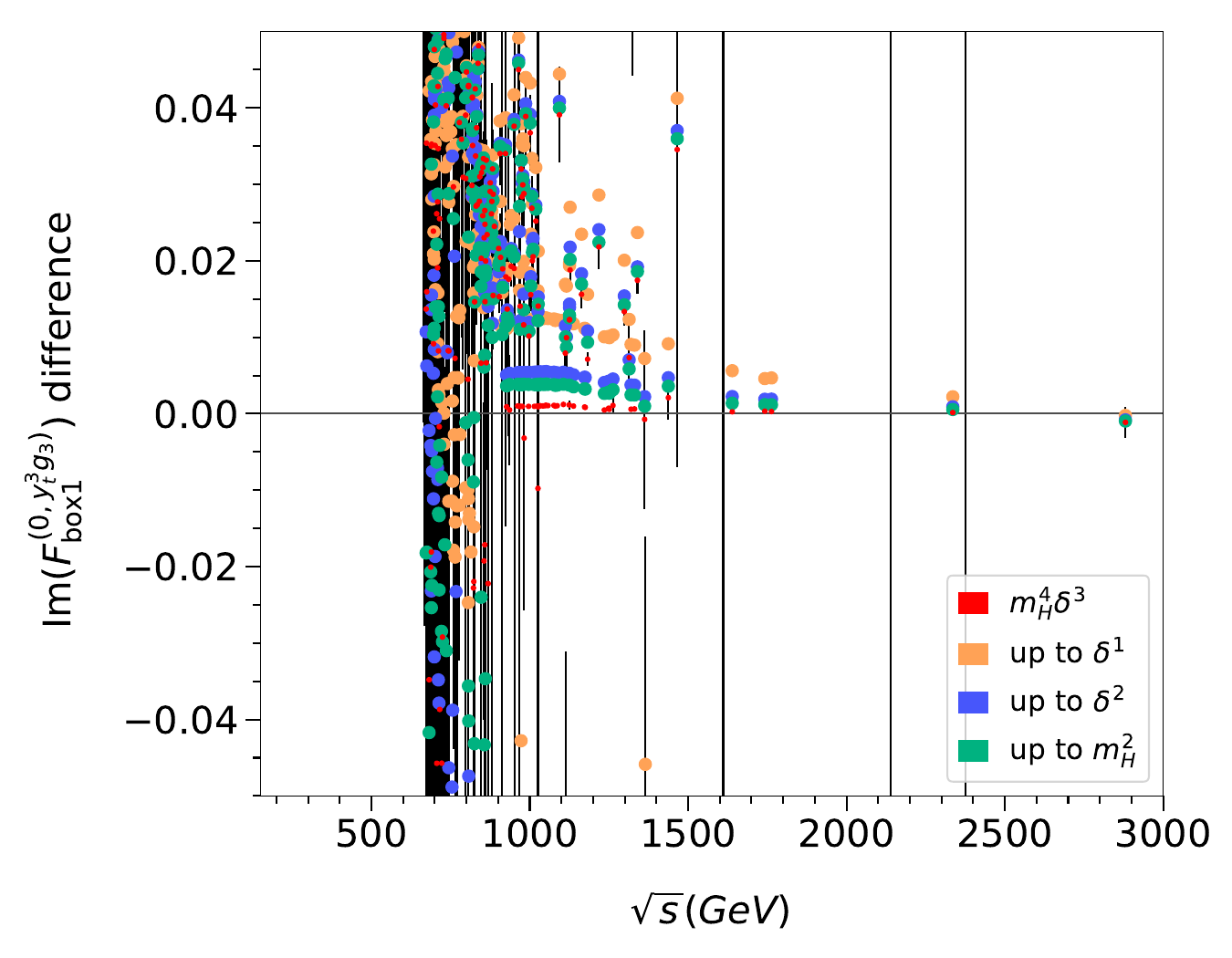}
  \includegraphics[width=0.35\textwidth]{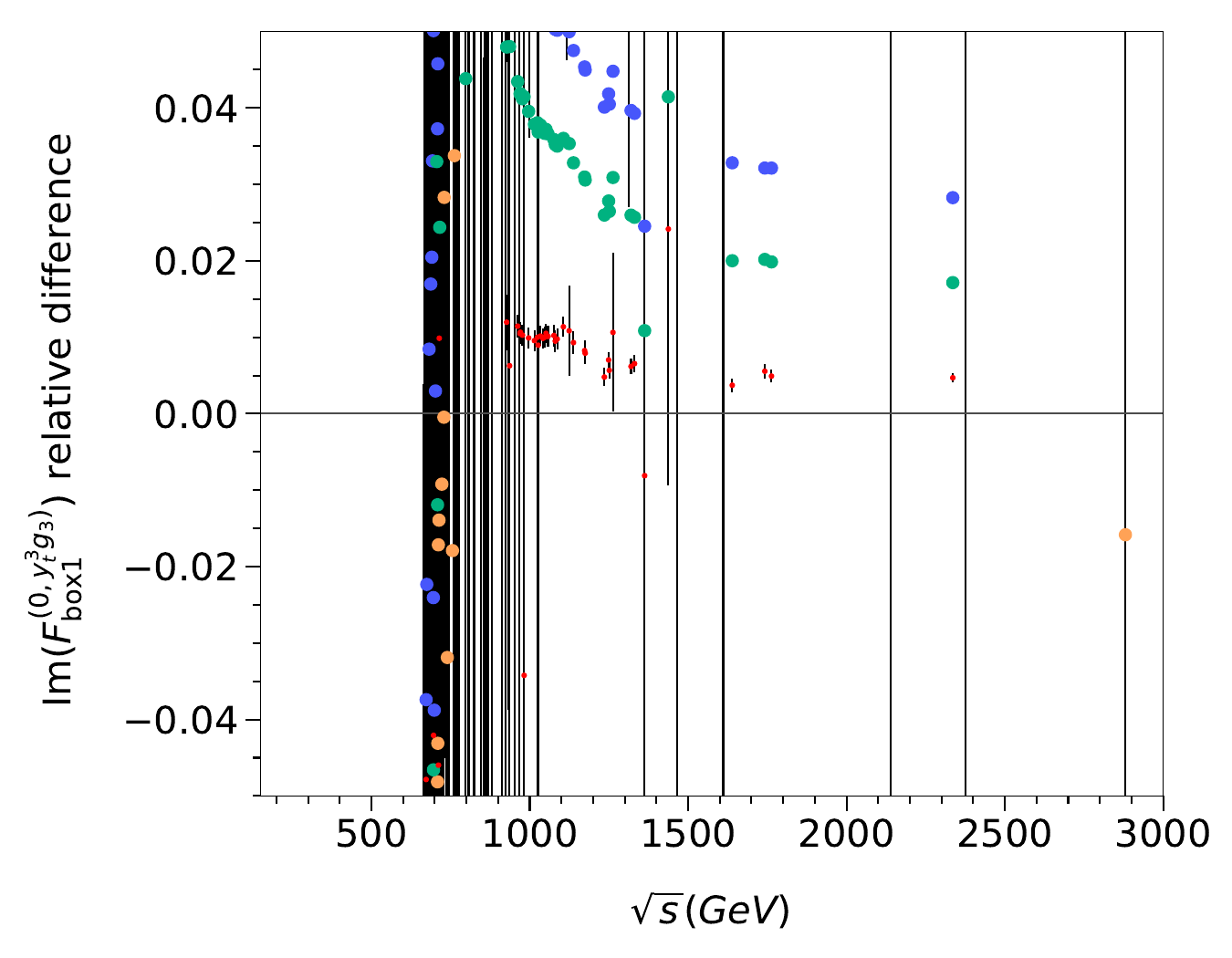}
   \end{tabular}
   \caption{\label{fig::diff_to_SD_yt3lam1}
   As Fig.~\ref{fig::diff_to_SD_yt4}, for $F_{\rm box1}^{(0,y_t^3 \, g_3)}$.
   }
\end{figure}

\begin{figure}[t]
  \centering
  \begin{tabular}{ccc}
  \includegraphics[width=0.35\textwidth]{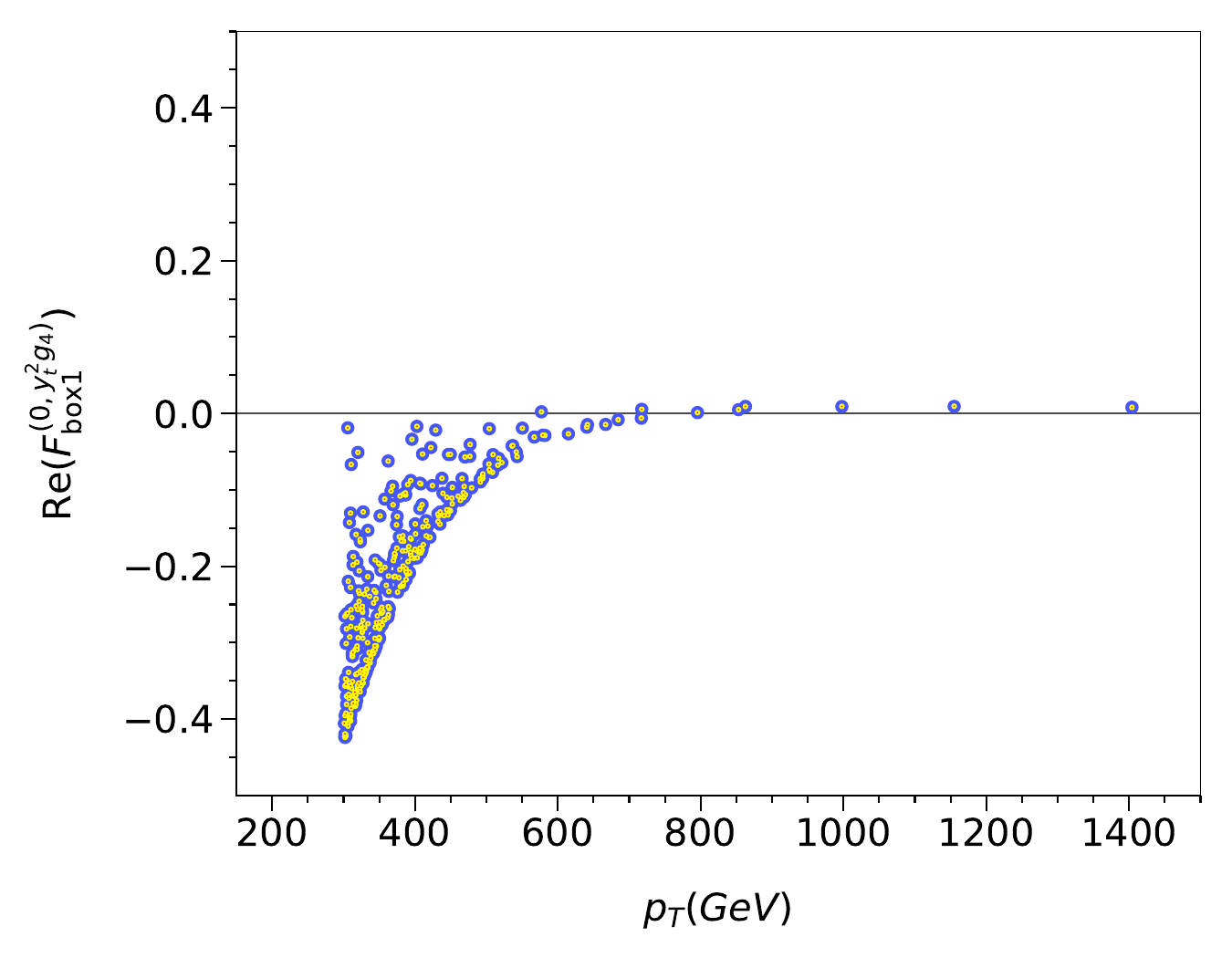}
  \includegraphics[width=0.35\textwidth]{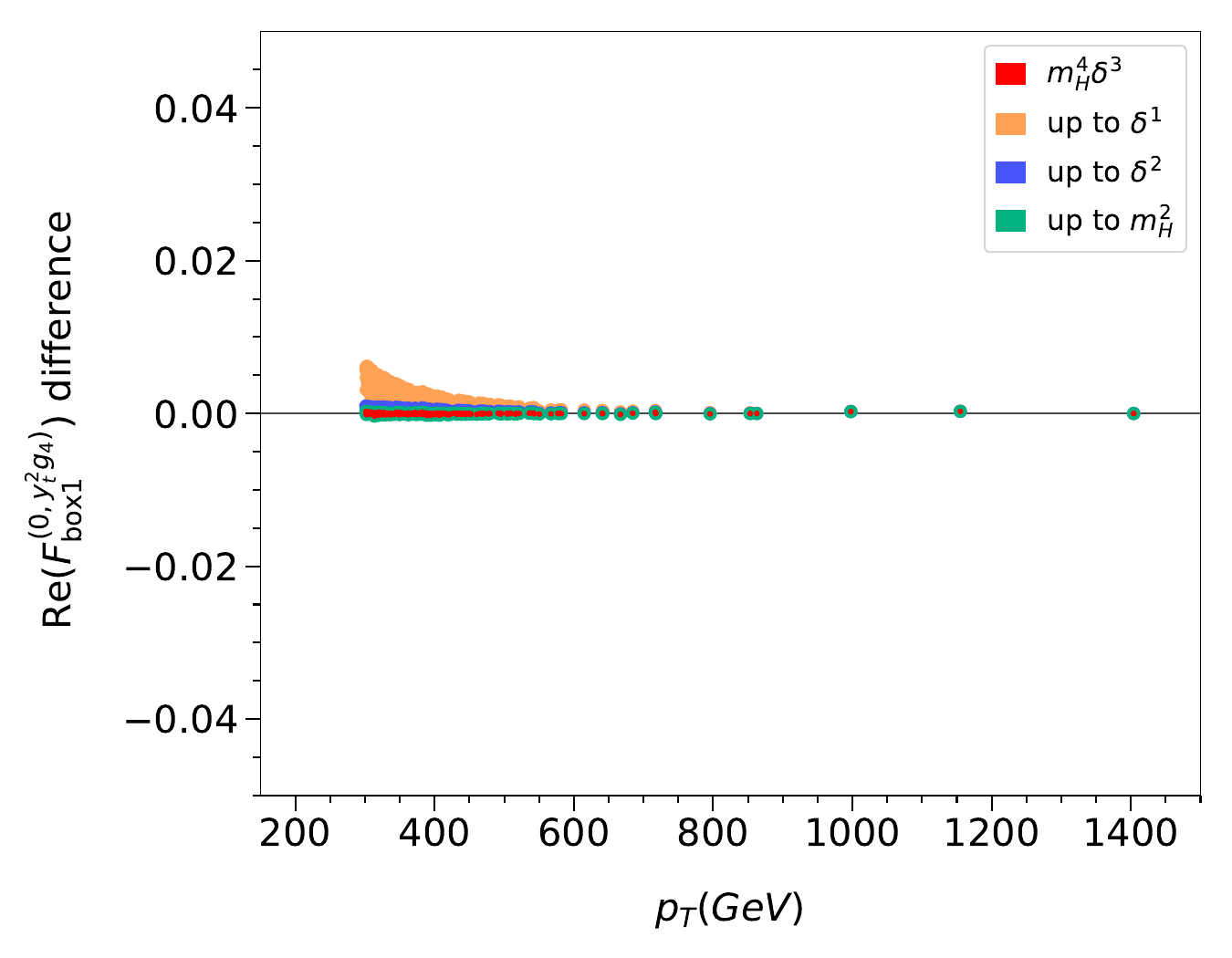}
  \includegraphics[width=0.35\textwidth]{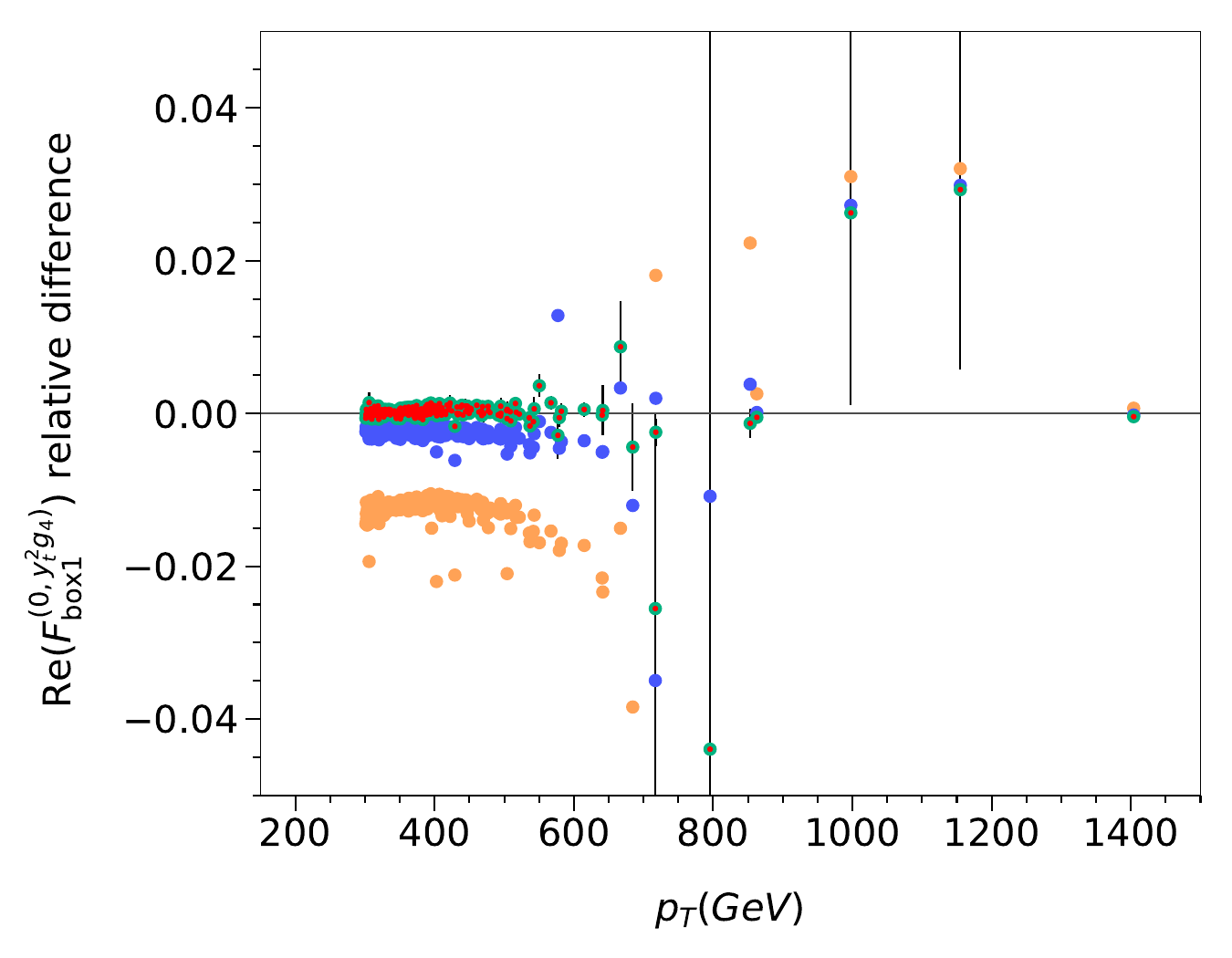}
\\
  \includegraphics[width=0.35\textwidth]{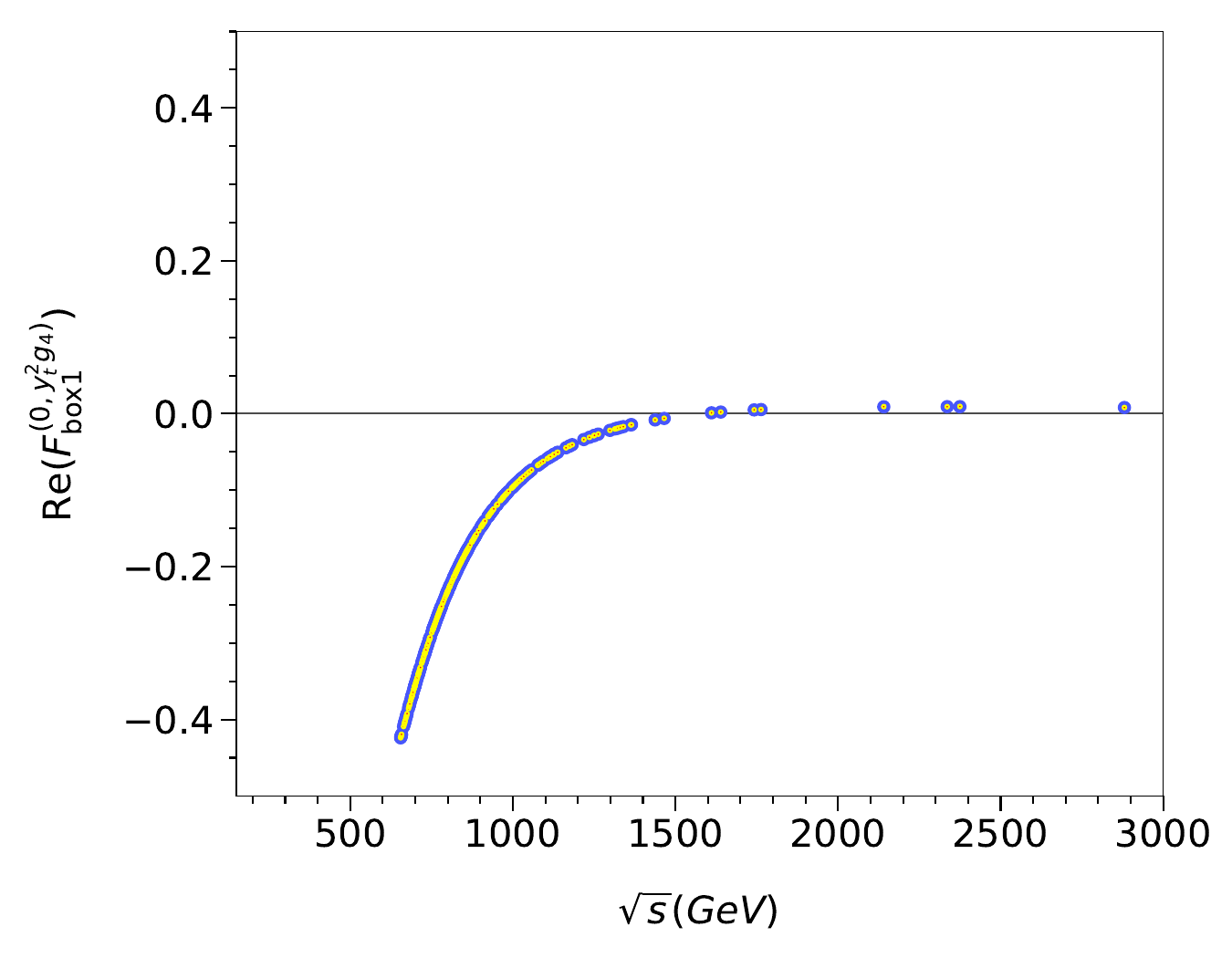}
  \includegraphics[width=0.35\textwidth]{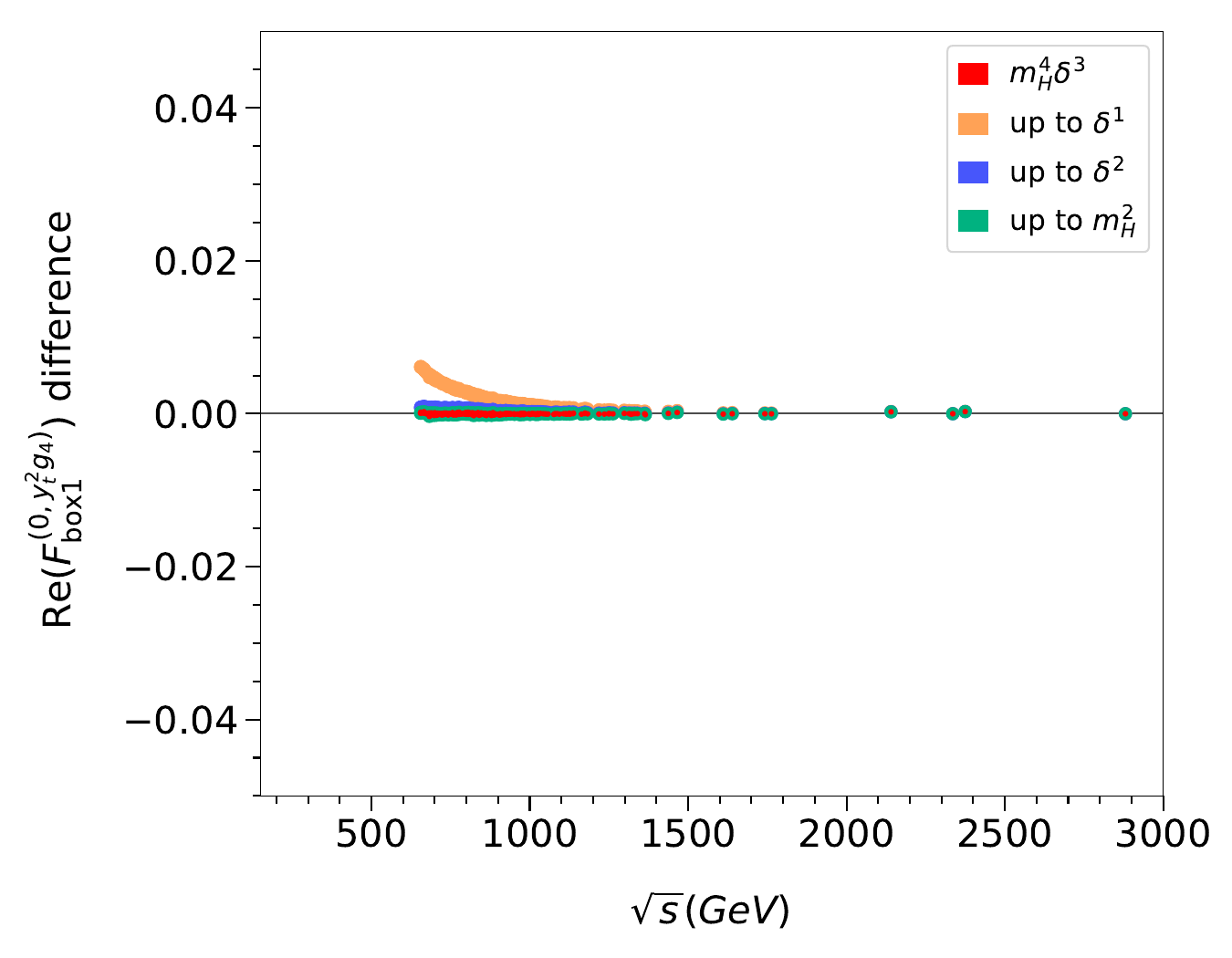}
  \includegraphics[width=0.35\textwidth]{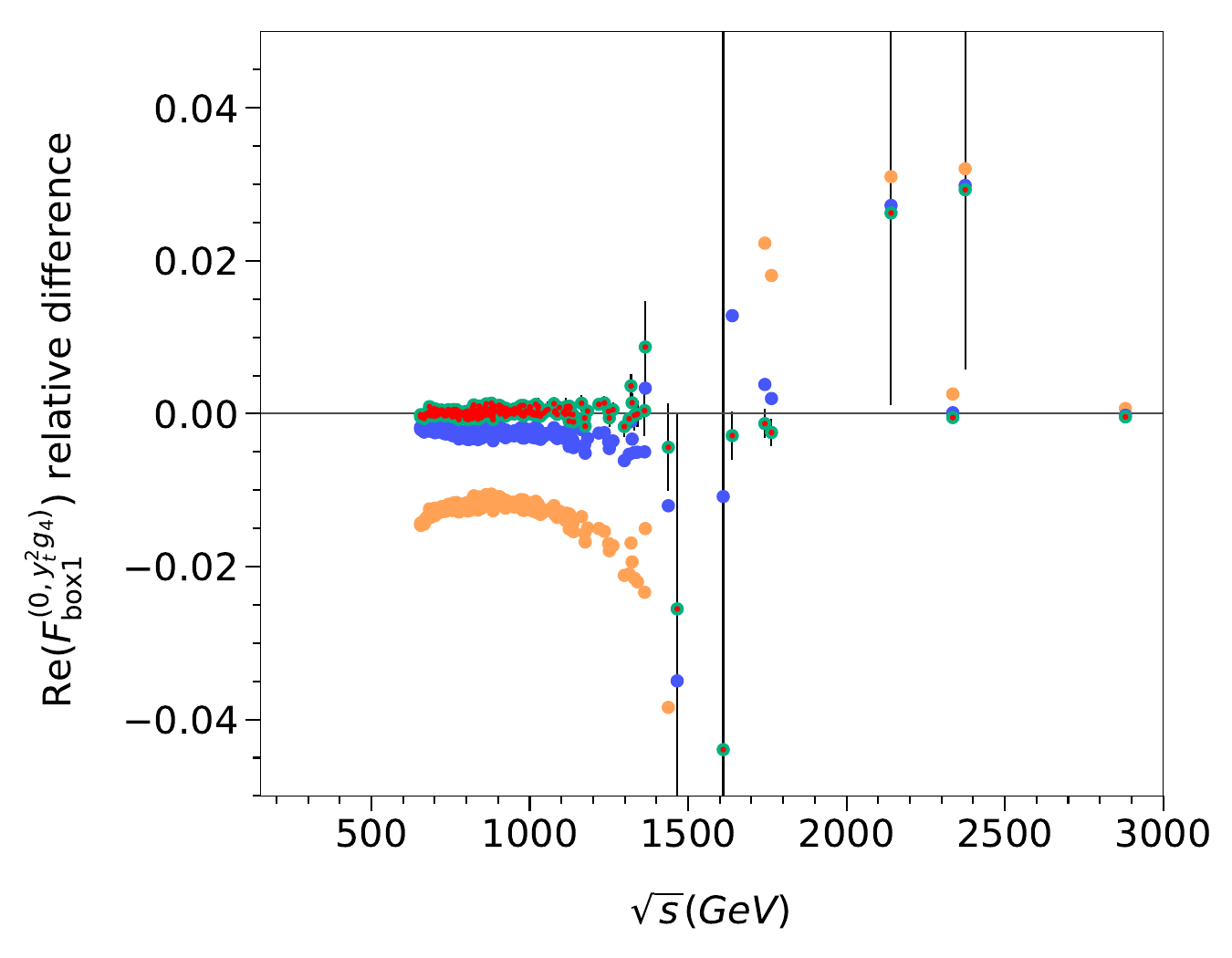}
\\
  \includegraphics[width=0.35\textwidth]{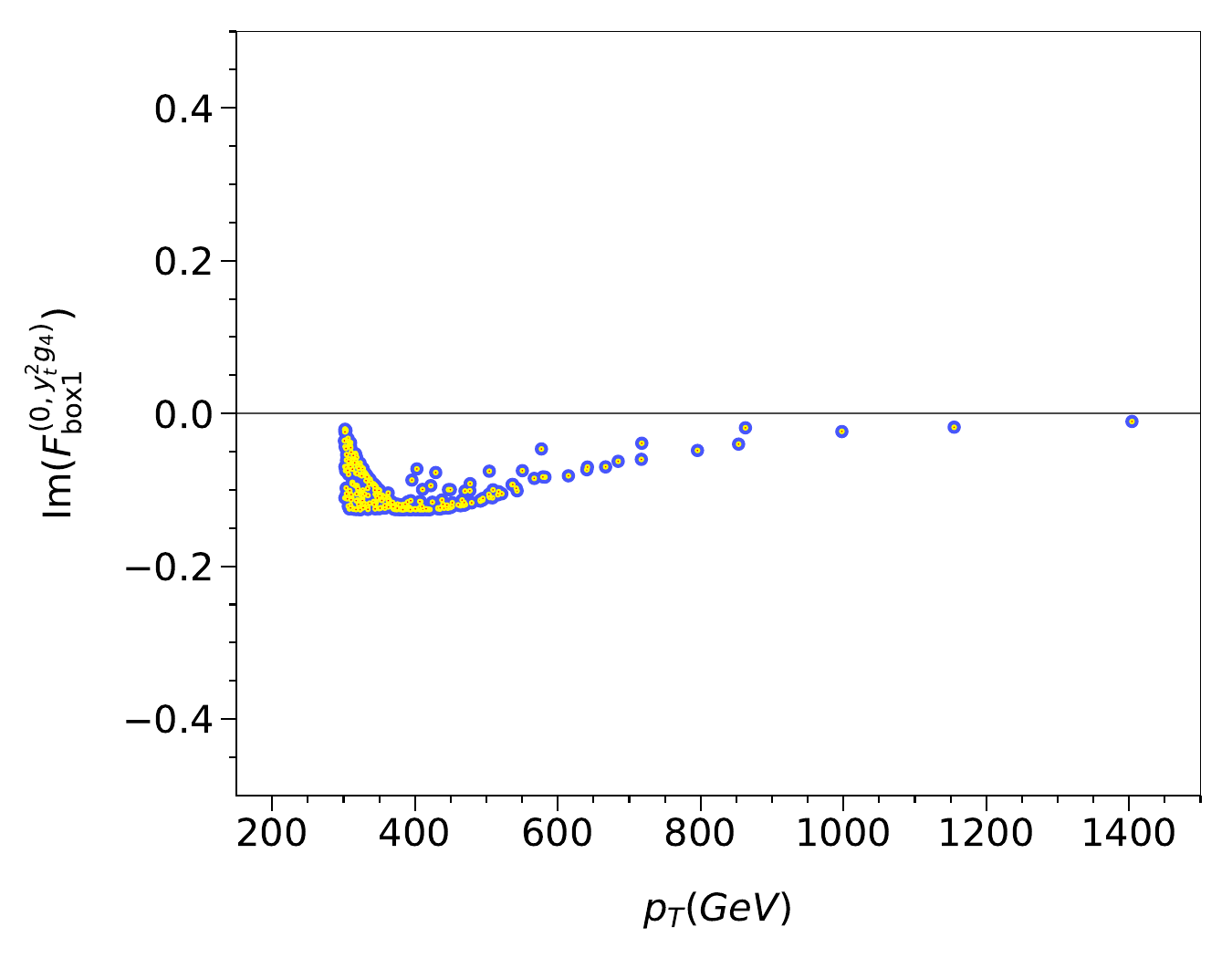}
  \includegraphics[width=0.35\textwidth]{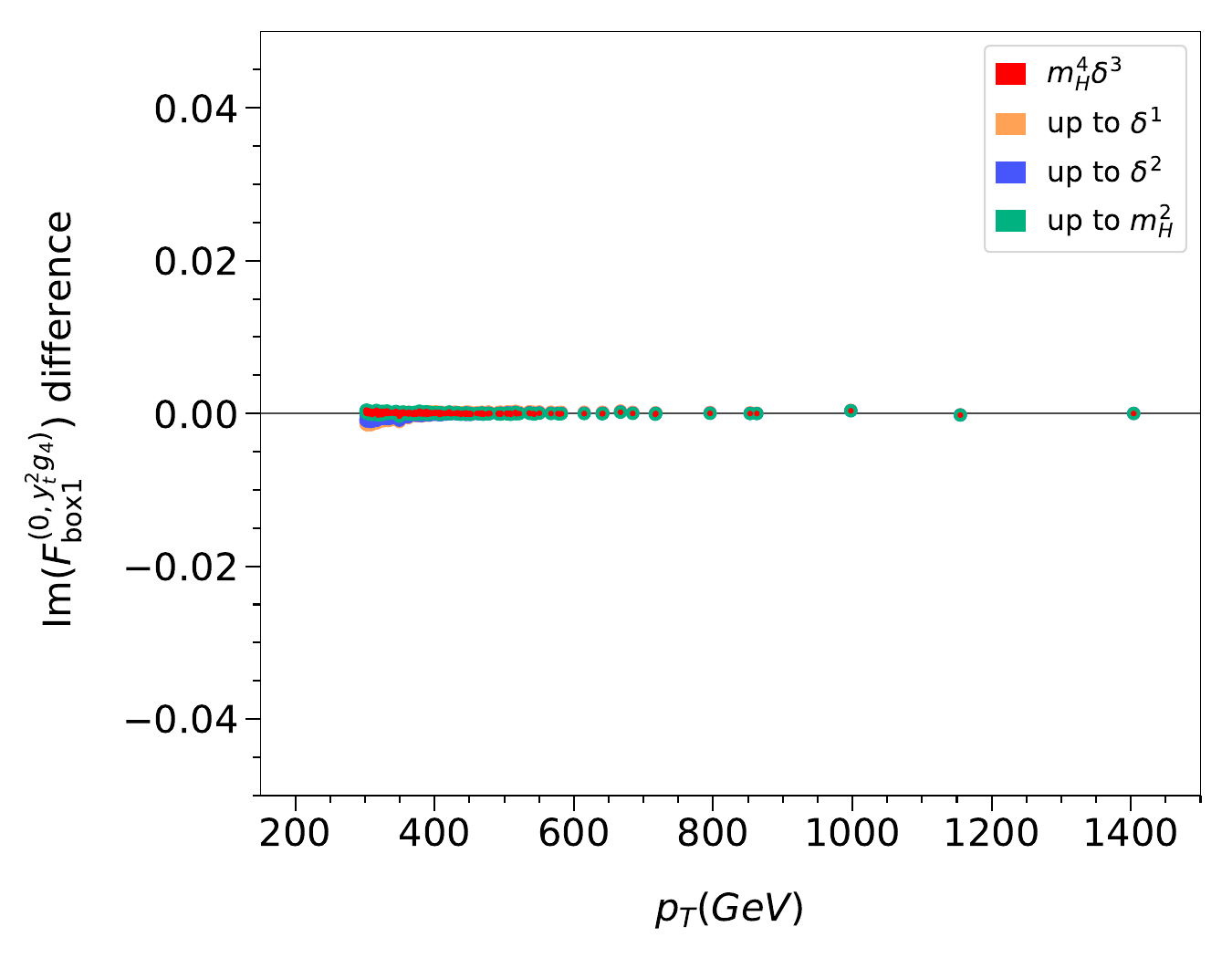}
  \includegraphics[width=0.35\textwidth]{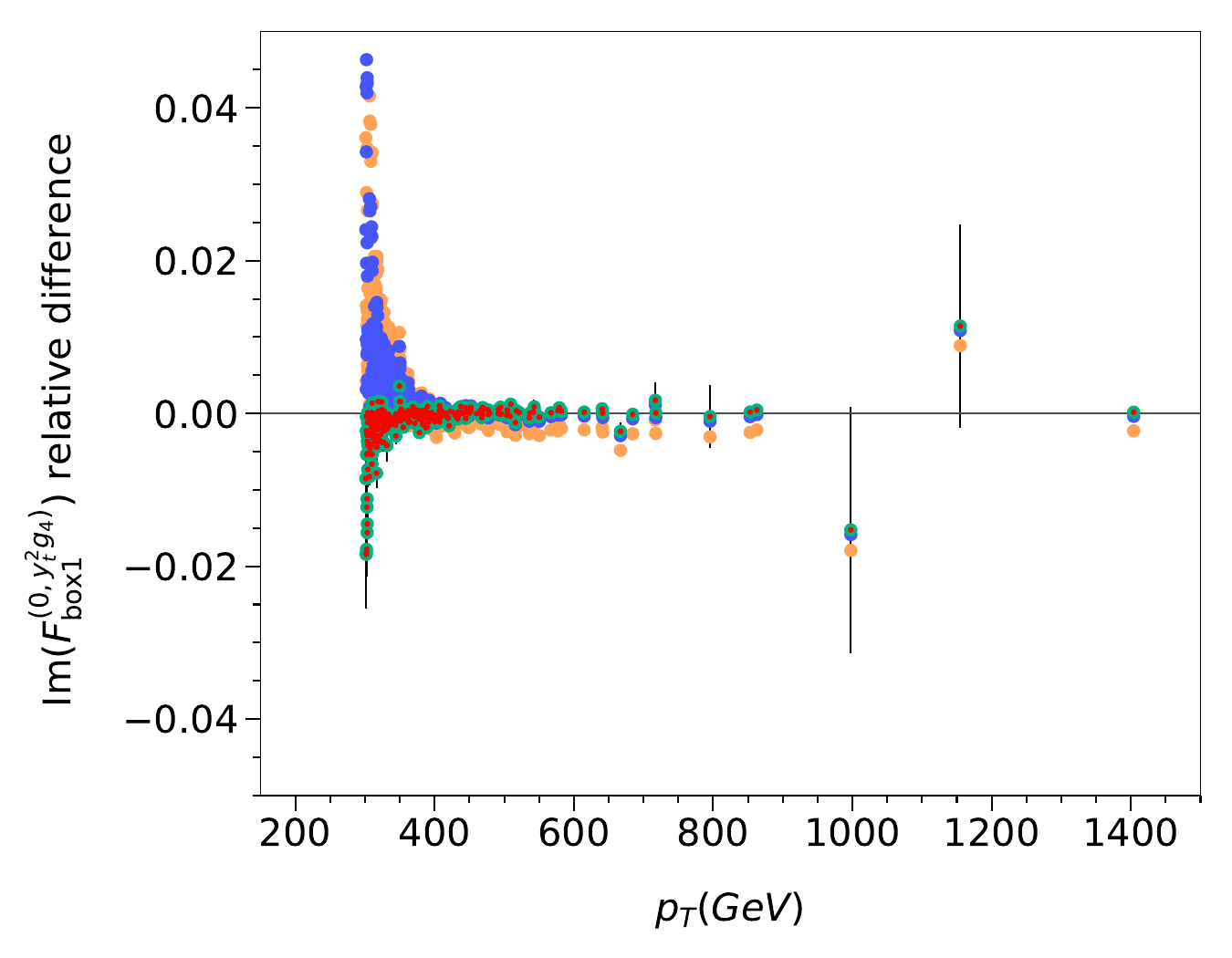}
\\
  \includegraphics[width=0.35\textwidth]{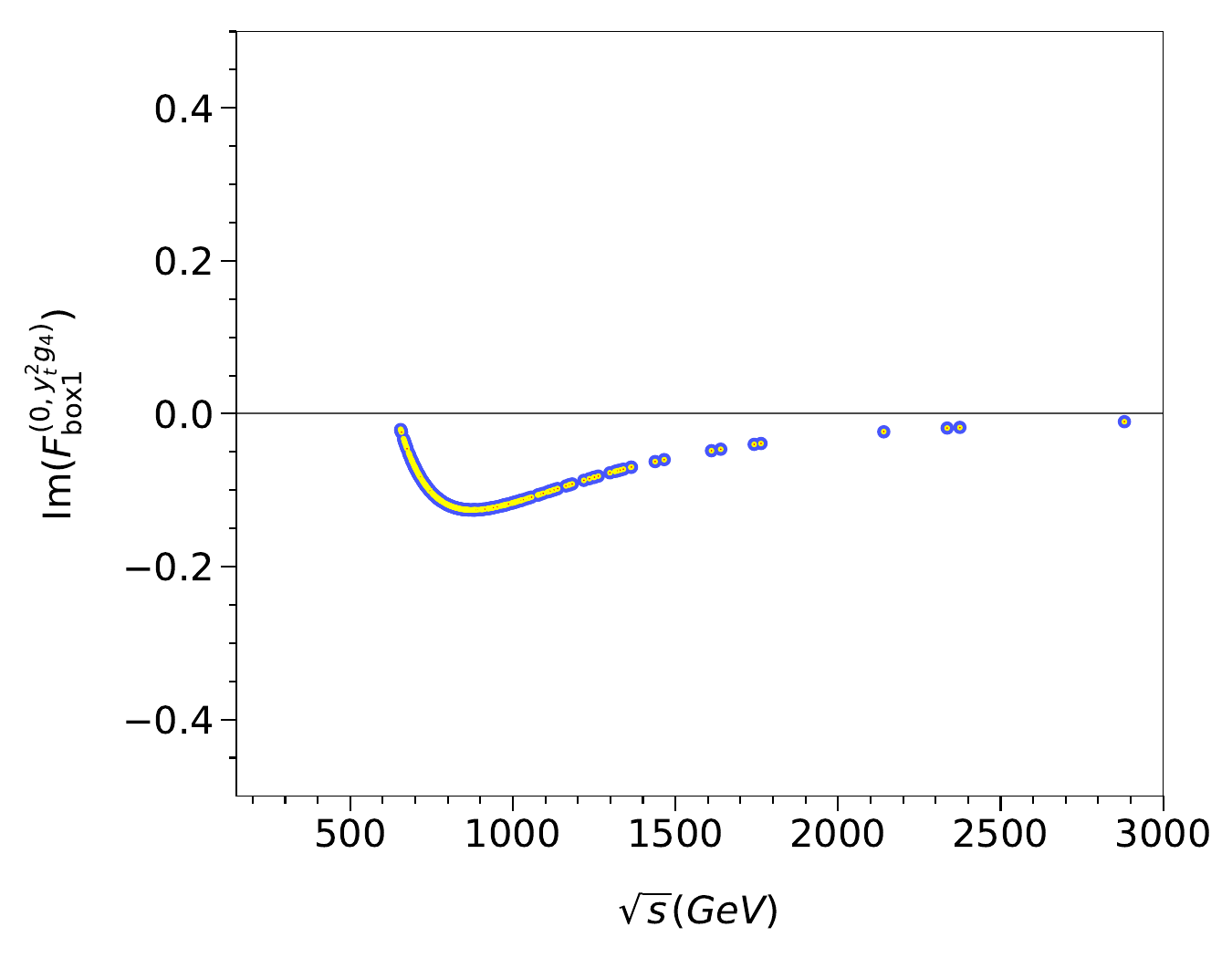}
  \includegraphics[width=0.35\textwidth]{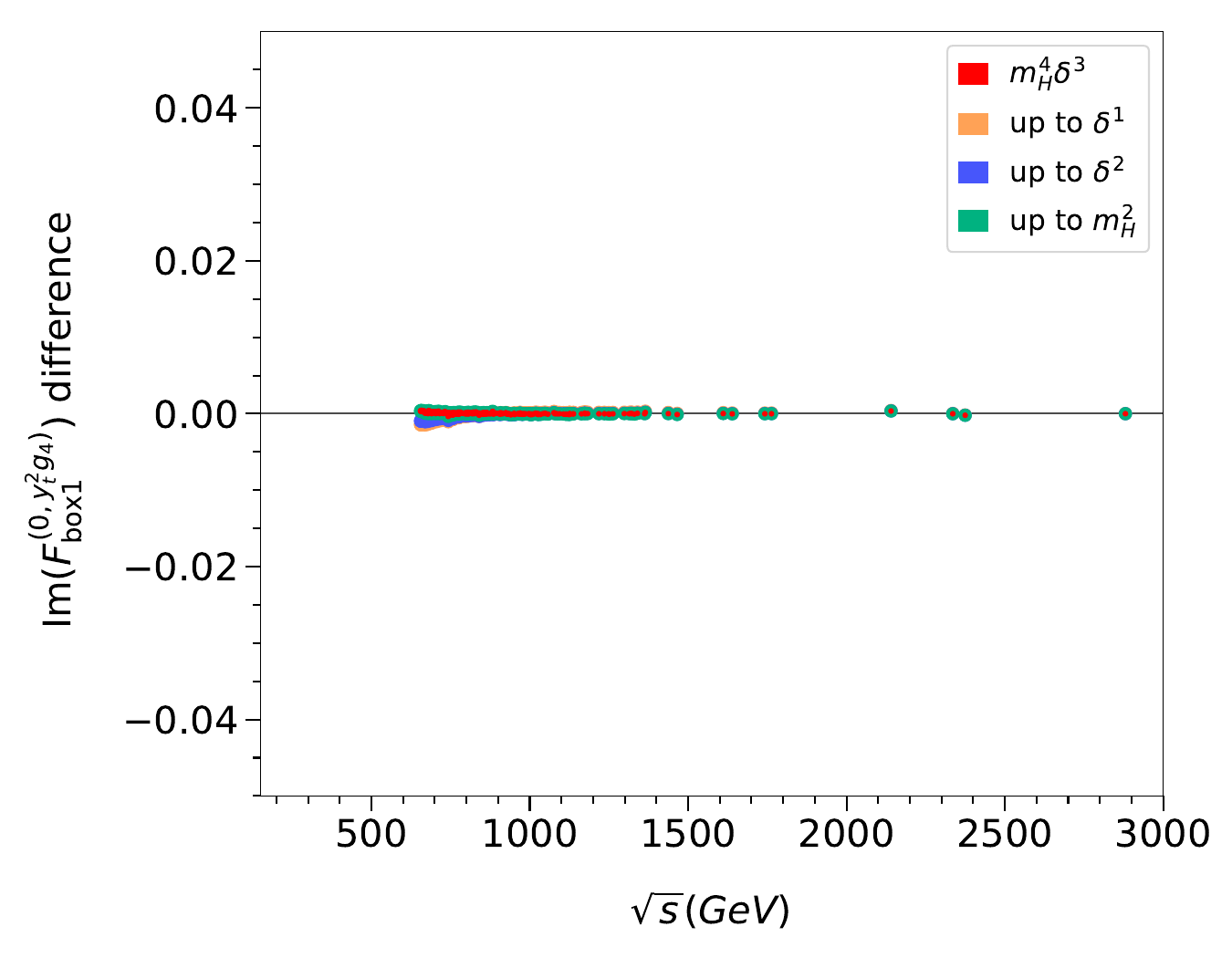}
  \includegraphics[width=0.35\textwidth]{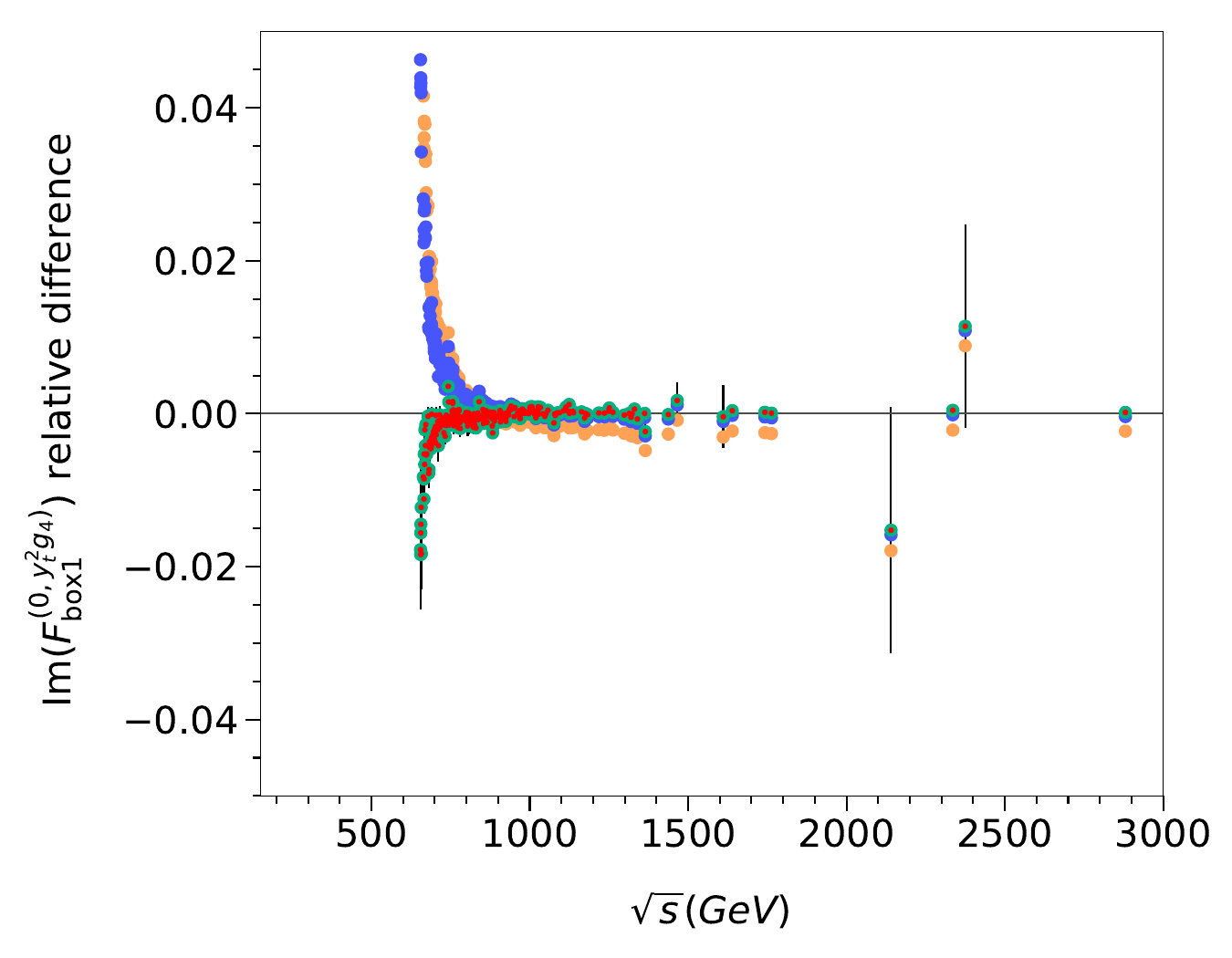}
  \end{tabular}
   \caption{\label{fig::diff_to_SD_yt2lam1}
   As Fig.~\ref{fig::diff_to_SD_yt4}, for $F_{\rm box1}^{(0,y_t^2 \, g_4)}$.
   }
\end{figure}

In Figs.~\ref{fig::diff_to_SD_yt4}-\ref{fig::diff_to_SD_yt2lam1} we compare
our results and the numerically evaluated form factors for the four two-loop
form factors of Eq.~(\ref{eq::conversionNLO}). We show the comparison both as
a function of the final-state Higgs boson transverse momentum $p_T$ and as a
function of $\sqrt{s}$.  The uncertainties displayed in the plots are due to
our uncertainties and those of Ref.~\cite{Heinrich:2024dnz}, combined in
quadrature. We show data points for values of $p_T \geq 300$~GeV, which imply
a $\sqrt{s}$ value of at least 650~GeV.  In this region our Pad\'e
uncertainties are extremely small; the uncertainties visible in the plots are
therefore almost entirely due to the numerical integration.  The exception to
this is $F_{\rm box1}^{(0,y_t^3\,g_3)}$, where our Pad\'e uncertainties
dominate for smaller values of $p_T$. In the case of the real part there are
only a few data points with uncertainties. Those with $p_T<500$~GeV and
$\sqrt{s}<1000$~GeV are dominated by the Pad\'e uncertainty. In the case of
the imaginary part all uncertainties below $p_T=470$~GeV or
$\sqrt{s}=1300$~GeV basically come from the Pad\'e approach.

In the plots we show several sets of data points. The light blue and yellow
points in the left-hand column show the form factor values for our expansions
and from Ref.~\cite{Heinrich:2024dnz}, respectively. In the central and
right-hand columns we show the differences and relative differences between
Ref.~\cite{Heinrich:2024dnz} and our expansions truncated at various orders:
the red, orange, dark blue and dark green points include terms to orders
$m_H^4\,\delta^3$, $m_H^4\,\delta^1$, $m_H^4\,\delta^2$ and $m_H^2\,\delta^3$
respectively. Thus the red points display our best approximation.

Fig.~\ref{fig::diff_to_SD_yt4} shows the comparison for the real and imaginary
finite parts of the (bare) form factor $F_{\rm box1}^{(0,y_t^4)}$. We find
good agreement between our expansions and the numerical evaluations of
Ref.~\cite{Heinrich:2024dnz}; that is, the difference or relative difference
between the data points is consistent with zero within the uncertainties and
the relative difference has a typical value below about 1\% and 0.5\% for the
real and imaginary parts, respectively.  The differences in values are both
positive and negative, suggesting agreement up to the (Gaussian distributed)
numerical integration uncertainties.

Fig.~\ref{fig::diff_to_SD_yt2lam2} shows the comparison for the real and
imaginary parts of the form factor $F_{\rm box1}^{(0,y_t^2 \, g_3^2)}$.  Here,
though the difference between our results is small in an absolute sense, the
relative difference is inflated due to the small absolute value of the form
factor and its zero crossing. Here we observe that our results are not
consistent with the numerical evaluation within the uncertainties but rather
show a systematic offset from the values of Ref.~\cite{Heinrich:2024dnz} at
a level below about 2\% or better, which we attribute to the truncation
of our expansions at $m_H^4\,\delta^3$. The comparison of the imaginary part
(lower two rows) is qualitatively rather similar, however the larger values of
the form factor itself leads to a smaller relative difference of about 1\% or
better. The comparison of the real part of $F_{\rm box1}^{(0,y_t^3 \, g_3)}$
in Fig.~\ref{fig::diff_to_SD_yt3lam1} also displays the same systematic offset
from the numerical evaluation due to the truncation of our expansion and a
broadly similar relative difference of about 2\% or better.

On the other hand, the imaginary part of $F_{\rm box1}^{(0,y_t^3 \, g_3)}$
(shown in the lower two rows of Fig.~\ref{fig::diff_to_SD_yt3lam1}) is not
well controlled by our expansion for $p_T \lesssim 500$GeV.  In contrast to
the other plots of this section, here the error bars are dominated by the
uncertainty estimate of our Pad\'e procedure.  For $p_T \gtrsim 500$GeV we
observe the same $1-2\%$-level agreement as for the other form factors.
This form factor contains contributions from the poorly-performing $I_{157}$
and $I_{164}$ master integrals discussed in Section \ref{sub::MIs}.

Finally, in Fig.~\ref{fig::diff_to_SD_yt2lam1} we compare values of
$F_{\rm box1}^{(0,y_t^2 \, g_4)}$.  The quartic self-coupling of the Higgs
boson means that this form factor has ``$2\to 1$'' kinematics and is thus
independent of $p_T$ (or $t$ and $u$). This leads to excellent agreement at
a level below $0.5\%$, since there are no kinematic
cuts which can spoil the validity of our results above $\sqrt{s} = 3\,m_t$.

While we have only discussed the $F_{\rm box1}$ form factors here for brevity,
the $F_{\rm box2}$ form factors show approximately the same level of agreement
with the numerical evaluations of Ref.~\cite{Heinrich:2024dnz}. That is,
agreement at the $1-2\%$ level for $F_{\rm box2}^{(0,y_t^4)}$ and
$F_{\rm box2}^{(0,y_t^2\,g_3^2)}$, and agreement for
$F_{\rm box2}^{(0,y_t^3 \, g_3)}$ only in the region $p_T \gtrsim 500$ GeV.

\FloatBarrier


\section{Conclusions and outlook}
\label{sec::conc}

Analytic expansions in different regions in phase space constitute a
complementary approach to purely numerical evaluations of master integrals. In
this paper we compute the high-energy expansion of the master integrals needed for
the NLO electroweak corrections to Higgs boson pair production in gluon
fusion.  First we perform Taylor expansions in the final-state Higgs mass
and in the difference of the top quark and internal-Higgs boson masses, including
three and four terms respectively.  This leads to fully-massive two-loop box
integrals where all external lines are massless.  Applying an asymptotic
expansion and exploiting the differential equations for the master integrals
we expand them for small top quark mass and compute more than 100
expansion terms. This is the main result of the paper and the results can be
downloaded from~\cite{progdata}.

The master integrals are used to compute analytic results for all box-like
form factors for $gg\to HH$ which contain Higgs and top quark propagators.  For
this subset the renormalization is particularly simple: all but one form
factors are finite; for the contribution with four Yukawa couplings, which has
already been treated in Ref.~\cite{Davies:2022ram}, only the top quark mass
has to be renormalized.

For the numerical evaluation of the form factors we apply a Pad\'e
approximation procedure which improves convergence for smaller values of $p_T$ and
$\sqrt{s}$.
For most of the form factors we obtain precise results for
$p_T\gtrsim 300$~GeV and the uncertainties remain small even for smaller values
of $p_T$, see also discussion in Refs.~\cite{Davies:2022ram,Davies:2023vmj}.
The exception is the imaginary part of $F_{\rm box1}^{(0,y_t^3g_3)}$, which is
only well approximated for $p_T \gtrsim 500$ GeV.
From the detailed comparison with
the numerical results from Ref.~\cite{Heinrich:2024dnz} we estimate that the
remaining uncertainty due to the expansion in $\delta$ and $m_H^{\rm ext}$ is
at most of the order of 1--2\%.

We mention that our analytic calculation produces fairly large expressions in
intermediate steps.  However, the final numerical evaluation, which includes
the construction of ${\cal O}(10-100)$ Pad\'e results, is very fast. This is a
crucial advantage as compared to the numerical approach, allowing the easy evaluation
of additional phase space points and the ability to adjust mass values.

The master integrals computed in this paper together with the deep high-energy
expansion of the QCD master integrals from Refs.~\cite{Davies:2018ood,Davies:2018qvx}
are sufficient to compute the complete top quark mass-induced Standard Model
electroweak corrections in the high-energy limit. Furthermore,
it will be interesting to compute the amplitude in the forward scattering
limit such that the combination of both analytic expansions
covers the whole phase space. These investigations are subject to
ongoing work.


\section*{Acknowledgments}
The work was supported by the Deutsche Forschungsgemeinschaft (DFG, German
Research Foundation) under grant 396021762 --- TRR 257 ``Particle Physics
Phenomenology after the Higgs Discovery'', the European Research Council (ERC)
under the European Union’s Horizon 2020 research and innovation programme
grant agreement 101019620 (ERC Advanced Grant TOPUP), the UZH Postdoc Grant,
grant no.~[FK-24-115], the Swiss National Science Foundation (SNSF) under
contract TMSGI2 211209, and the STFC Consolidated Grant ST/X000699/1.  We
would like to thank the authors of Ref.~\cite{Heinrich:2024dnz} for providing
the data in Sec.~\ref{sub::comp} by private communication.

\bibliographystyle{jhep}
\bibliography{inspire.bib,extra.bib}

\end{document}